\documentclass[aps,12pt,nofootinbib]{revtex4}

\usepackage[english]{babel}
\usepackage{amsmath}
\usepackage{amssymb}
\usepackage{graphicx}
\usepackage{xcolor}
\usepackage{hyperref}
\usepackage{braket}
\usepackage{ulem}

\newcommand{\be}{\begin{eqnarray}}
\newcommand{\ee}{\end{eqnarray}}
\newcommand{\ba}{\begin{array}}
\newcommand{\ea}{\end{array}}
\newcommand{\ds}{\displaystyle}

\newcommand{\pa}[1]{\left(#1\right)}
\newcommand{\paq}[1]{\left[#1\right]}
\newcommand{\pag}[1]{\left\{#1\right\}}

\makeatletter\AtBeginDocument{\let\@elt\relax}\makeatother

\begin{document}
\normalem

\title{Luminosity distance uncertainties from gravitational wave detections of binary neutron stars by third generation observatories}

\author{Josiel Mendon\c{c}a Soares de Souza}
\affiliation{Departamento de F\'\i sica Te\'orica e Experimental, Universidade Federal do Rio Grande do Norte, Natal-RN 59072-970, Brazil}
\email{josiel.mendonca.064@ufrn.edu.br}

\author{Riccardo Sturani}
\affiliation{Instituto de F\'\i sica Te\'orica, UNESP-Universidade Estadual Paulista
  \& ICTP South American Institute for Fundamental Research,
  S\~ao Paulo 01140-070, SP, Brazil}
\email{riccardo.sturani@unesp.br}

\begin{abstract}
  A new generation of terrestrial gravitational wave detectors is
  currently being planned for the next decade, and it is expected to detect
  most of the coalescences of compact objects in the universe with masses up to
  a thousand times the solar mass.
  Among the several possible applications of current and future detections, we
  focus on the impact
  on the measure of the luminosity distance of the sources, which is an invaluable
  tool for constraining the cosmic expansion history of the universe.\\
  We study two specific detector topologies, triangular and $L$-shaped,
  by investigating how topology and relative orientation of up to three
  detectors can minimize the uncertainty measure of the luminosity distance.
  While the precision in distance measurement is correlated with several
  geometric angles determining the source position and orientation, focusing
  on bright standard sirens and assuming redshift to be
    measured
  with high accuracy, we obtain
  analytic and numerical results for its uncertainty depending on
  type and number of detectors composing a network, as well as on the inclination angle
  of the binary plane with respect to the wave propagation direction.
  We also analyze the best relative location and orientation of two third
  generation detectors to minimize luminosity distance uncertainty, showing that
  prior knowledge of the inclination
  angle distribution plays an important role in precision recovery of luminosity
  distance, and that a suitably arranged network of detectors can reduce
  drastically the uncertainty measure, approaching the limit imposed by lensing
  effects intervening between source and detector at redshift $z \gtrsim 0.7$.
\end{abstract}


\maketitle

\section{Introduction}
\label{sec:intro}

While still in its infancy, gravitational wave (GW) astronomy
is already providing observational data
\cite{LIGOScientific:2018mvr,LIGOScientific:2020ibl,LIGOScientific:2021djp}
of invaluable importance also for testing the fundamental nature of gravity
\cite{LIGOScientific:2020tif,LIGOScientific:2021sio}
and cosmology
\cite{LIGOScientific:2017adf,LIGOScientific:2019zcs,LIGOScientific:2021aug}.

Second generation (2G) detectors LIGO \cite{TheLIGOScientific:2014jea} and Virgo
\cite{TheVirgo:2014hva} collected signals
from coalescing binaries at the rate of $O(1)$ event per week during their third
observation run, and a similar or large rate is expected for future observations
runs to happen presently or in the near future \cite{KAGRA:2013rdx}, when the Japanese
detector KAGRA will also be part of the observational network \cite{KAGRA:2020tym}.

The GW detections from coalescing binaries impacted on several fields in
physics, and in this work we focus on the determination of the
\emph{luminosity distance} of their sources, which is a crucial
ingredient to reconstruct the cosmic expansion history.

As well known \cite{Schutz:1986gp,Holz:2005df}, coalescing binaries are
\emph{standard sirens}, i.e. their characteristic chirp signal enables an
absolute calibration of the gravitational luminosity leading to an unbiased
determination of the luminosity distance, which, together with
redshift, are the two observables necessary to
determine the cosmic expansion history of the universe.
However redshift is in general not provided by GW detections,
but it can be obtained by the host galaxy identification, which is possible for
electromagnetically (EM) \emph{bright} standard sirens.
The most likely case, even though not exclusive, see e.g. \cite{10.1111/j.1365-2966.2012.21486.x,Bartos_2017,PhysRevLett.124.251102,McKernan_2019,Kimura:2021xxu,Palmese:2021wcv,East:2021spd}, of
GW signal accompanied by an EM counterpart is given by binary neutron star
systems with mass ratio close to unity, for which tidal forces are larger
\cite{Flanagan:2007ix}\footnote{Tidal forces in the
  final stage of the inspiral they are inversely proportional to the (square)
  mass of the object sourcing the tidal field.}.
In the case the neutron star is tidally disrupted outside the effective
innermost stable orbit of
the binary, material is ejected from the system and it is expected to
produce an EM counterpart \cite{East:2021spd}, in particular a
short gamma ray burst and a kilonova, beside
lower frequency emission lasting up to years
\cite{Metzger:2011bv,East:2021spd}.

For reference, GW-determined sky localisation areas encompassed $\Delta\Omega$
of, say, $\sim 16-{\rm few}\times 10^3$ deg$^2$
in recent detections \cite{LIGOScientific:2018mvr,LIGOScientific:2020ibl,LIGOScientific:2021djp}.
Up to a distance of $100$ Mpc the number of galaxies $N_{100}$ included in a
sky region of area $\Delta\Omega$ is roughly 
$N_{100}\sim 5\times \Delta\Omega/(10\, {\rm deg}^2)$
assuming the average milky-way-like galaxy density to be
$\sim 5\times 10^{-3}{\rm Mpc}^{-3}$ \cite{Dalya:2018cnd}.

Events collected so far by 2G detectors show an overwhelming
majority of binary black holes over binary systems involving at least one
neutron star. In only one case, the notable binary neutron star (BNS)
system that
sourced GW170817 \cite{LIGOScientific:2017vwq}, the GW signal has been
accompanied by EM counterparts, with consequent host galaxy identification
and redshift determination.
While several methods have been proposed and used to get redshift information
from EM-dark GW detections, using e.g.~statistical identification
of host galaxy \cite{DelPozzo:2011vcw,DES:2019ccw,Gray:2019ksv},
full cross-correlating with galaxy catalogues \cite{Mukherjee:2020hyn,Diaz:2021pem,LIGOScientific:2021aug},
statistical redshift distribution of sources \cite{Zhu:2021aat,Leandro:2021qlc},
features in the black hole mass spectrum \cite{Ezquiaga:2020tns},
neutron star tidal effects imprinted in the waveform \cite{Messenger:2011gi},
the golden events having the individually highest constraining power over
the cosmic expansion history are EM-bright standard sirens,
where host galaxy identification enable redshift determination with negligible
error \cite{Chen:2017rfc}.

\begin{figure}
  \begin{center}
    \includegraphics[width=.6\linewidth]{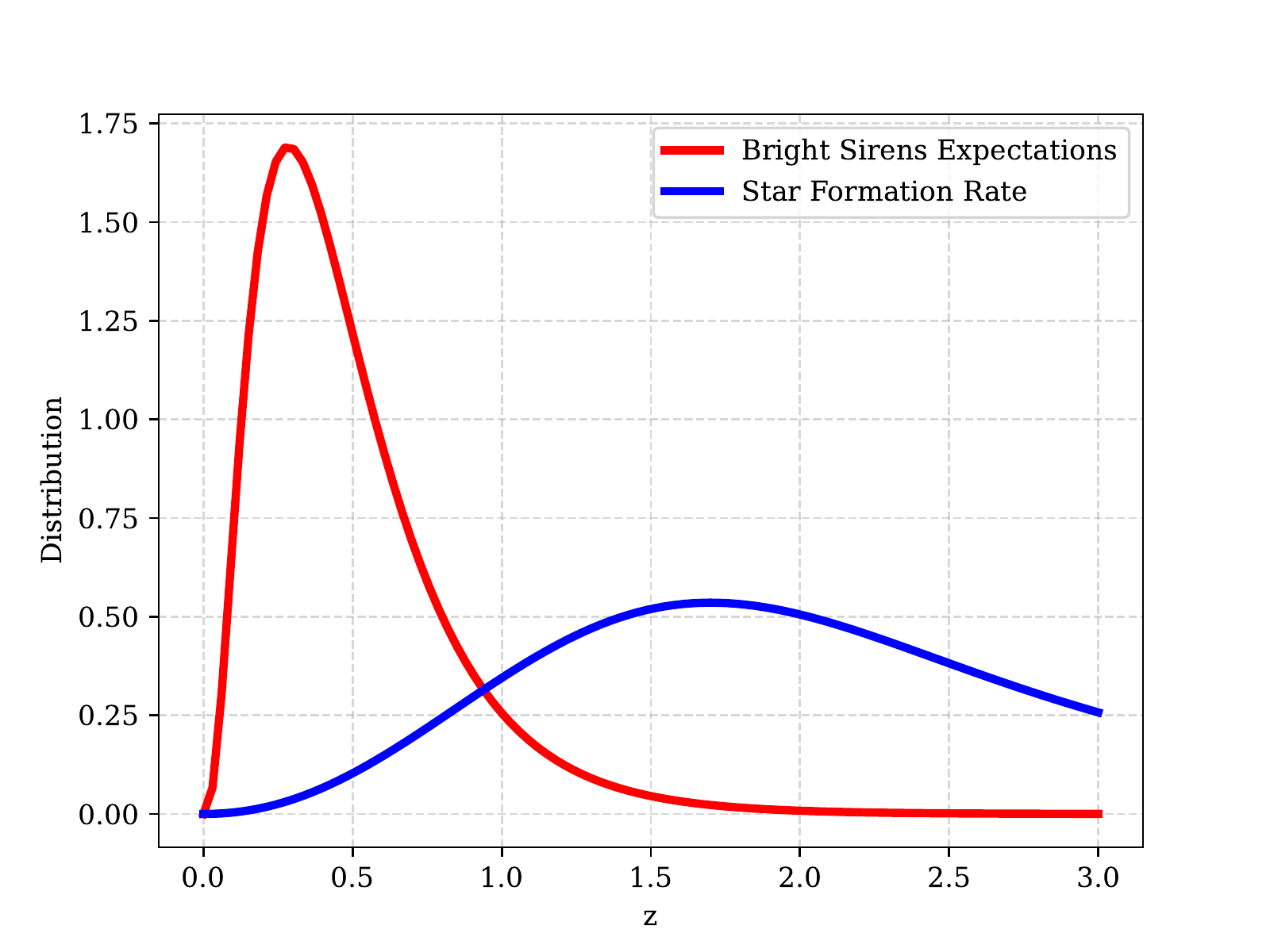}
    \caption{Expected redshift distribution of bright standard sirens
      assuming an electromagnetic counterpart is detected by Theseus
      \cite{Belgacem:2019tbw}, compared with observed star formation rate
      \cite{Madau:2014bja}.}
    \label{fig:histo_embsc}
  \end{center}
\end{figure}

For third generation (3G) detectors most of the BNS coalescences
will be visible \cite{Regimbau:2016ike}, but EM counterparts are
not expected to be detected beyond a limiting redshift $\bar z\simeq 0.7$, with
the bulk of the distribution of bright standard sirens expected around
$z\sim 0.3-0.4$ \cite{Belgacem:2019tbw,deSouza:2021xtg}, making most of them
invisible to 2G detectors \cite{Vitale:2018nif}, see Figure \ref{fig:histo_embsc}.

\begin{figure}
  \begin{center}
    \includegraphics[width=.49\linewidth]{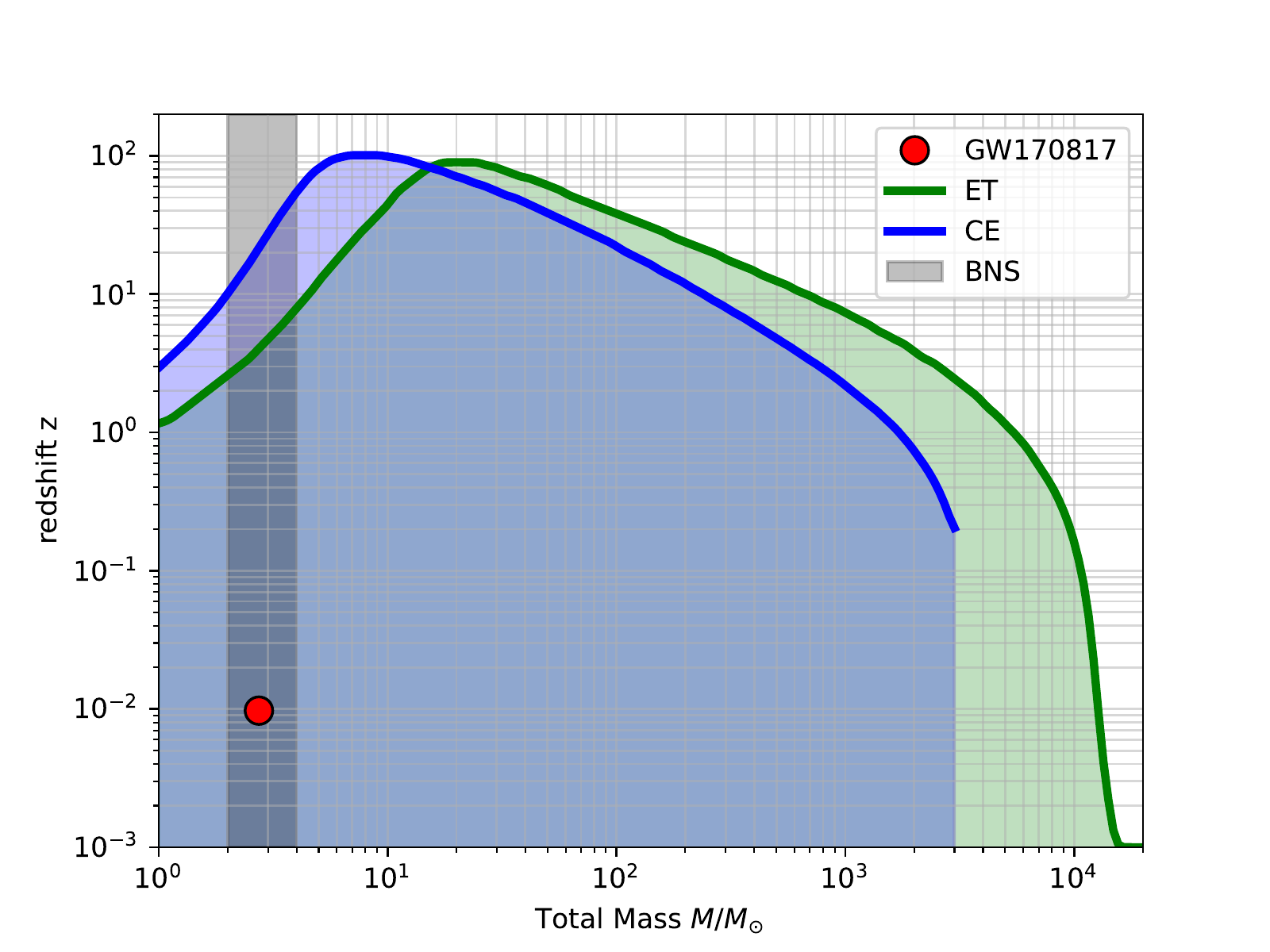}
    \includegraphics[width=.49\linewidth]{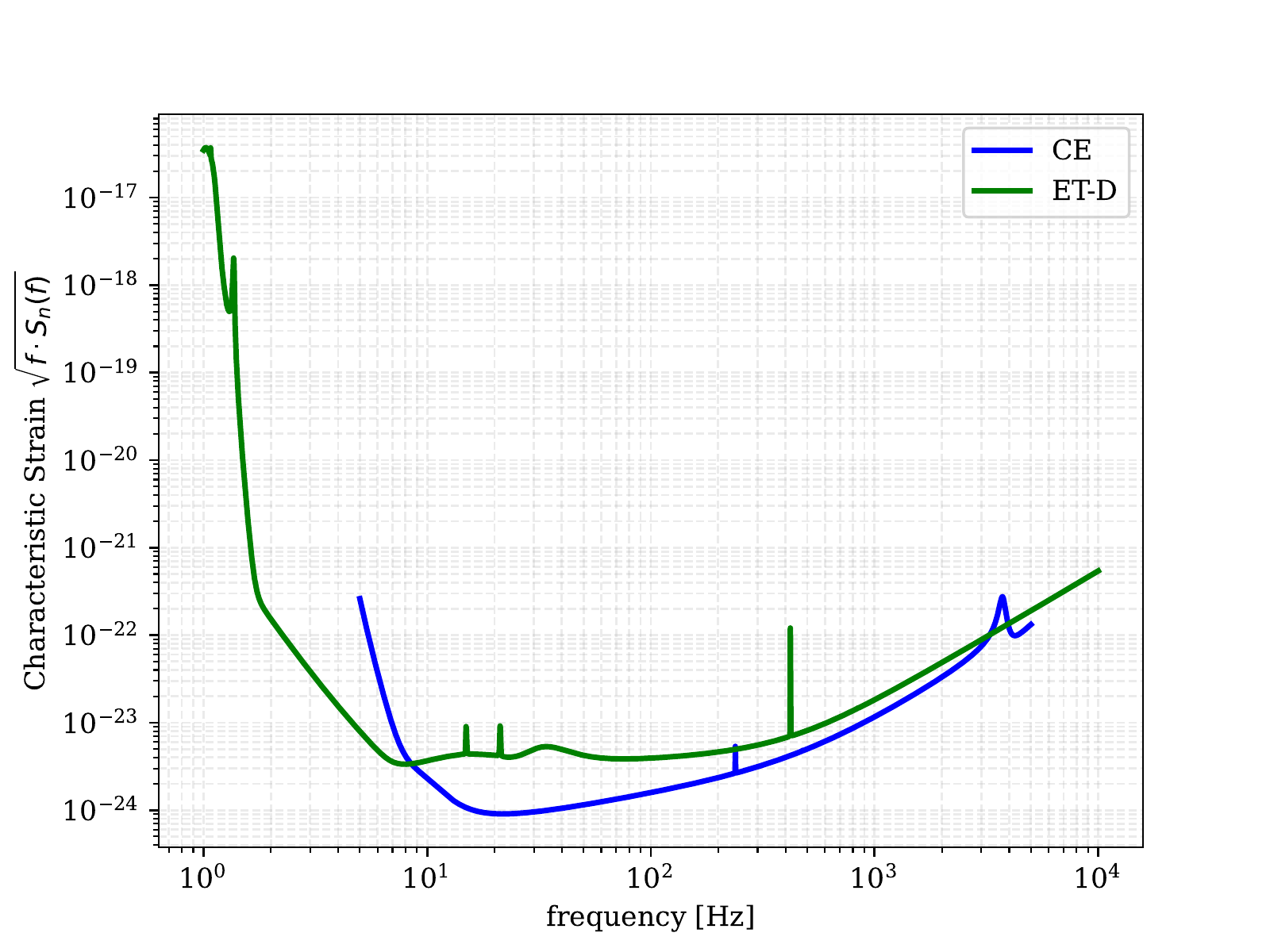}
    \caption{(Left) Luminosity distance reach for equal mass, non-spinning system, assuming
      fundamental $l=m=2$ mode only, for optimally oriented binaries, given
      the spectral noise density $S_n$ \cite{Srivastava:2022slt,ce_official} for CE and
      [ET-D] from \cite{etd} for ET. The mass and luminosity distance of
      GW170817 are highlighted, as well as the mass region where BNSs are
      expected. (Right) Dimensionless noise characteristic strain $h_c$, defined
      in terms of spectral noise $h_c\equiv \sqrt{f S_n}$ for $L$-shaped CE and triangle-shaped ET.}
    \label{fig:reach3G}
  \end{center}
\end{figure}

3G detectors are currently under active research and development, and in the
present work we will assume that the Einstein Telescope (ET) will be
a triangular interferometer with arms at $60^o$ degrees \cite{Punturo:2010zz},
and Cosmic Explorer (CE) a single $L$-shaped
interferometer \cite{Evans:2021gyd} with arms at $90^o$, with dimensionless
noise characteristic strain $h_c=\sqrt{fS_n}$
displayed in Figure \ref{fig:reach3G}, $S_n$ being the standard
single sided noise spectral density, see e.g. ch.7 of \cite{maggiore2008gravitational}, while analogue quantities
for 2G detectors can be found in \cite{Chen:2017wpg}.

Within this context we investigate the relative configuration of ET-like and
CE-like detectors that maximizes the accuracy on the luminosity distance
determination of binary neutron stars, taking into account statistical
features of the sources like inclination angle distribution.

The paper is organised as follows. In Section \ref{sec:method} we lay out the
tools employed to quantitatively analyze luminosity distance measures
in EM-bright GW detections by the 3G detectors mentioned above, with the
results presented in Section \ref{sec:results}.
Section \ref{sec:conclusions} contains the conclusions that can be drawn
from our study.

\section{Method}
\label{sec:method}

\subsection{Basics}
Gravitational radiation in general relativity are endowed with two polarisations,
conventionally indicated by $h_+$ and $h_\times$, which can be suitably
decomposed into modes using spin-weighted spherical harmonics ${}^{-s}Y_{lm}$ of
weight $s=-2$, according to
\be
\label{eq:hlms}
h_+-ih_\times=\frac {GM}r\sum_{l\geq 2}\sum_{|m|\leq l}H_{lm}{}^{-2}Y_{lm}(\iota,\phi)\,,
\ee
where Newton constant $G$, total rest mass $M$ of the source, and coordinate
distance of the source to the observer $r$ have been factored out.
The luminosity distance $d_L$ is related to $r$ and the redshift $z$
via $d_L=(1+z)r$.
Applying Equation (\ref{eq:hlms}) to the case of a binary system, $\iota$
denotes the angle between the unit vector perpendicular to the binary
plane $\hat L$ and the radiation direction parameterized by the unit vector
$\hat N$, $\phi$ parameterizes a rotation in the binary plane.

The expansion coefficients $H_{lm}$ are complex functions of the
intrinsic parameters and the retarded time $t-r$.
Detector $d_i$ output contain GW signals $h_{d_i}$ which are linear combinations
of the GW polarisations weighted by the \emph{pattern functions} $F_{+,\times}$
\be
\label{eq:hdi}
h_{d_i}=F_+(\alpha_i,\beta_i,\psi_i)h_++F_\times(\alpha_i,\beta_i,\psi_i) h_\times\,,
\ee
where $\alpha_i,\beta_i$ are detector dependent angles determined by the sky
position of the source,
and the polarisation angles $\psi_i$ can be interpreted
as the additional angle (beside $\iota$) relating $\hat L$ to $\hat N$
\cite{Apostolatos:1994mx}.
Together $\psi_i$, $\iota$, and $\phi$ compose the \emph{Euler angles} determining the
relative orientation between \emph{source frame} (defined by the orbital plane
and its normal $\hat L$) and the \emph{radiation frame} whose
$\hat z_{rad}$ axis is the unit vector $\hat N$, and whose
$\hat x_{rad}-\hat z_{rad}$ plane contains the normal to the detector's plane
$\hat z_i$, see Appendix \ref{app:psi} for detailed definition and properties of
the polarisation angle.

\begin{figure}[ht]
  \includegraphics[width=.48\linewidth]{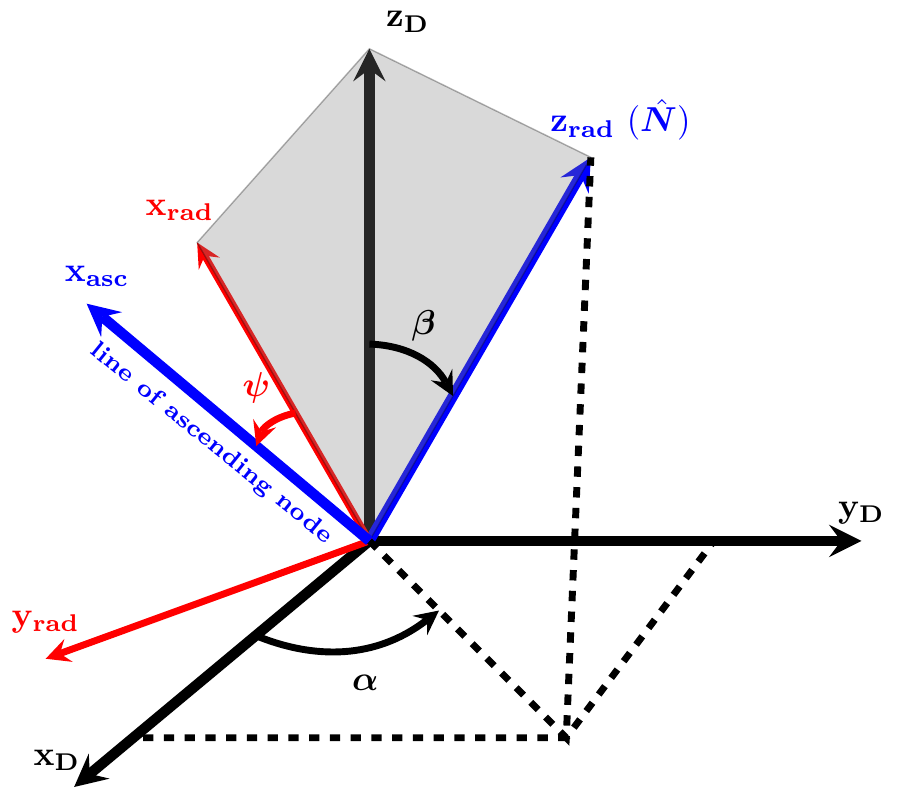}
  \includegraphics[width=.48\linewidth]{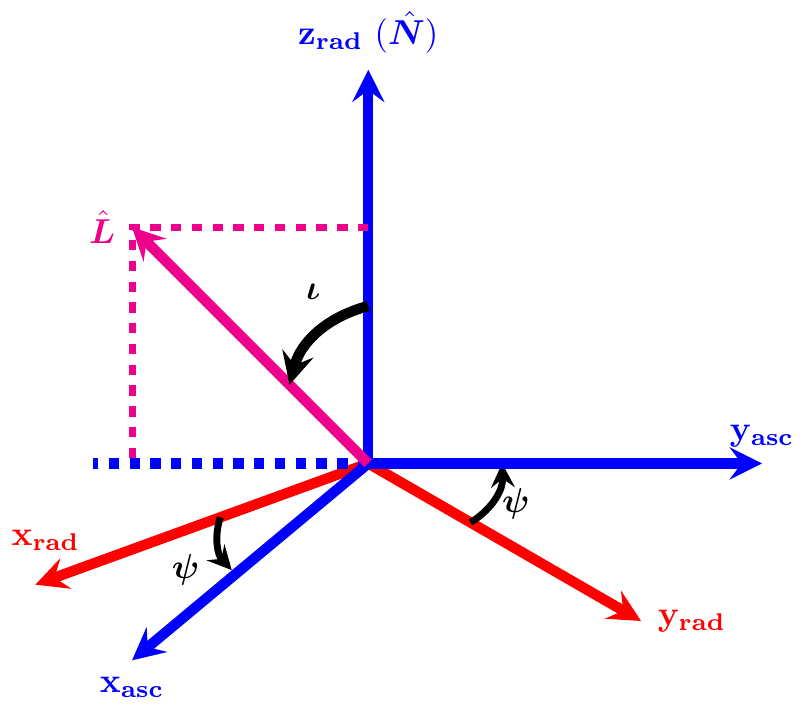}
  \caption{Schematic representation of detector geometry and of radiation frame.}
  \label{fig:det_angles}
\end{figure}

The pattern functions $F_{+,\times}$ can be written as
\be
\label{eq:Fpc}
\ba{rcl}
F_+(\alpha_i,\beta_i,\psi_i)&=&\ds \cos(2\psi_i)f_+(\alpha_i,\beta_i)-\sin(2\psi_i)
f_\times(\alpha_i,\beta_i)\,,\\
F_\times(\alpha_i,\beta_i,\psi_i)&=&\ds \cos(2\psi_i)f_\times(\alpha_i,\beta_i)+\sin(2\psi_i)f_+(\alpha_i,\beta_i)\,,
\ea
\ee
with $f_{+,\times}$ defined as
\be
\label{eq:fpc}
\ba{rcl}
f_+(\alpha_i,\beta_i)&\equiv&\ds-\sin(\Omega)\frac 12\pa{1+\cos^2\beta_i}\sin(2\alpha_i)\,,\\
f_\times(\alpha_i,\beta_i)&\equiv&-\sin(\Omega)\cos\beta_i\cos(2\alpha_i)\,,
\ea
\ee
where $\beta_i$ is (the complement of) the source elevation and the azimuth angle
  $\alpha_i$ is measured with respect to the bisector of the angle formed by the interferometer's arms.
The pattern functions (\ref{eq:fpc}) can be obtained by projecting the gravitational perturbation tensor onto the
interferometer response tensor $\frac 12\pa{u^iu^j-v^iv^j}$, begin
$\hat u, \hat v$ the unit vector pointing along the detector's arms, 
and we allow the possibility of a variable opening angle
$\Omega$ between them, see Figure \ref{fig:det_angles}.\footnote{We understand
  the $\Omega$ dependence in the notation of $f_{+,\times}$. For all applications we
  will use $\Omega=\pi/2$ for $L$-shaped (CE-like) interferometers and $\Omega=\pi/3$ for triangle-shaped (ET-like) ones.}

As per standard treatment, the detectors' output $h_{d_i}$ are processed via \emph{matched-filtering}
\cite{vainshtein1970extraction}, which consists in taking a noise-weighted
correlation of the data with a pre-computed waveform model,
or \emph{template} $h_{t}$, according to\footnote{We adopt the convention
$\tilde g(f)=\int g(t)e^{2i\pi ft}dt$.}
  
\be
\label{eq:mf}
\langle h_{d_i},h_{t_i}\rangle(t_i)\equiv 2\int_0^\infty \frac{
  \paq{\tilde h_{d_i}(f)\tilde h^*_{t_i}(f)e^{-2i \pi ft_i}+
    \tilde h^*_{d_i}(f)\tilde h_{t_i}(f)e^{2i \pi ft_i}}}{S_{n_i}(f)}df\,,
\ee
whose output is the time-dependent correlation between data and the specific
template $h_{t_i}$ translated in time by the detector dependent quantity
$t_i$. $S_n(f)$ is the noise spectral
density defined in term of detector noise $\tilde n(f)$ averaged over many realizations
\be
\langle \tilde n(f)\tilde n(f')\rangle=\frac 12S_n(f)\delta(f+f')\,.
\ee
The correlation in Equation (\ref{eq:mf}) can be used to define a scalar product
$\braket{h_1|h_2}\equiv \langle h_1,h_2\rangle(0)$ and consequently
a norm $\|h\|\equiv\sqrt{\braket{h|h}}$.

Searches for maximum matched-filtering output by varying the templates lead to
the determination of the best fit waveform and, in a Bayesian inference scheme,
to probability distribution functions for all waveform template parameters.
Note that the binary constituent masses $m_{i}$ that can be recovered are the
so-called \emph{redshifted}, or \emph{detector} ones,
related to intrinsic, or source ones $m_i^{(s)}$ via $m_i=m_i^{(s)}(1+z)$\cite{Schutz:1986gp}.
Optimal matched-filtering leads to the definition of signal-to-noise ratio
of signal $h$ ($SNR_h$):
\be
\label{eq:snr2}
SNR^2_h\equiv \langle h|h\rangle=4\int^\infty_0 \frac{|\tilde h(f)|^2}{S_n(f)}df\,.
\ee

Focusing on the fundamental mode, i.e. taking the contribution only from
$l=2=|m|$ in (\ref{eq:hlms}), the signal from the inspiral phase admits
a simple analytic description:
\be
\label{eq:hpc22}
\ba{rcl}
\ds \tilde h_+&=&\ds \frac{\pa{1+\cos^2\iota}}2h_0(f)e^{i\Phi_{gw}(f)}\,,\\
\ds \tilde h_\times&=&\ds i\cos\iota\, h_0(f)e^{i\Phi_{gw}(f)}\,,
\ea
\ee
where for $h_0(f)$ the analytic expression is known analytically for the
inspiral in the \emph{stationary phase approximation}
\cite{Finn:1992xs}
\be
h_{0insp}(f)\equiv\pa{\frac 5{24}}^{1/2} \pi^{-2/3}\frac{\pa{GM_c}^{5/6}f^{-7/6}}{d_L}
\,,
\ee
being $M\equiv m_1+m_2$, $M_c\equiv \nu^{3/5}M$, $\nu\equiv m_1m_2/M^2$.
Similarly the $f$-domain phase $\Phi_{gw}(f)$ has a well known
analytic, perturbative representation for the \emph{inspiral} phase of the $i$-th detector: 
\be
\label{eq:psif}
\Phi_{gw_i-insp}(f)-2\pi f t_i+\phi_0+\frac \pi 4\simeq \frac 3{128\nu v^5}\paq{1+O(v^2)}=\frac 3{128\pa{\pi G M_cf}^{5/3}}\paq{1+O(v^2)}\,,
\ee
where the quantity $(\Phi_{gw_i-insp}-2\pi ft_i)$ is independent of the
detector ($t_i$ is the arrival time at the $i$-th detector), the small
parameter of expansion in Equation (\ref{eq:psif}) is $v\equiv (\pi GMf)^{1/3}$,
and $\phi_0$ a constant phase.

Out of the 15 parameters needed to determine the signal imprinted into a
detector by a GW source made of a binary system in circular motion, see Tab.~\ref{tab:pars},
we are interested in the precision of luminosity distance determination of EM
bright standard sirens, for which we assume that sky position, and consequently
host galaxy and then redshift, can be determined with
negligible uncertainty.

\begin{table}[t]
  \begin{tabular}{||c|c||}
    \hline
    Intrinsic parameters & Extrinsic parameters\\
    \hline
    $M_c$, $\nu$, $\vec S_1$, $\vec S_2$ &
    ${\bf d_L}$, $\boldsymbol{\psi}$, $\boldsymbol{\iota}$, ${\color{gray} \phi}$, $\alpha$, $\beta$, $t$\\
    \hline
  \end{tabular}    
  \caption{Parameters defining the observation of a binary system observation
    (whose constituents
    are treated as point-like object),  divided between intrinsic ($\vec S_{1,2}$
    are the binary constituent spin vectors) and extrinsic ($t$ denotes the arrival time),
    according to the distinction introduced in \cite{Owen:1995tm}.
    In bold are those we searched over via Bayesian inference
    ($d_L,\psi,\iota$), in grey the one marginalised over
    (${\color{gray}\phi}$).}
  \label{tab:pars}
\end{table}

As explained in the introduction, we focus our analysis on equal
  mass binary neutron stars, which are obvious candidates (even if not exclusive) to produce GW signals with EM counterpart.
We will make the additional simplifying assumptions that binary constituent spins
can be neglected, as neutron stars in binaries are observed to have in general
negligible spins, with values $\lesssim 0.05 m_i^2$ \cite{Burgay:2003jj}.
The remaining intrinsic parameters, the individual masses, are expected to be
measured with sub-percent accuracy, as it happened for GW170817
\cite{LIGOScientific:2017vwq},\footnote{For reference, the luminosity distance of GW170817 has been measured
  with $\sim 20\%$ accuracy and it had a network $SNR_{net}$, i.e.~the $SNR$
  summed over three detectors of $SNR_{net}=(\sum_{i=1}^3SNR^2_{d_i})^{1/2}\sim 32$.}
and as confirmed in general
by Fisher matrix analysis for 3G detectors \cite{Iacovelli:2022bbs}\footnote{
  In particular in Fig.~13 of \cite{Iacovelli:2022bbs} it is shown that about
  99\% of binary neutron star detections by both 2G and 3G detectors will have individual masses accuracy below 0.1\%.}.

Matched-filtering technique can provide a very accurate
determination in general of the intrinsic parameters, in particular the chirp
mass $M_c$ can be determined with accuracy \cite{Flanagan:1997kp}
\be
\frac{\delta M_c}{M_c}\sim \frac 1{N_{cyc}}\times \frac{10}{SNR}\,,
\ee
where beside the factor $SNR^{-1}$ usually obtained in a Fisher matrix
approximation, one has a $N^{-1}_{cyc}$ uncertainty decrease with increasing
number of observed cycles $N_{cyc}$ from the phase $\Phi_{gw}$ dependence on $M_c$,
and the numerical factor $\sim 10$ is due to correlation with the other mass
parameter $\nu$ which enters Equation (\ref{eq:psif}) beyond leading order.

We further neglect in our Bayesian inference search the arrival time $t_i$,
which is usually obtained with $\sim msec$ accuracy \cite{LIGOScientific:2020ibl}
and concentrate on the parameters that have larger correlation with the
luminosity distance, hence have stronger impact on its recovery value precision.

Note that also calibration errors can affect the measure of luminosity distance,
  and a proxy for the threshold at which relative calibration error $\Delta_c$
  become comparable with statistical ones is $\Delta_c\sim 1/SNR$ \cite{Goncharov:2022dgl}.
  While GW signals from binary neutron stars can reach $SNR$ of $10^3$ in 3G
  detectors \cite{2019CQGra..36v5002H},
the bulk of their distribution will lie at $SNR\lesssim{\rm few}\times 10$
\cite{Iacovelli:2022bbs}. For this reason, projecting to 3G detectors the
Advanced LIGO systematic calibration error in the first half of the third observation run, estimated $< 2\%$ \cite{Sun:2020wke},
one can assume that calibration uncertainty should not affect the overwhelming
majority of signals we are discussing.

According to Equation (\ref{eq:hdi}), detectors with different orientations
measure different combinations of $h_+$ and $h_\times$, hence in principle with two or
more detectors it is possible to disentangle the $\iota-d_L$ degeneracy.
However the two LIGOs are oriented to have very similar pattern functions (apart from a sign) \cite{LIGOScientific:2003vzz}
and in the GW170817 case very little SNR was present in the remaining detector
of the network, Virgo \cite{LIGOScientific:2017vwq}.

Another element that can break the degeneracy is the presence in the signal of a significant contribution
from sub-dominant modes with $l>2$, which are weighted by different function of $\iota$ than the
$l=2$, $|m|=2$ mode determining Equation (\ref{eq:hpc22}).
However sub-dominant modes are not expected to be seen in GWs emitted by
binary neutron stars, whose part of the signal visible in the detectors is in the inspiral phase\footnote{For
  reference, the inner most stable circular orbit for a spin-less, equal mass
  binary system, corresponds to a GW frequency
  $f_{GW}\simeq 730{\rm Hz}\paq{M/\pa{3M_\odot}}^{-1}$.} for which subdominant modes are suppressed by powers of the relative binary constituent velocity as
$v^{l-2}$.
Moreover odd higher modes vanish in the limit $m_1=m_2$, hence they are suppressed
for comparable masses and in general sub-dominant modes are more important
for edge-on viewing angles, i.e.~$\iota\sim \pi/2$ \cite{Varma:2014jxa}.
We will come back on source $\iota$ distribution in Subsection\ref{ssec:impct_iotad}.

\subsection{Cutler-Flanagan parameterization}

Crucial roles to estimate the luminosity distance measurement uncertainty
are played by detector topology, location and orientation and we find
convenient to investigate this issue using the parameterization of the signal
introduced in \cite{Cutler:1994ys}. For a single detector $d_i$, denoting
$\upsilon\equiv \cos\iota$, $\chi_+\equiv (1+\upsilon^2)/2$
(and $\chi_\times\equiv -i\upsilon$ for future reference) one can write the
$SNR$ as
\be
\label{eq:cf1}
\ba{rcl}
SNR_i^2&=&\ds 2\pa{f_{+i}^2+f_{\times i}^2}
\Big\{\pa{\chi_+^2+v^2}+\pa{\chi_+^2-v^2}\cos\paq{4\pa{\psi_i+\bar\psi_i}}
\Big\}\int_0^\infty \frac{|h_0(f)|^2}{S_{n_i}(f)}df\\
&=&\ds 2\pa{f_{+i}^2+f_{\times i}^2}\pa{\chi_+^2+v^2}
\paq{1+f(v)\cos\pa{4\pa{\psi_0+\delta\psi_i+\bar\psi_i}}}
\int_0^\infty \frac{|h_0(f)|^2}{S_{n_i}(f)}df\,,
\ea
\ee
where we have used the detector's output Equation (\ref{eq:hdi}) in
the $SNR$ expression (\ref{eq:snr2}).
The newly introduced quantity $\bar\psi_i$ is defined via
\be
\label{eq:bpsii}
\cos\pa{4\bar\psi_i}\equiv
\frac{f_{+ i}^2-f_{\times i}^2}{f_{+ i}^2+f_{\times i}^2}\,,\qquad
\sin\pa{4\bar\psi_i}\equiv
\frac{2f_{+ i}f_{\times i}}{f_{+i}^2+f_{\times i}^2}\,,
\ee
and in the second line of Equation (\ref{eq:cf1}) we have written
$\psi_i=\psi_0+\delta\psi_i$, where $\psi_0$ is the polarisation angle relative
to the earth north pole unit vector $\hat z_0$ (i.e. using $\hat z_0$ for $\hat
z_i$ in Equation (\ref{eq:pol}), $\delta\psi_i$ being defined as a consequence).
Finally we adopted the notation
\be
f(v)\equiv \frac{\chi_+^2-v^2}{\chi_+^2+v^2}=\frac{\pa{1-v^2}^2}{1+6v^2+v^4}\,.
\ee
The pattern functions $f_{+i,\times i}$ depend on detector location via their
arguments $(\alpha_i,\beta_i)$ as per Equation (\ref{eq:fpc}), $\delta\psi_i$
depends on the source location with respect to the detector but it
has the non-trivial property of not depending on the polarisation angle, see
Appendix \ref{app:psi}.

The main advantage of the $SNR$ formulation in Equation (\ref{eq:cf1}) is that it
separates the contribution to the $SNR$ in a part that depends on the
polarisation angle $\psi$ and a part that is $\psi$-independent. In general
it is difficult to recover $\psi$ and its uncertainty
affects the measure of the $SNR$, see Equation (\ref{eq:cfn}),
hence jeopardizing the precision of $d_L$.

Given that the polarisation angle parameterizes rotations in the $+,\times$
space, the signal in each detector can be elegantly written in terms of
quadratic forms via
\be
\label{eq:hdi_CF}
\ba{rcl}
\ds\tilde h_{d_i}&=&\ds
h_0(f) e^{i\Phi_{gw_i}}V_A(\upsilon)R_{AB}\pa{2\psi_0}R_{BC}\pa{2\delta\psi_i}f_{Ci}\\
&=&\ds h_0(f)\mathcal{A}_AR_{AB}\pa{2\delta\psi_i}f_{Bi}e^{2\pi i ft_i}
\ea
\ee
where capital Latin indices $A,B,C$ run over $\pag{+,\times}$, $R_{AB}$ is the
standard $2\times 2$ rotation matrix
\be
\label{eq:rpsi}
R_{AB}(\alpha)\equiv\pa{\ba{cc}
\cos\alpha & -\sin\alpha\\
\sin\alpha & \cos\alpha
\ea}\,,
\ee
we have collected in a 2-vector the dependence of the GW polarisations on
$\iota$, i.e. $V_A(\upsilon)\equiv \pa{\chi_+,\chi_\times}$,
and in the second line of Equation (\ref{eq:hdi_CF}) we have defined the
detector independent quantity
$\mathcal{A}_A(\upsilon,\psi_0)\equiv V_B(\upsilon)R_{BA}(2\psi_0)e^{i\Phi_{gw}}$.
The rationale of this parameterization is to separate parameters which
  for a given signal are common to all detectors ($h_0(f)\cal{A}_A$),
  from those depending on the detector$R_{AB}(2\delta\psi_i) f_{Bi}e^{2\pi ift_i}$.

Following \cite{Flanagan:1997kp}, it is possible to generalize the $SNR$
parameterization of Equation (\ref{eq:cf1}) to the case of a network made of $n_{det}$
detectors
\be
\label{eq:snr_multi}
\ba{rcl}
\ds SNR^2_{net}&\equiv&\ds \sum_{i=1}^{n_{det}}SNR^2_i\\
&=&\ds SNR_0^2\mathcal{R}eal[\mathcal{A}_A^*(\upsilon,\psi_0)\mathcal{A}_B(\upsilon,\psi_0)]
\sum_{i=1}^{n_{det}} \Xi_{ABi}\,,
\ea
\ee
with
\renewcommand{\arraystretch}{1.4}
\be
\label{eq:xidefs}
\ba{rcl}
      \ds\Xi_{ABi}&\equiv&\ds R_{AC}(2\delta\psi_i)R_{BD}(2\delta\psi_i)f_{Ci}f_{Di}
      \omega_i\,,\\
      \ds\omega_i&\equiv&\ds \frac{\int_0^\infty |h_0(f)|^2S_{n_i}^{-1}(f) df}
      {\int_0^\infty |h_0(f)|^2S_{n_{avg}}^{-1}(f)df}\,,\\
      \ds SNR^2_0&\equiv&\ds 4\int_0^\infty df \frac{|h_0(f)|^2}{S_{n_{avg}}(f)}\,,\\
      \ds S_{n_{avg}}^{-1}(f)&\equiv&\ds\frac 1{n_{det}}\sum_i S_{ni}^{-1}(f)\,.
      \ea
\ee
\renewcommand{\arraystretch}{1.}
One can then define a symmetric $2\times 2$ matrix $\Xi_{AB}$, which can be
diagonalized by a suitable rotation matrix of the type (\ref{eq:rpsi})
with angle $\bar\psi$, as:
\be
\label{eq:xi_diag}
\Xi_{AB}\equiv\sum_i\Xi_{ABi}=\sum_iR_{AC}(2\bar \psi)\bar\Xi_{CD} R_{DB}(2\bar\psi)\,,
\ee
whose explicit expression is reported in Appendix \ref{app:psi},
and it is convenient to parameterize the two degrees of freedom of diagonal matrix $\bar\Xi_{AB}$ as
\be
\label{eq:Xi_AB}
\bar\Xi_{AB}=\sigma\pa{\ba{cc}
1+\epsilon & 0\\
0 & 1-\epsilon
\ea}
\ee
with $\sigma=\frac 12\pa{\Xi_{++}+\Xi_{\times\times}}$
and $\epsilon=\sqrt{\pa{\Xi_{++}-\Xi_{\times\times}}^2+4\Xi_{+\times}^2}/\pa{\Xi_{++}+\Xi_{\times\times}}$, which is bounded by $0\leq \epsilon\leq 1$.
In the particular case of a single detector one has that $\epsilon=1$
(and $\omega_1=1$), as each $\Xi_{ABi}$ has vanishing determinant, being the
outer product of two copies of the same vector.

For a generic detector network, the combined $SNR_{net}^2$ of
Equation (\ref{eq:snr_multi}) can then be written as
\be
\label{eq:cfn}
\ba{rcl}
\ds SNR^2_{net}&=&\ds SNR^2_0 \mathcal{R}eal[{\cal A}_C^*
R_{CA}\pa{2\bar\psi}\bar\Xi_{AB}R_{BD}\pa{2\bar\psi}{\mathcal A}_D]\\
&=&\ds SNR_0^2 \sigma\paq{
  \pa{\chi_+^2+v^2}+\epsilon\pa{\chi_+^2-v^2}\cos\pa{4\pa{\psi_0+\bar\psi}}}\,.
\ea
\ee

The quantities $\sigma,\epsilon$ depends on the detector network and on the
  direction of propagation $\hat N$, but they are independent of other angles parameterizing
  the binary plane orientation ($\iota,\psi,\phi$).
  For a fixed detector configuration $\epsilon,\sigma$ parameterize
  in a simple way the $SNR$ dependence which respectively depend and does not
  depend on the polarisation angle $\psi$.
  Note that an analogue but not equivalent parameterization has been introduced in
 \cite{Usman:2018imj}, whose parameterization allows to pinpoint the \emph{dominant polarisation} mode, i.e.
 the combination of polarizations that contributes the most to the $SNR$, see Appendix \ref{app:dpf}.

\subsection{Expected rates}

For cosmological applications it is crucial to have an accurate
measure of the luminosity distance, which one can expect to be
obtained by using multiple observatories sensitive to different
polarisation combinations.
The rate of EM bright standard siren is not supposed to exceed $O(1)$ per year
with current generation detectors \cite{KAGRA:2013rdx},
estimate for 3G detectors can lead to $O(100)$ per year \cite{Belgacem:2019tbw}.
As for the expected redshift distribution, an example for EM-bright standard
sirens is reported in Figure \ref{fig:histo_embsc}, where as a reference is
also reported the star formation rate
\be
R_{sfr}=\frac 1{1+z}\frac{dV_c}{dz}\psi_{DM}(z)\,,
\ee
where $\psi_{DM}$ is the star formation rate taken from \cite{Madau:2014bja} and $V_c$ is the comoving volume.\footnote{Given the moderate range of expected
  bright standard sirens, our $d_L$ recovery prior is uniform in comoving volume.}

In Figure \ref{fig:reach3G} we report the luminosity distance reach of BNS for optimally oriented, equal
mass, spin-less systems
(i.e. the distance at which $SNR_i=8$) and the design noise curves of CE
\cite{Evans:2021gyd} and ET \cite{Hild:2010id}.

In the following Section we show non-trivial consequences that can be deduced
from the parameterization in Equation (\ref{eq:cfn}), supporting them with
numerical results obtained with Bayesian inference methods.

\section{Results}
\label{sec:results}

\subsection{The Bayesian setup}

In a standard Bayesian inference framework one has to consider the likelihood
\be
{\cal L}=e^{-\frac 12\sum_i\|h_{d_i}-h_{t_i}\|^2}\,,
\ee
where the norm has been inherited by the scalar product defined implicitly in
Equation (\ref{eq:mf}).
We will consider the above likelihood for fixed values of the masses (setting the spins to zero), sky-position
angles, and time of the event.
This is a reasonable simplification of the problem, assuming that the EM
counterpart allowed a precise sky localization, and that the correlation
of the detected signal with long templates allowed a precise determination
of the arrival time and the masses which determine the chirping phase of the
signal.

By using only the dominant modes $l=|m|=2$, one can cast the likelihood $\cal L$
for data $d_i$ into the form
 \be
 \ba{rcl}
 \log \cal{L}&=&\ds -\sum_{i=1}^{n_{det}}\int_{-\infty}^\infty df \frac{|\tilde h_{d_i}(f)-\tilde h_t(f)|^2}{S_{n_i}(f)}\\
 &=&\ds -\sum_{i=1}^{n_{det}}\int_{-\infty}^\infty \frac{df}{S_{n_i}(f)}
 \paq{|\tilde h_{d_i}(f)|^2+|\tilde h_{t_i}(f)|^2-2\tilde R(f)\cos\pa{2\phi}
   -2\tilde I(f)\sin\pa{2\phi}}\,,
 \ea
 \ee
 where we have defined $\tilde R(f), \tilde I(f)$ as, respectively, the real and imaginary part of $\tilde h_{d_i}(f)\tilde h_{t_i}^*(f)$.
 
 Assuming a flat prior, marginalization over $\phi$ can be performed
 analytically using \cite{margphi,Veitch:2014wba}
 \be
 \frac 1{2\pi}\int_0^{2\pi} dx\ e^{a \cos(x)+ b \sin(x)}=I_0(\sqrt{a^2+b^2})\,,
 \ee
 where $I_\nu(x)$ is the Bessel function of the first kind of order $\nu$,
 and the marginalized likelihood ${\cal L}_\phi$ will depend only on the extrinsic
 parameters $d_L$, $\iota$ and $\psi$:
 \be
    {\cal L}_\phi= \frac 1{2\pi}\int_0^{2\pi}{\cal L}\,d\phi=
    \exp\paq{-\frac 12\sum_i^{n_{det}}\pa{\|h_{d_i}\|^2+\|h_{t_i}\|^2}}\,
    I_0\pa{2\left|\sum_i^{n_{det}}\int_{-\infty}^\infty df\frac{h_{d_i}{h^*_{t_i}}_{\phi=0}}
      {S_{n_i}(f)}\right|}\,.
    \ee

 When considering the injection data to correlate with templates, we will
 work in the zero noise approximation \cite{Rodriguez:2013oaa}, as usually done
 it literature to estimate average uncertainties in Gaussian noise.

 We run \texttt{Bilby} \cite{Ashton:2018jfp} with the \texttt{Nestle} sampler
 \cite{Mukherjee:2005wg}, which implements the nested sampling algorithm
 \cite{Skilling:2006gxv}, with 300 live points,
 searching over 3 parameters $d_L,\psi,\iota$.
 Results for every $d_L$ injection are averaged over the 300 injections
 simulating random values of $\alpha,\beta,\iota,\psi$,
 which survives the $SNR$ cutoff at 8 in each detector.

For simulation efficiency reason, we used \emph{TaylorF2} waveform model
\cite{Sathyaprakash:2009xs} with fixed total mass $M=3M_\odot$, equal
binary component masses, no spins. We have also verified in a few cases that no significant deviations occur in the result
when replaced with waveform complete with merger-ringdown model, like
\emph{IMRPhenomD} \cite{Husa:2015iqa,Khan:2015jqa}, see right plot in
Figure \ref{fig:2Gnoise} for qualitative reference, and we neglected tidal effects.

\subsection{Impact of {\boldmath{$\iota$}} and source location on
  {\boldmath{$d_L$}} uncertainty}
\label{ssec:impct_iotad}

Expressions (\ref{eq:Xi_AB},\ref{eq:cfn}) permit to highlight the following
fundamental features:
\begin{enumerate}
\item  For a single detector $\epsilon=1$, $\sigma=\frac 12\pa{f_+^2+f_\times^2}$
  and we recover Equation (\ref{eq:cf1}) (and also Equation (\ref{eq:bpsi}) reduces
  to Equation (\ref{eq:bpsii})).
\item For co-located detectors and a source on the top of them one has
  $f_+=f_\times=\sin\Omega$, implying $\sigma=n_{det}/2\times \sin^2\Omega$,
  which is its maximum value.
  
\item
  In the case of a single $L$-shaped detector ($\epsilon=1$), the detected
  signal is a single combination of the two polarisations and one has no information on
  the polarisation angle.
  Considering that for large argument the Bessel function has $I_0(x)\sim e^x/\sqrt{2\pi x}$, the
  marginalized likelihood can then be written as
  \be
  \ba{rcl}
  \ds -2 \log {\cal L}_\phi&=& \sum_{i=1}^{n_{det}}\pa{\|h_{d_i}\|^2+\|h_{t_i}\|^2}
  -2\paq{x-\frac 12\log\pa{2\pi x}}\,,\\
  \ds x&\equiv&\ds
  \left|\sum_{i=1}^{n_{det}}\int_{-\infty}^\infty \frac{h_{d_i}{h_{t_i}^*}_{\phi=0}}{S_{n_i}}df
  \right|\,,
  \ea
  \ee
  and the scalar product between data and template can be decomposed analogously
  to what done in Equation (\ref{eq:snr_multi}) for the SNR:
  \be
  \braket{h_{d_i}|h_{t_i}}=SNR_0^2{\cal R}eal\paq{{{\cal A}^*_{d_i}}_A{{\cal A}_t}_B{\Xi_{dt_i}}_{AB}}\,,
  \ee
where
  ${\Xi_{dt_i}}_{AB}=R_{AC}(2\delta\psi_{d_i}) R_{BD}(2\delta\psi_{t_i}) f_C f_D\omega_i$ is
  the matched-filter analogue of the $SNR$-related quantity $\Xi$ defined in (\ref{eq:snr_multi}), to
  which it reduces when the template equals the data.
For a
single detector ($\omega_i=1=\epsilon$, see Equations (\ref{eq:xidefs},\ref{eq:xi_diag})) the matrix $\Xi_{dt_i}$ is not symmetric but can still be diagonalized
into the form
  \be
  \label{eq:xibar}
     \bar{\Xi}{{}_{dt_i}}_{AB}=\pa{F_+^2+F_\times^2}\cos\pa{2(\psi_{d_i}-\psi_{t_i})}\pa{
       \ba{cc}
       1 & 0\\
       0 & 0
       \ea
     }\,,
  \ee
  each detector giving a contribution to the log-likelihood
  \be
  \label{eq:lmarg}
  \ba{rcl}
  \ds{\log{\cal L}_i}_\phi&\propto&\ds 2\sigma\cos\pa{2(\psi_{d_i}-\psi_{t_i})}
  \pa{|{\cal A}_{d_i+}|^2+|{\cal A}_{t_i+}|^2-2|{\cal  A}_{d_i+}
    {\cal A}_{t_i+}|}\,,
  \ea
  \ee
  where here with $+,\times$ we denoted the ``principal'' polarisations
     diagonalizing ${\Xi_{dt_i}}_{AB}$ (see Appendix \ref{app:psi} for
     the matrix diagonalising ${\Xi_{dt_i}}_{AB}$, and
     \cite{Chassande-Mottin:2019nnz} for the version of Equation
     (\ref{eq:lmarg}) not marginalized over $\phi$).
     Equation (\ref{eq:lmarg}) indicates that for $\epsilon\sim 1$ the
     likelihood can constrain only one polarisation, leading to the well known
     bimodal degeneracy between $\upsilon$ and $d_L$, as shown in top plots of Figure \ref{fig:di_deg} for any value of the
  inclination angle sufficiently away from the $\pi/2$ value.

  For a network of interferometers things are qualitatively different as they
  are in general sensitive to more than one combination
  of the two GW-polarisations and $\epsilon$ in Equation (\ref{eq:Xi_AB})
  can assume values between $0$ and
  $1$, depending on the source location, see Figure \ref{fig:se_3G}.\\
  For instance for a triangle-shaped detector the
  condition $\epsilon\sim 1$ is realized only by sources located in a small
  region of the sky corresponding to the blind (or almost blind) regions
  of the individual interferometer composing the triangle. They correspond to
  directions in the plane of the interferometers bisecting their arms, i.e.
  $\alpha=\Omega/2$, $\beta=\pi/2$ in Equation (\ref{eq:fpc}). See also
  Figure \ref{fig:scatter_es_3G} and the additional material in Appendix
  \ref{app:dL_iota}, showing that for a value of $\epsilon\sim 1$ and
  a value of $\iota$ sufficiently distant from the symmetric point $\iota=\pi/2$,
  bimodality appears in the $d_L$-$\iota$ 2-dimensional
  probability distribution function (PDF).
  As expected, adding detectors into the network reduces the sizes of $\epsilon\sim 1$ regions, see bottom of Figure \ref{fig:se_3G}.
  
\begin{figure}
\begin{center}
  \includegraphics[width=.193\linewidth]{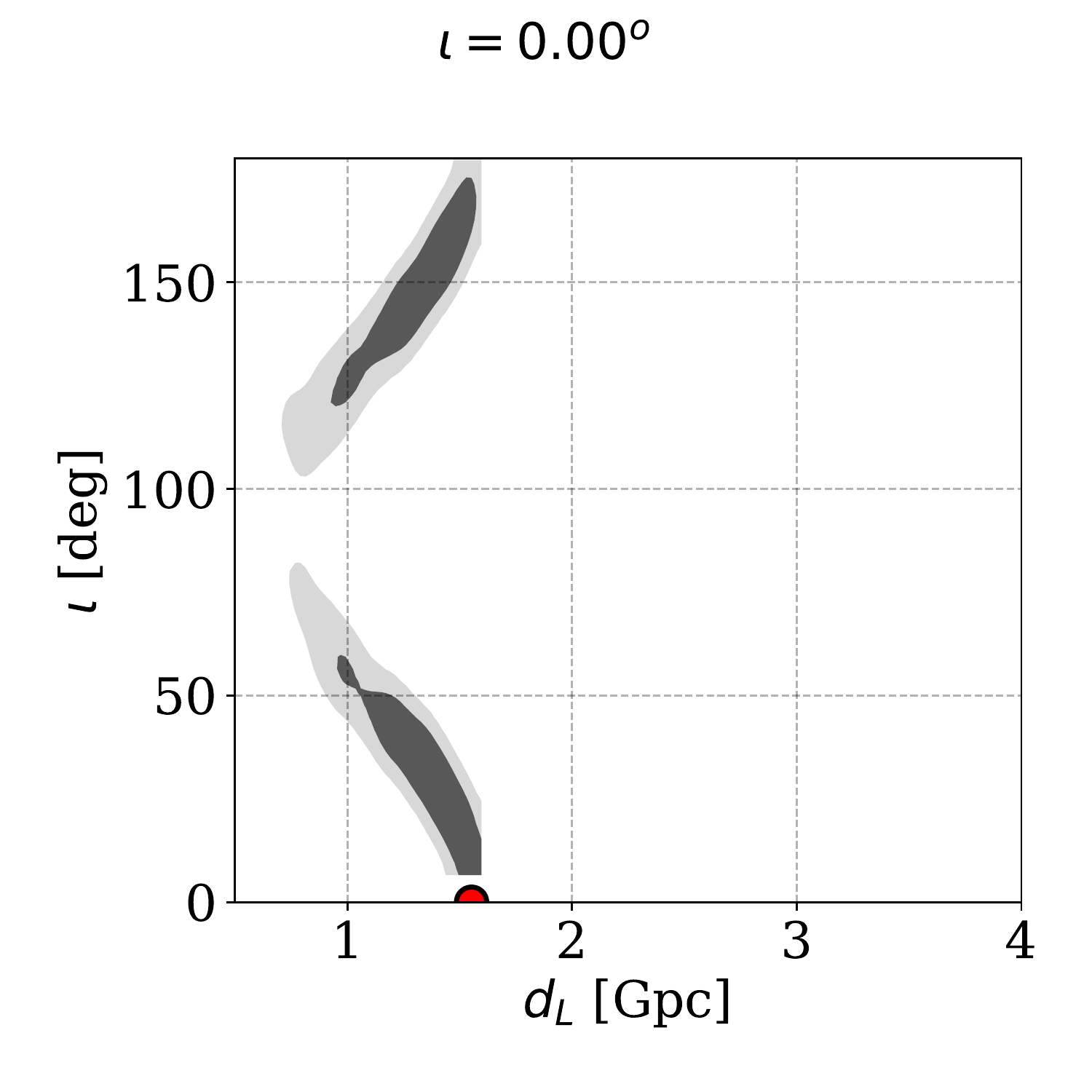}
  \includegraphics[width=.193\linewidth]{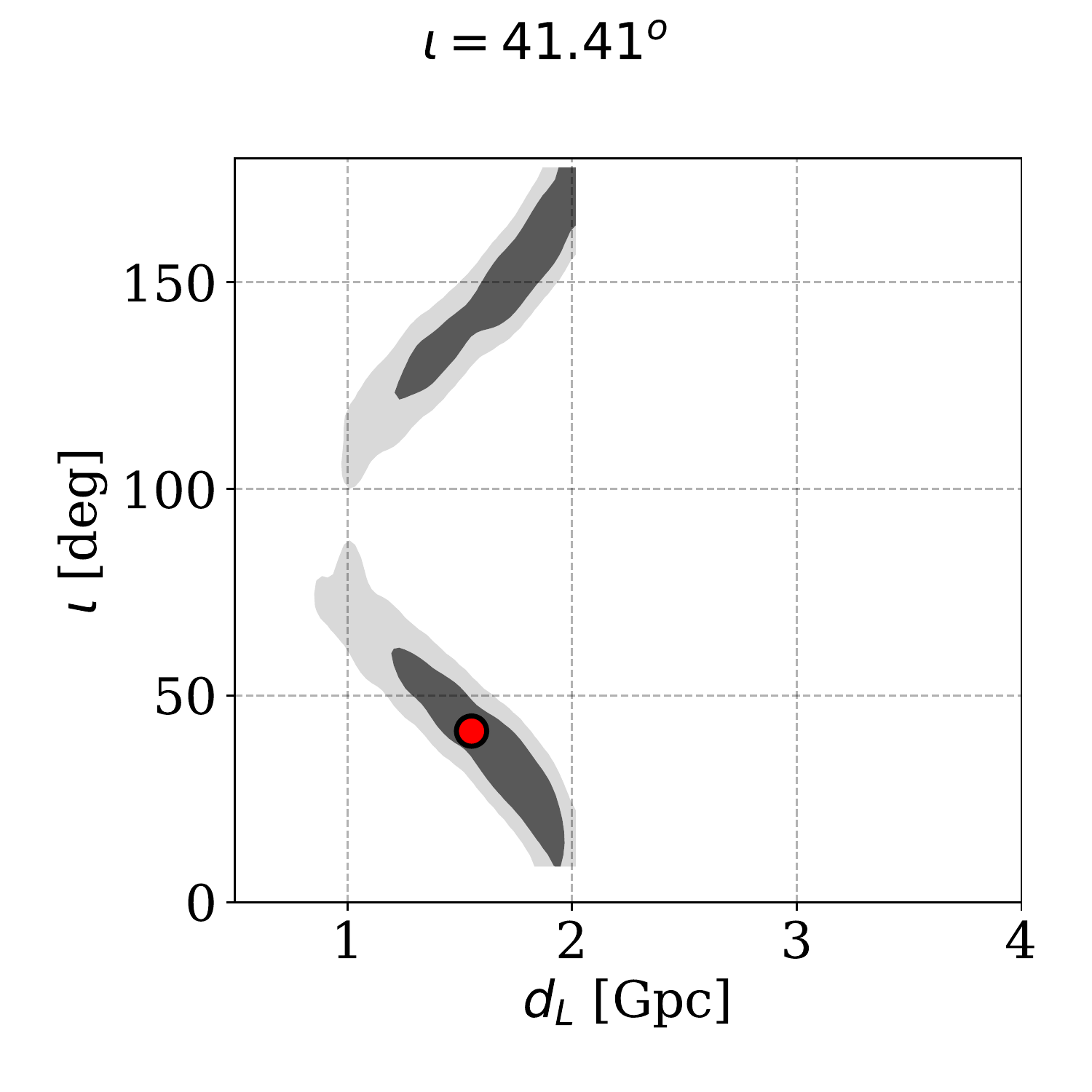}
  \includegraphics[width=.193\linewidth]{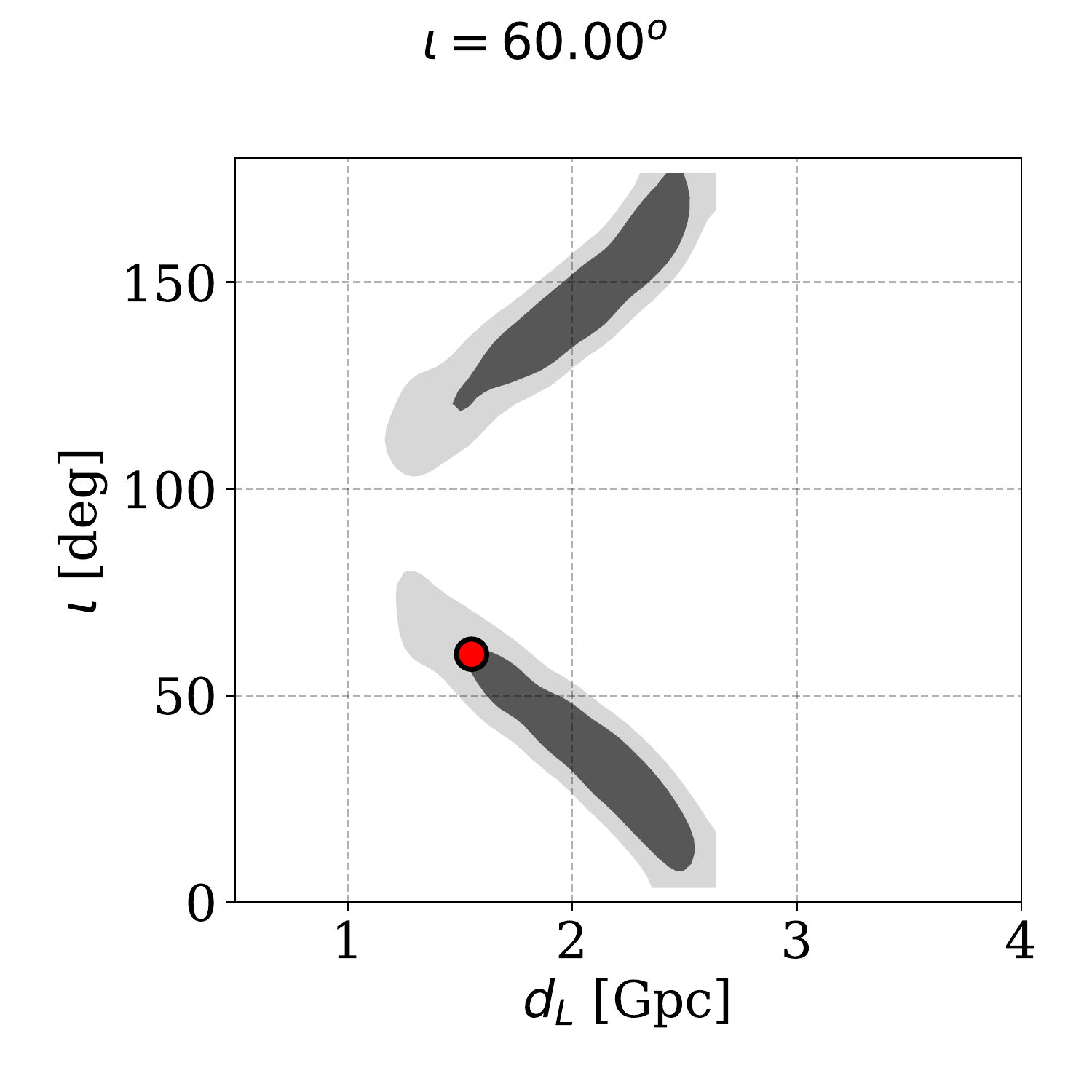}
  \includegraphics[width=.193\linewidth]{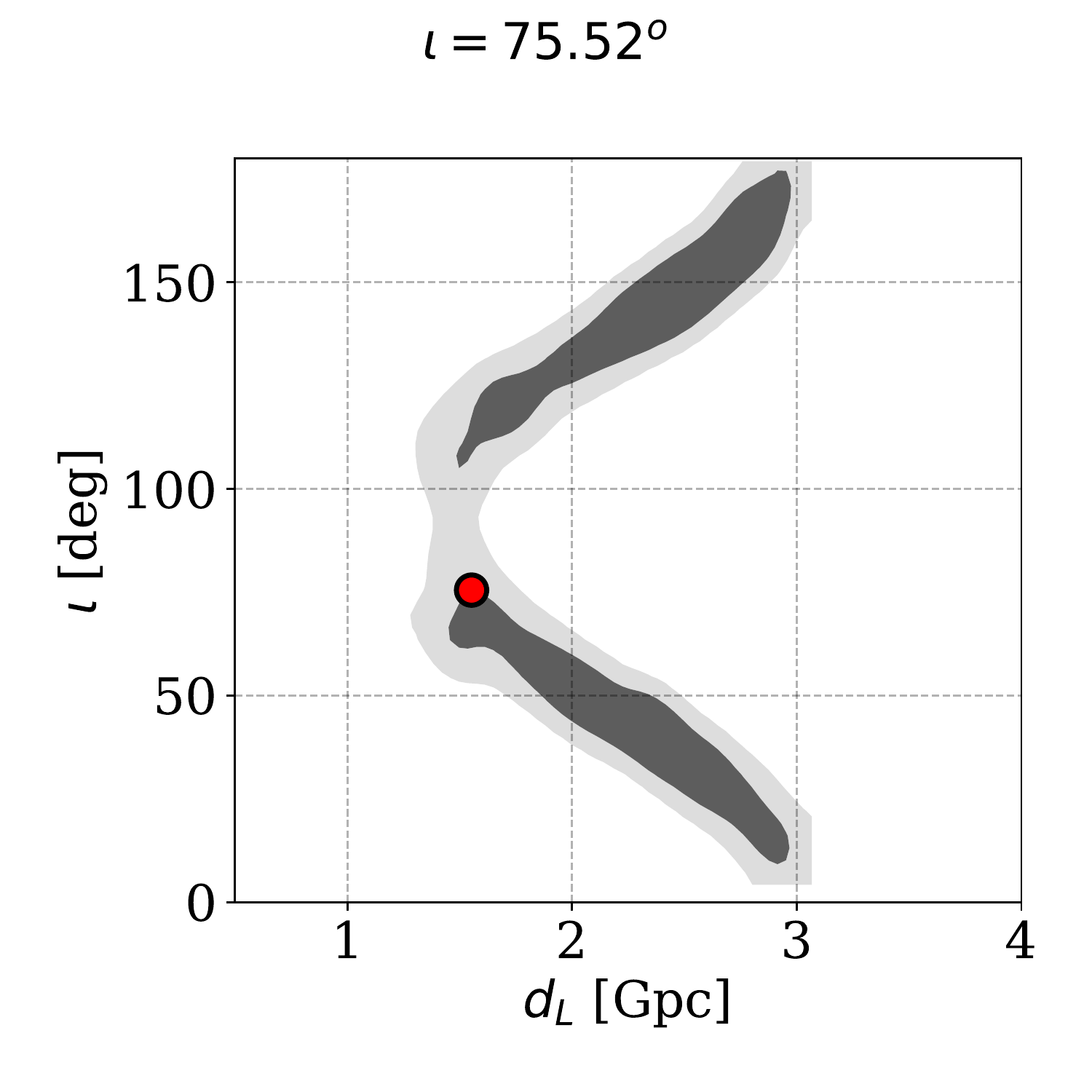}
  \includegraphics[width=.193\linewidth]{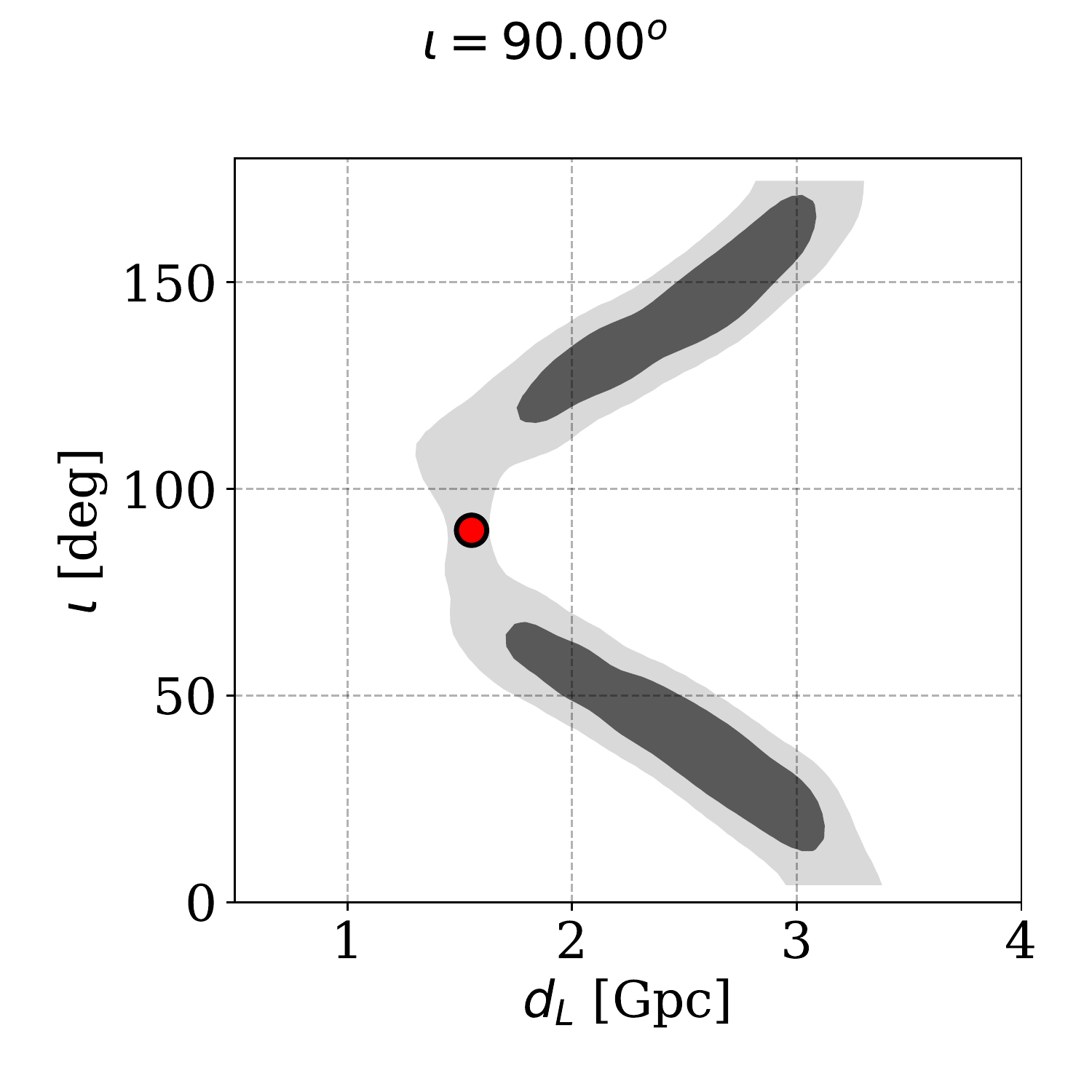}\\
  \includegraphics[width=.193\linewidth]{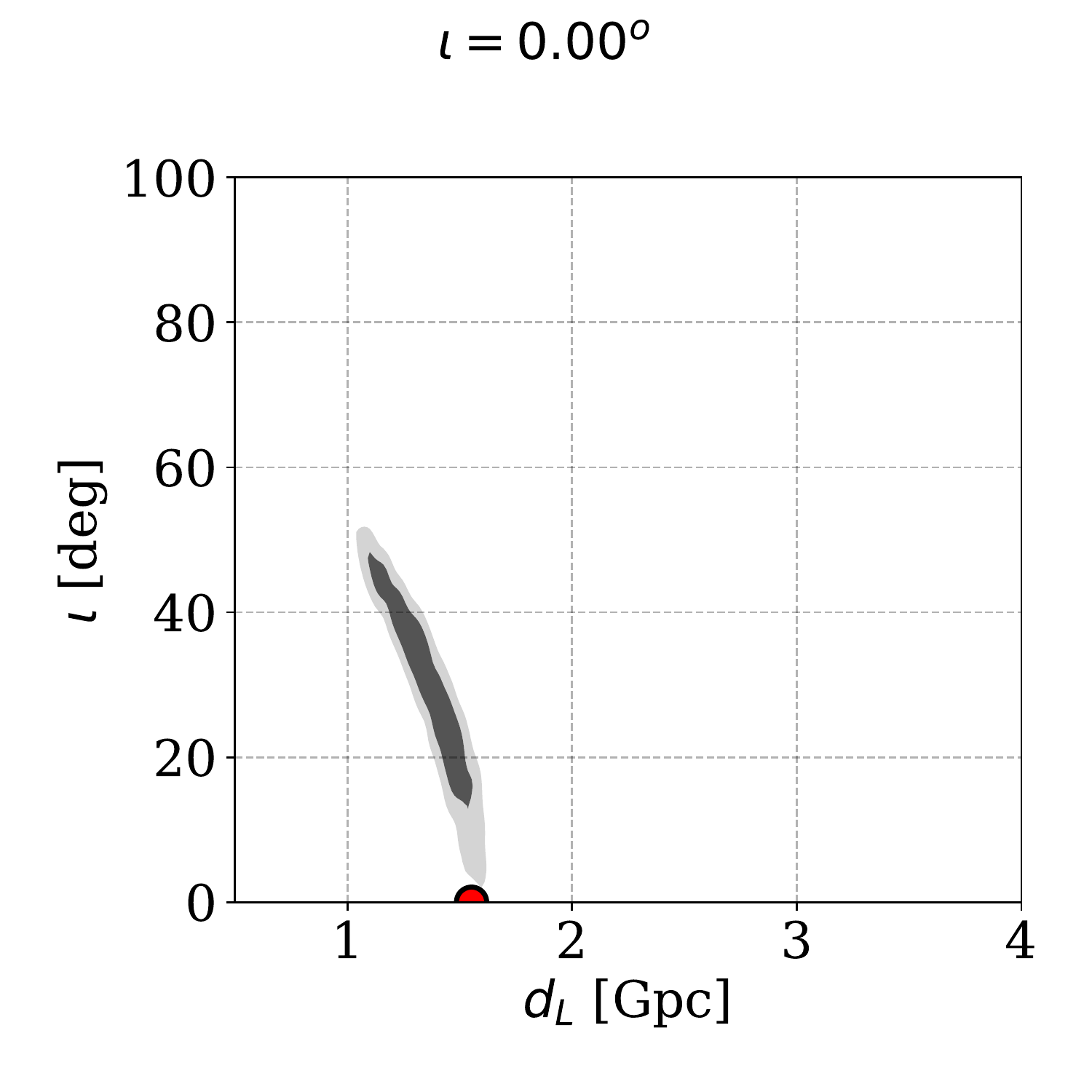}
  \includegraphics[width=.193\linewidth]{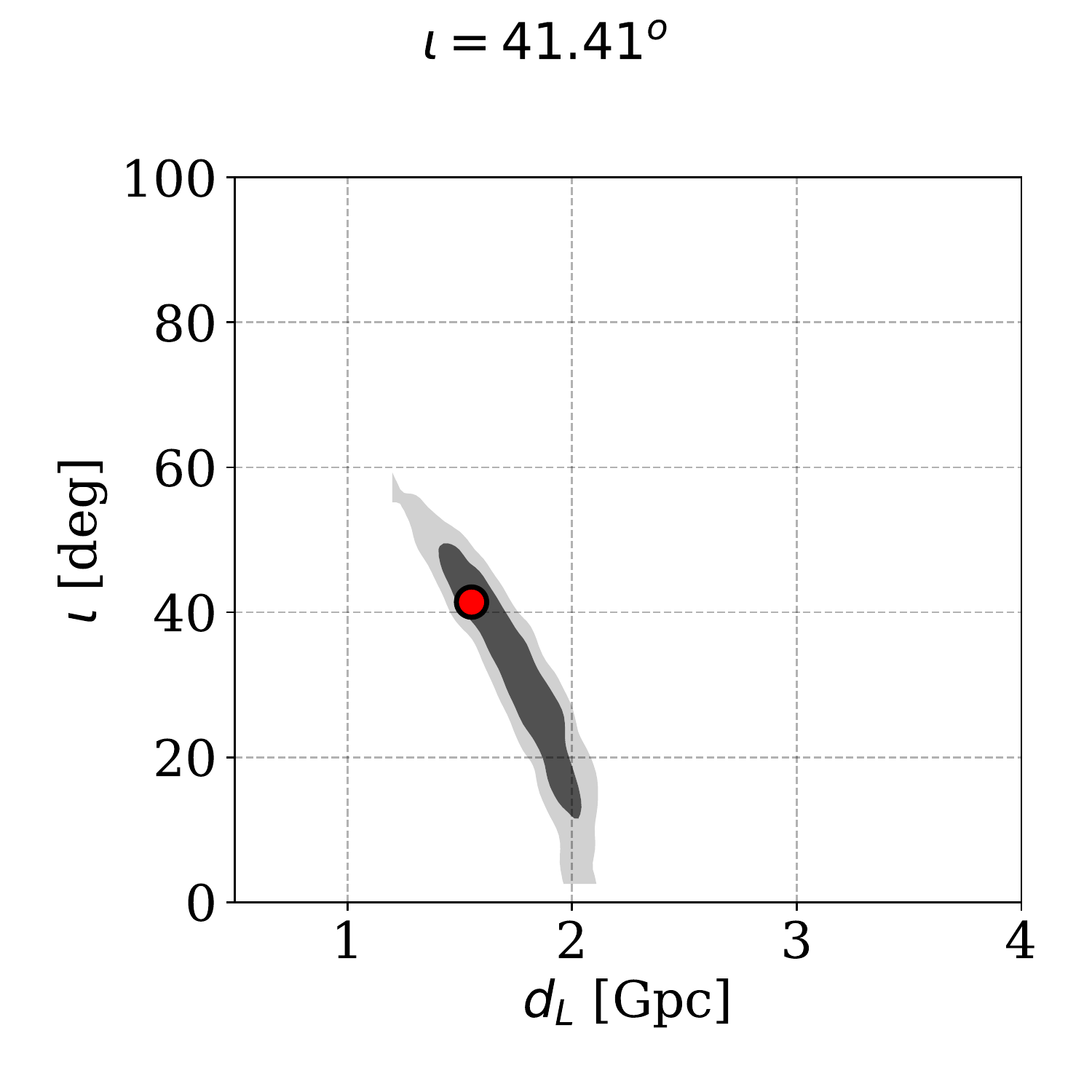}
  \includegraphics[width=.193\linewidth]{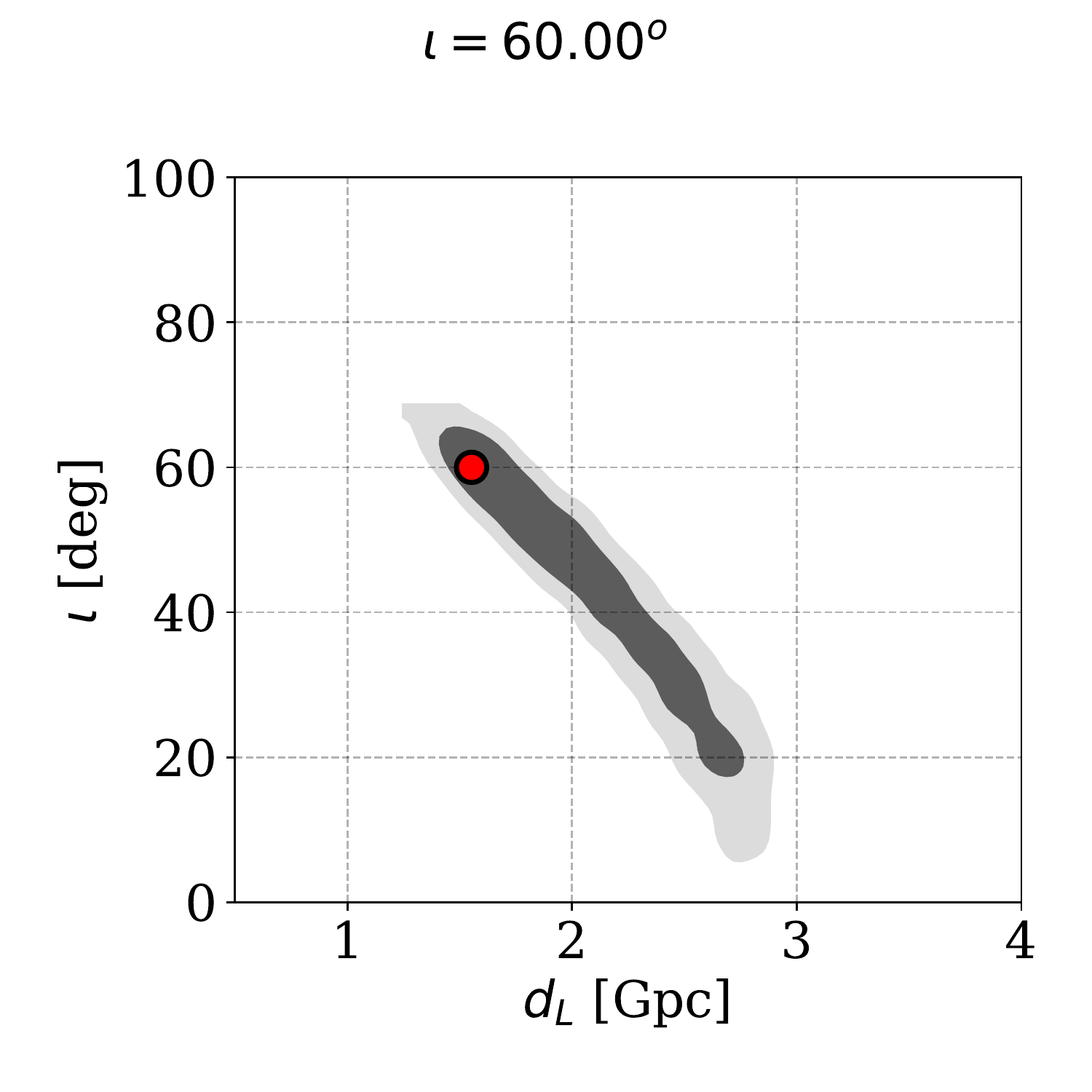}
  \includegraphics[width=.193\linewidth]{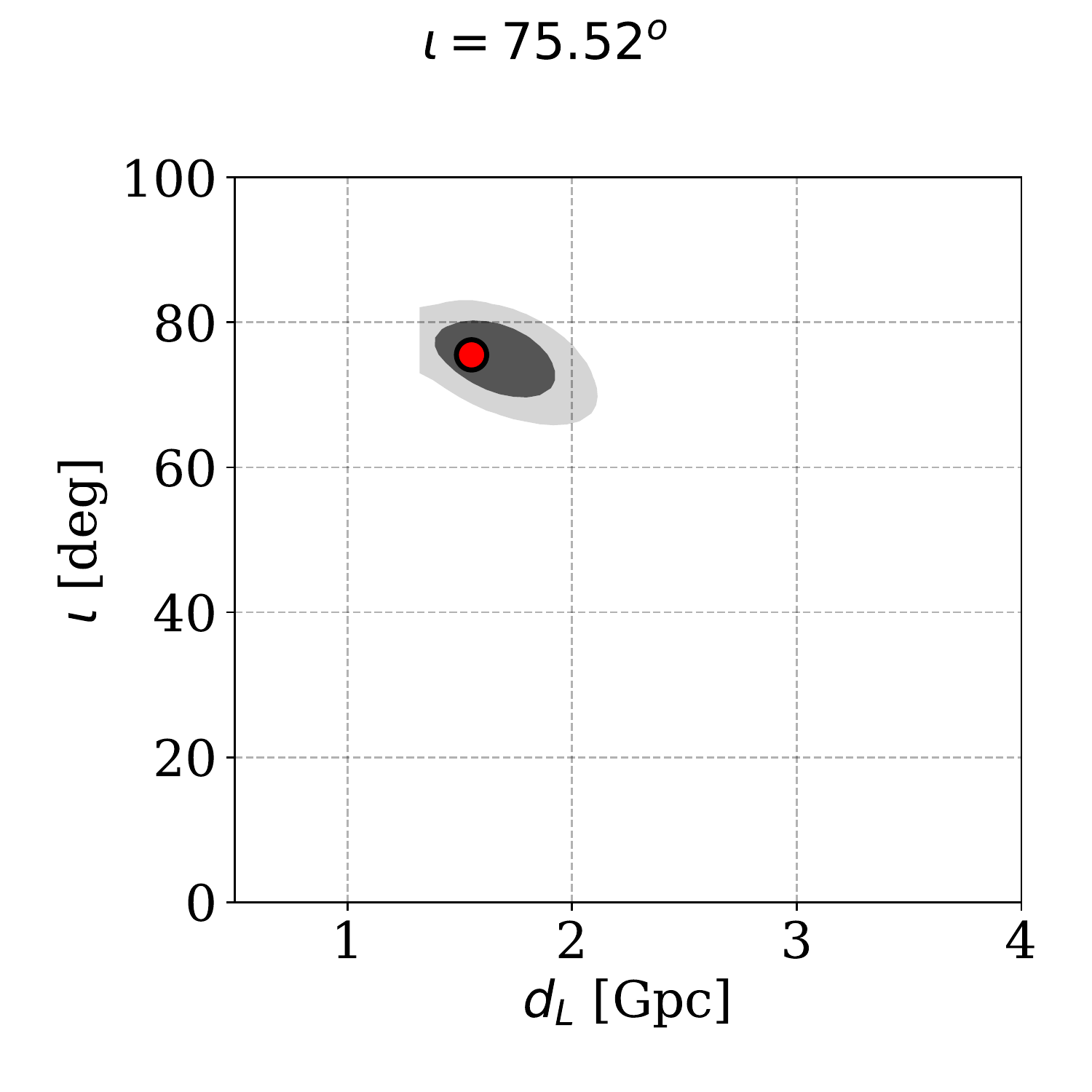}
  \includegraphics[width=.193\linewidth]{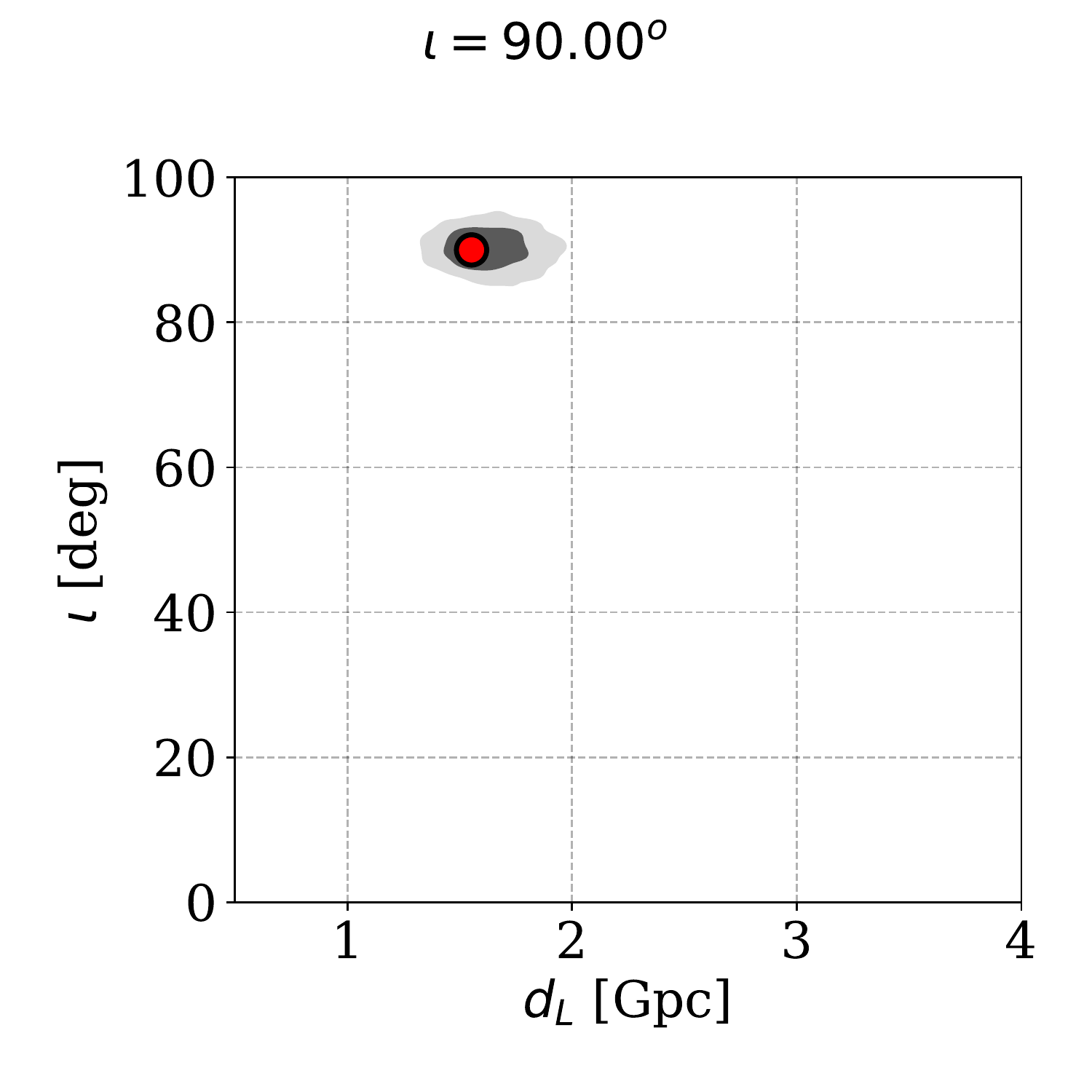}\\
  \caption{Examples of two-dimensional PDF for $\iota$ vs. $d_L$ for a single CE-like, $L$-shaped interferometer (top) and a triangle-shaped one (bottom) for source
    parameters giving $\epsilon=0.0089$ \cite{Alfradique:2022tox}.
  Note that the volumetric prior at recovery tends to disfavour $\iota\to\pi/2$ for $L$-shaped detectors.}
  \label{fig:di_deg}
\end{center}
\end{figure}

\begin{figure}
  \begin{center}
    \includegraphics[width=.48\linewidth]{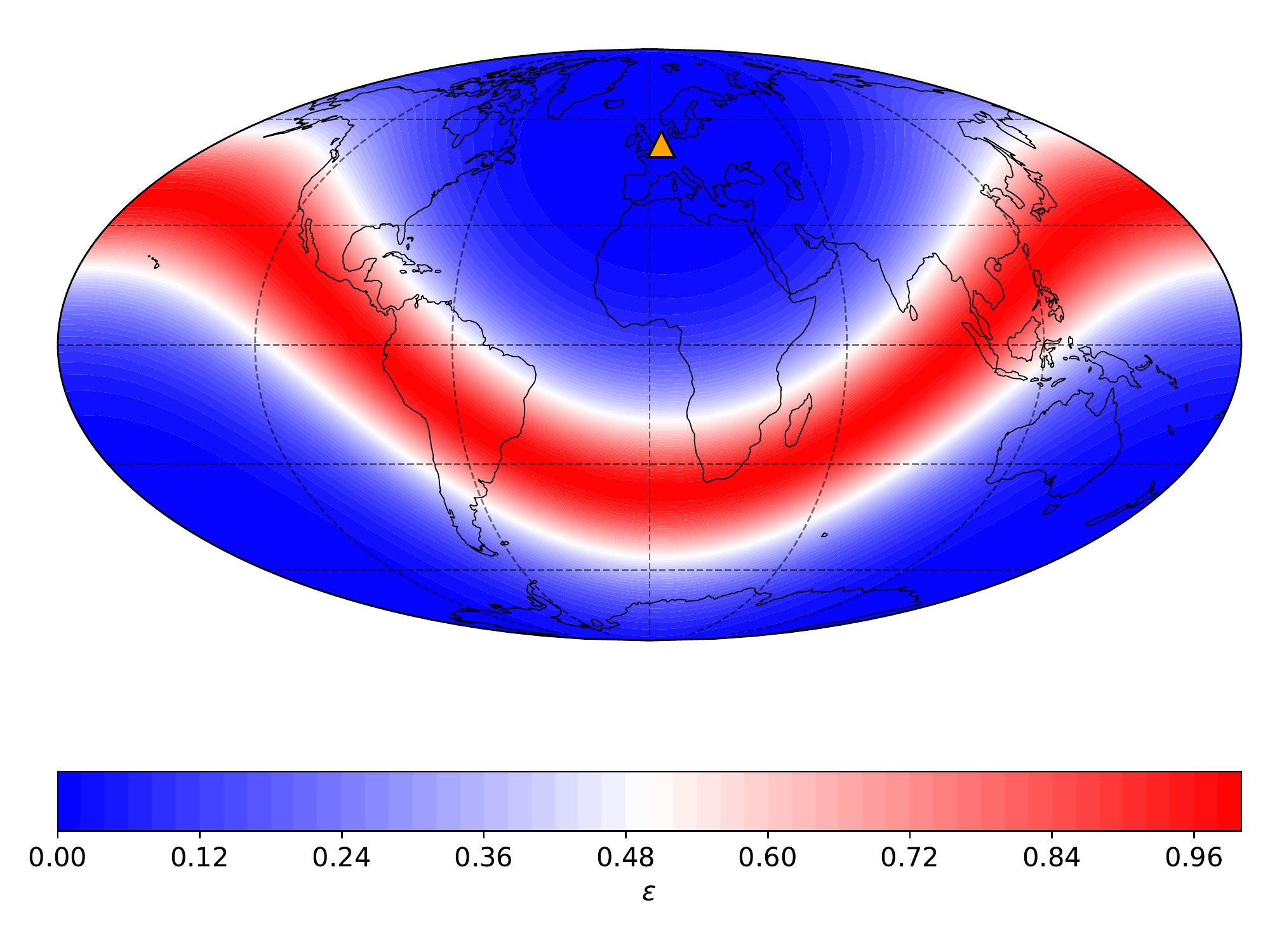}
    \includegraphics[width=.48\linewidth]{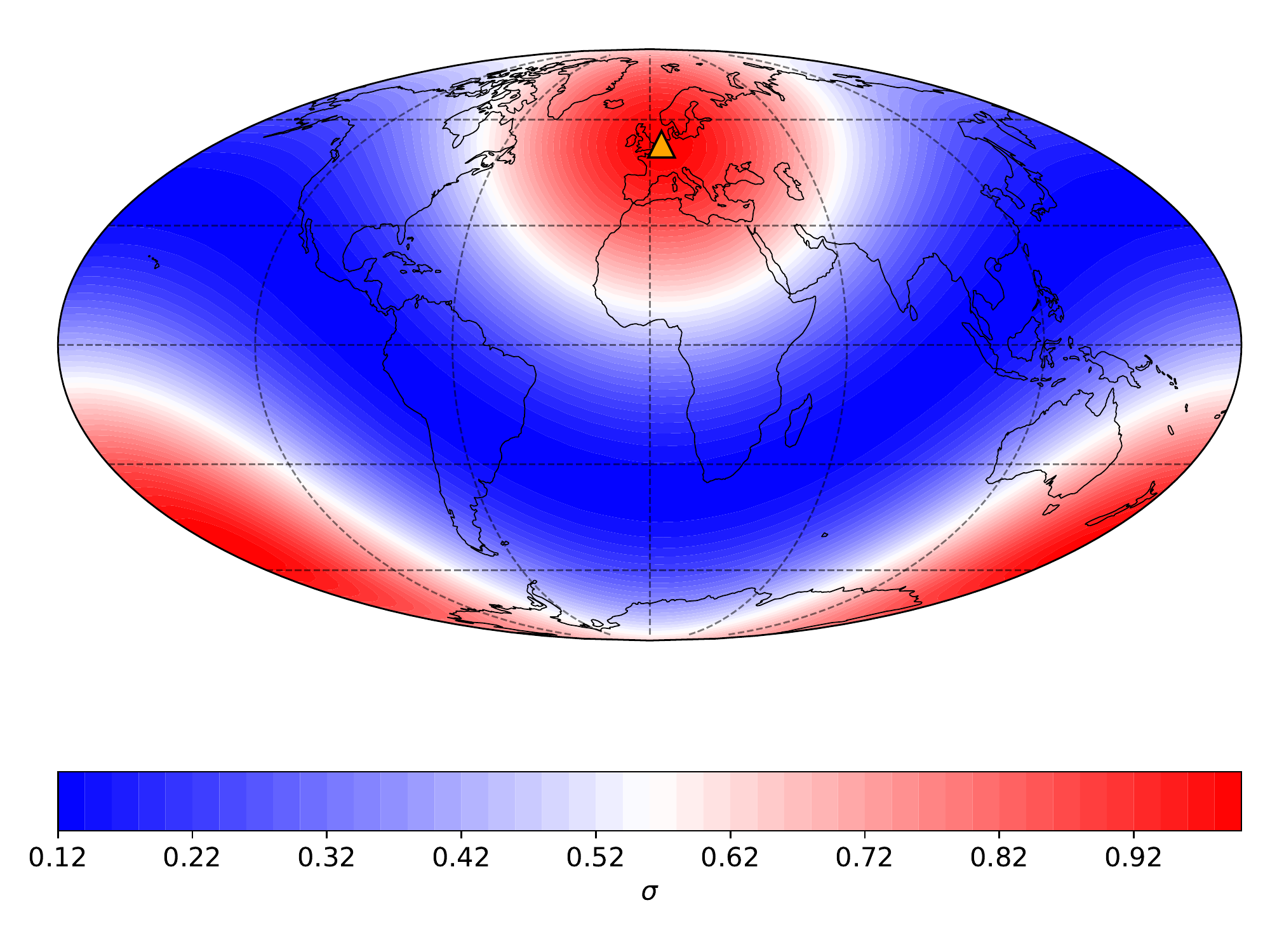}\\
    \includegraphics[width=.98\linewidth]{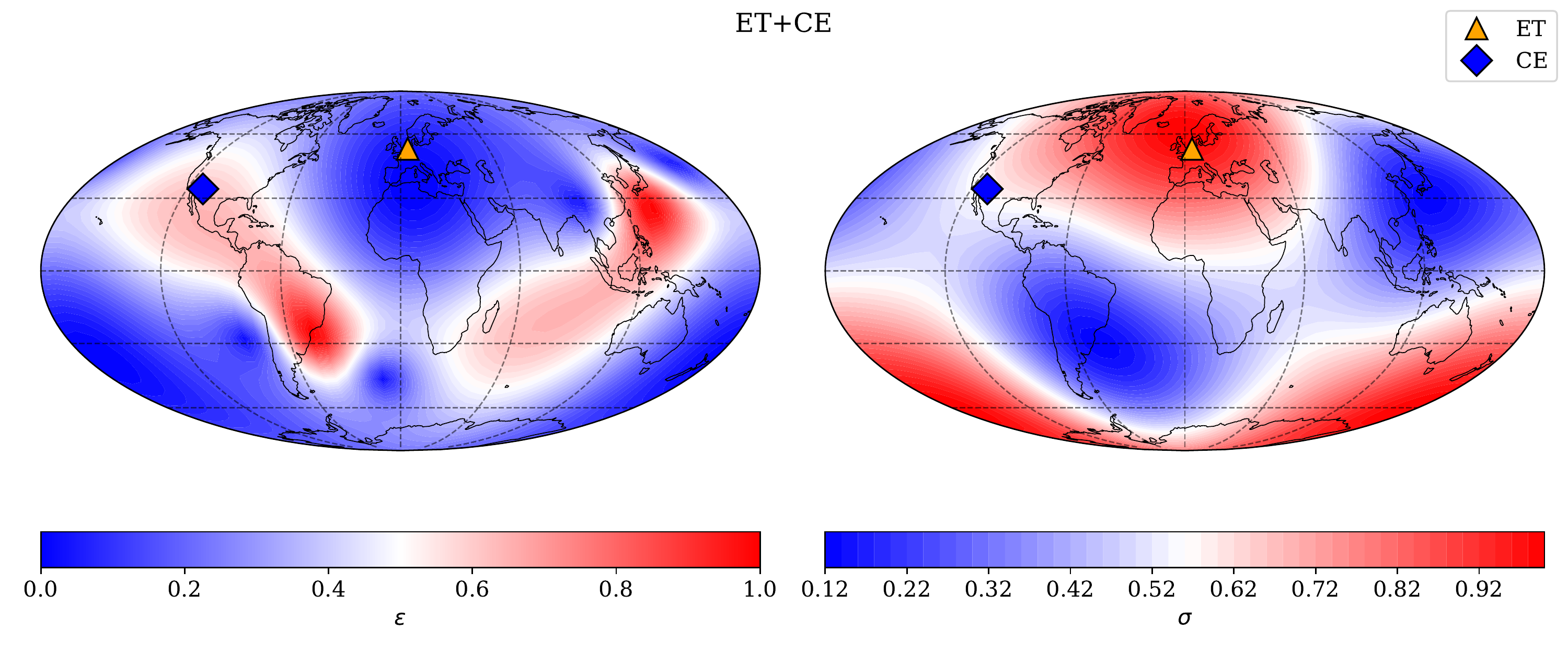}
    \caption{Values of $\epsilon$ (left) and $\sigma$ (right) for a single ET
      detector (top, ET location marked with a triangle) and for
      an ET-CE network (bottom, CE location marked with a diamond).}
    \label{fig:se_3G}
  \end{center}
\end{figure}

\begin{figure}
  \begin{center}
    \begin{minipage}[h]{.495\linewidth}
      \includegraphics[width=.98\linewidth]{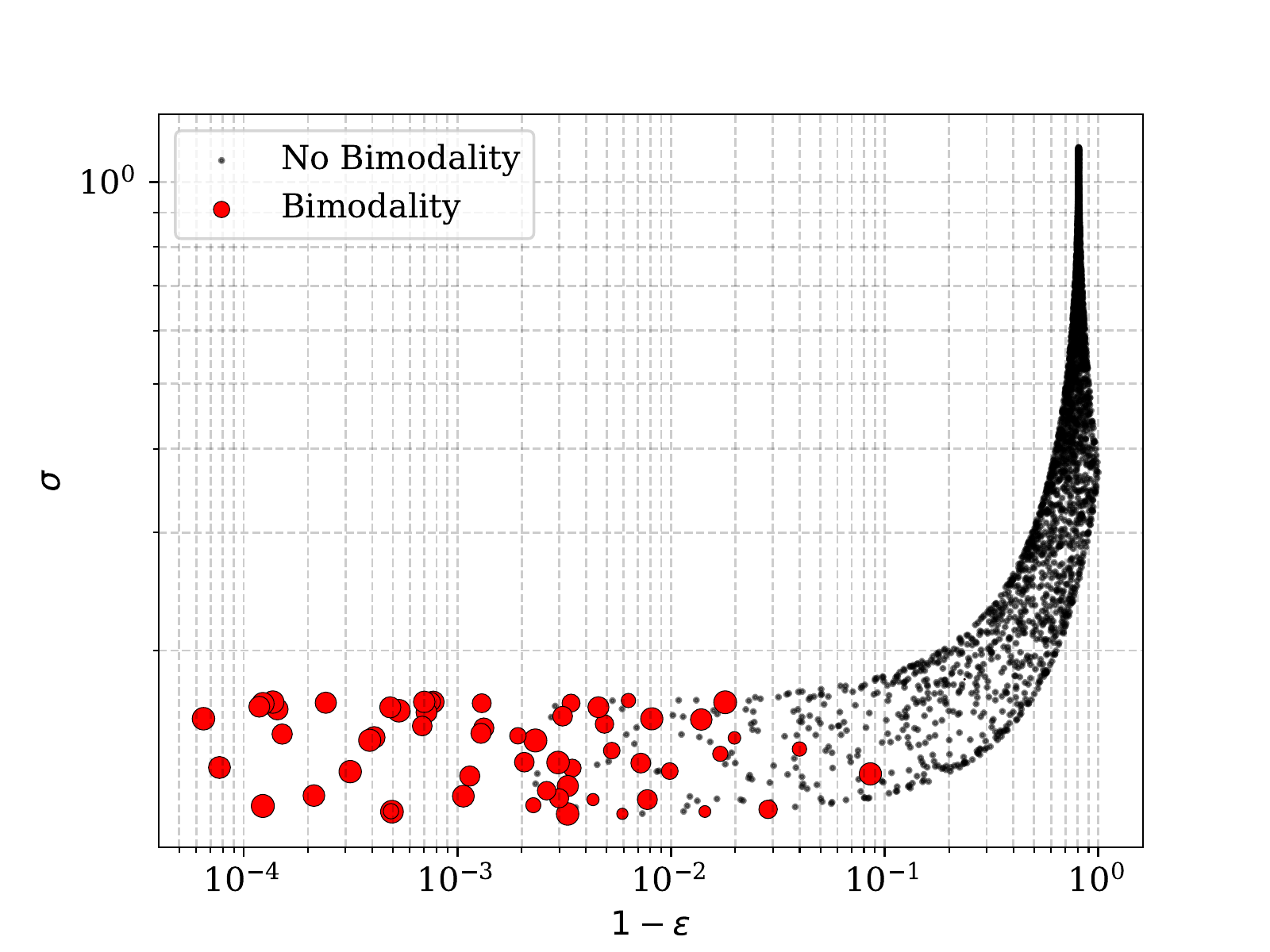}
    \end{minipage}
    \begin{minipage}{.495\linewidth}
      \includegraphics[width=.98\linewidth]{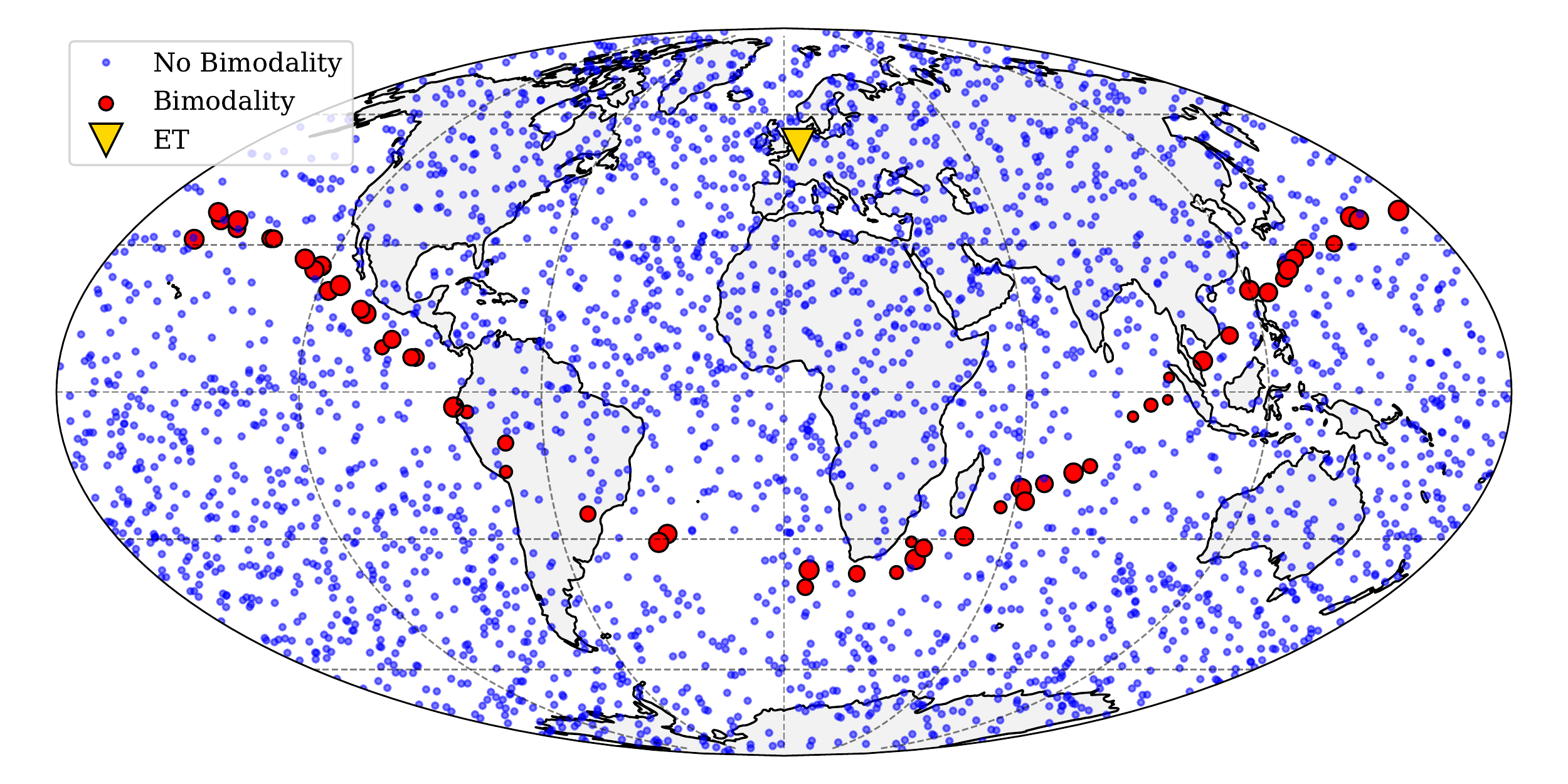}\\
      \mbox{}
      \end{minipage}
    \caption{(Left) Distribution of $\sigma$ and $\epsilon$ values,
      defined in Equation (\ref{eq:Xi_AB}), for a triangle-shaped
      interferometer. (Right) Points in the sky presenting bimodality
      are confined to the plane of the detector, where blind directions
      to individual interferometer appear.
      Operationally, we defined bimodality to be present
      when the ratio (smaller or equal to 1) of the height of the peaks of the
      $\iota$ PDF is larger than the PDF value at $\iota=\pi/2$.}
    \label{fig:scatter_es_3G}
  \end{center}  
\end{figure}

\item It has been empirically noted, e.g. in \cite{Chassande-Mottin:2019nnz,Vitale:2018wlg}
  where sky position is fixed, and \cite{Chen:2018omi} where sky localization
  angles are among the parameters searched for in the Bayesian inference, 
  that for  $\iota\sim\pi/2$ the uncertainty in $d_L$ usually drops
  for triangle-shaped detector, see Figure \ref{fig:dd_1dL}.
  Note that the drop in uncertainty while varying $\iota$ at fixed redshift
  for the triangle-shaped detector is not due to an increase in $SNR$,
  which rather \emph{decreases} as
  $\iota\to \pi/2$, as shown in Figure \ref{fig:dd_1dT}.
  Actually, it is due to
  $\upsilon\to 0$, leading to the polarisation
  dependent term to become equally important as the non-$\psi$ dependent
  term in (\ref{eq:cfn}). In turn, this leads to a better individuation of the
  polarisations, as exemplified by Figure \ref{fig:psid_deg}.
  For $z\sim 1$ detections disappear for the triangle-detector as they go below
  the $SNR$ threshold, while this happens for the $L$-detector at larger distances
  because of its better sensitivity, see Figure \ref{fig:reach3G}.

For ``tropical'' inclination angles ($\iota\sim \pi/2$) the $\psi$-dependent
term is as important as the $\psi$-independent one,
with the consequences that while for a $L$-shaped detector $\psi$ cannot be
constrained, for a triangle-shaped one a bimodality $d_L$-$\psi$ appears,
see Figure \ref{fig:psid_deg}.

Note that while it is more difficult for a CE-like detector to
  determine $\iota$ than for a triangle one, Figure \ref{fig:dd_1dL}
  shows that for specific cases
  CE can achieve a better precision due its better sensitivity, see
  Figure \ref{fig:reach3G} and top line in Figure
  \ref{fig:dd_1dT}.
  Note that we used a volumetric prior on $d_L$, which tends to perform better for $\iota\sim 0$, but which
  can introduce bias for a $L$-shaped detector for ``tropical'' $\iota$ (i.e.
  $\iota \sim \pi$), as shown by the last top graph in Figure \ref{fig:di_deg}.

\begin{figure}
  \begin{center}
    \includegraphics[width=.49\linewidth]{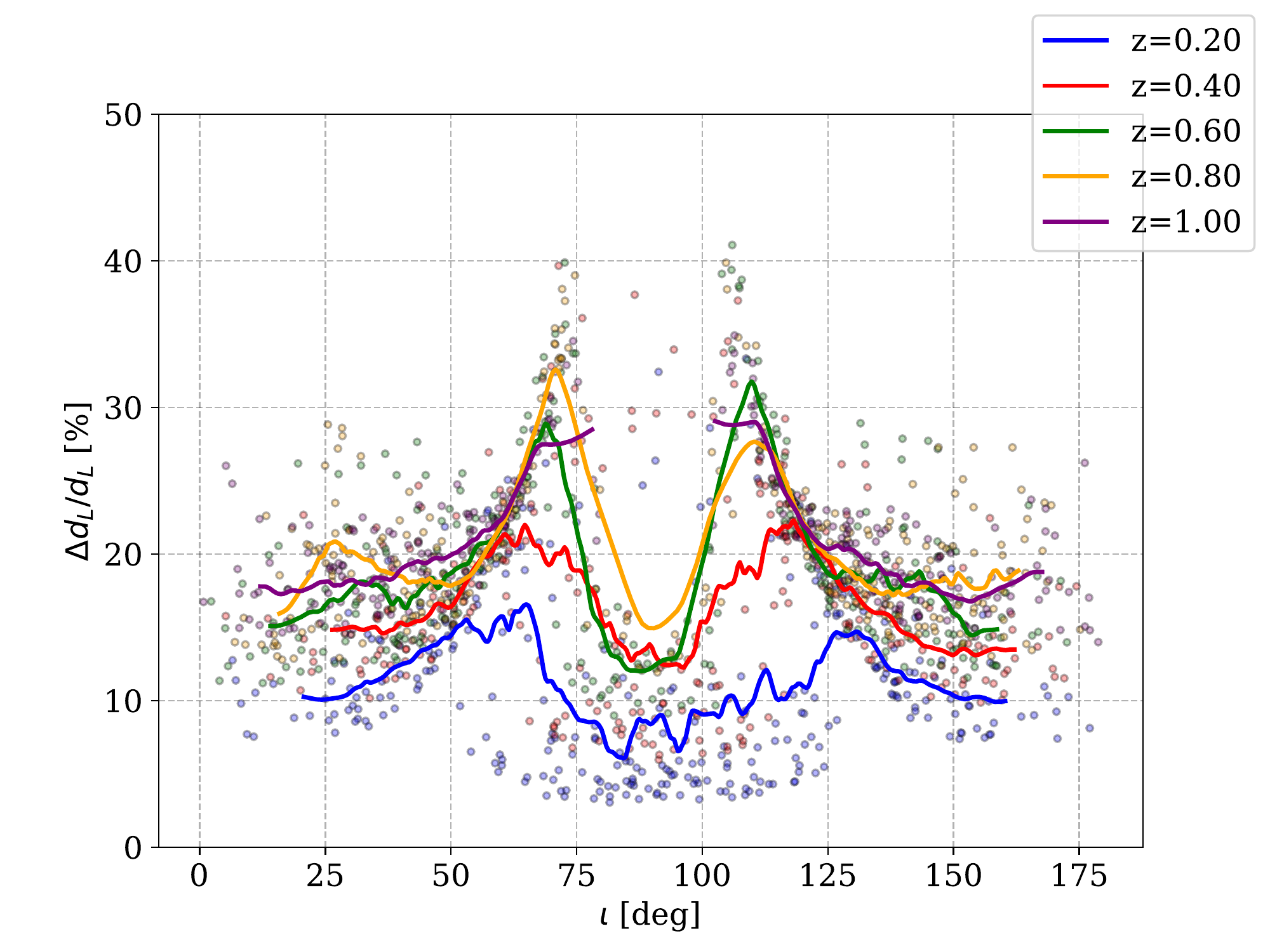}
    \includegraphics[width=.49\linewidth]{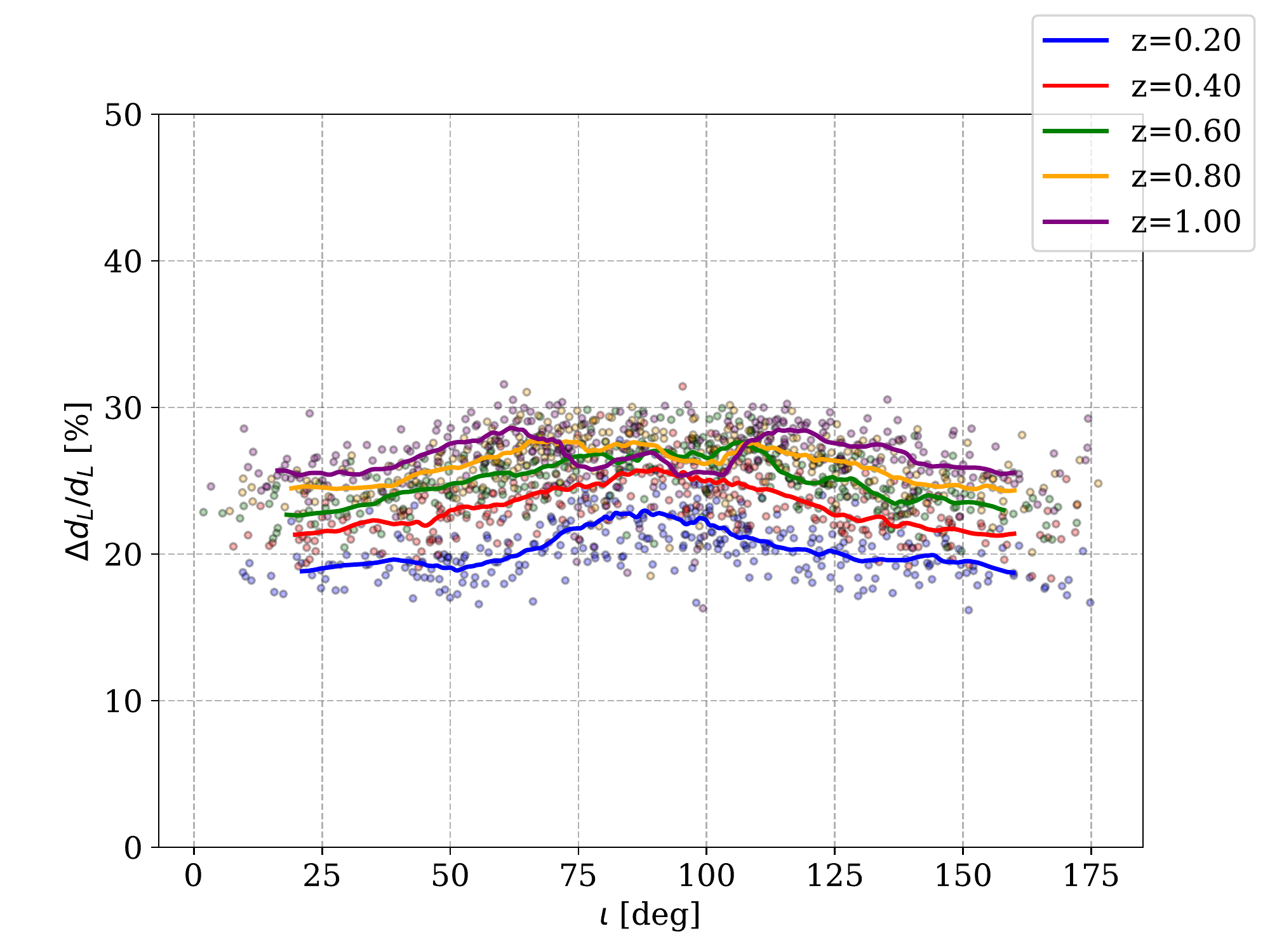}\\
    \includegraphics[width=.49\linewidth]{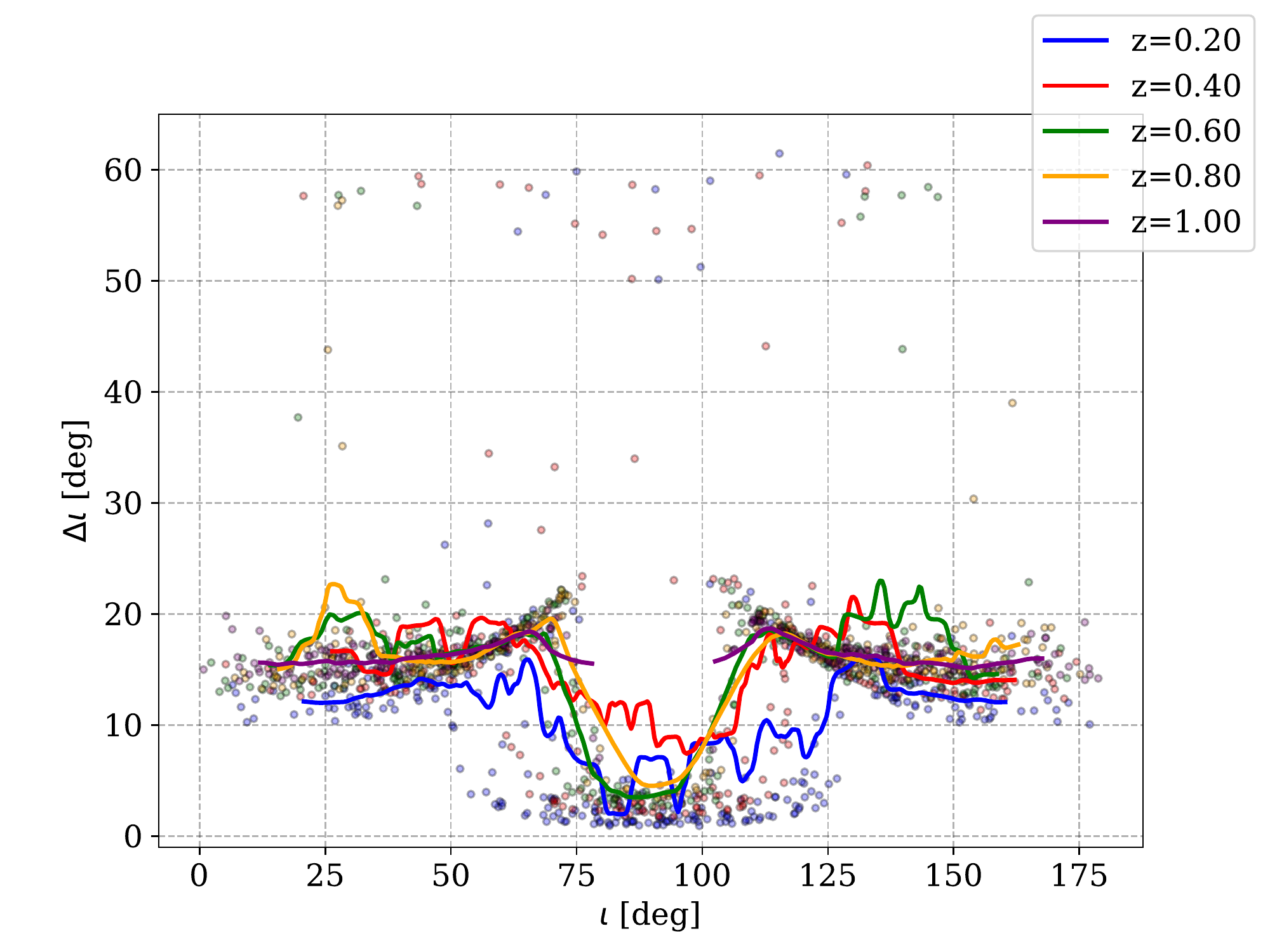}
    \includegraphics[width=.49\linewidth]{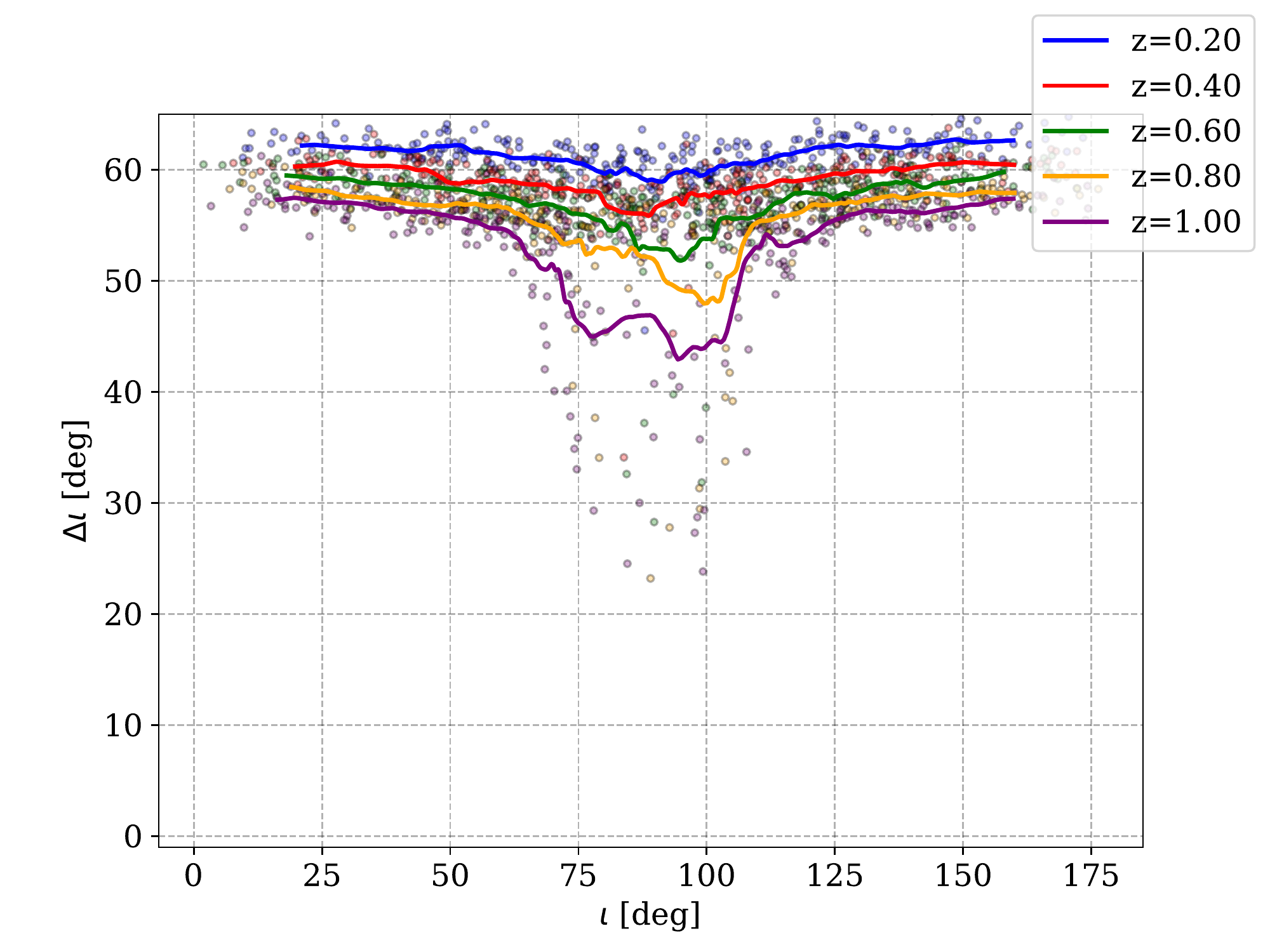}\\
  \caption{Scatter values and averages for relative uncertainty of $d_L$ (top) and absolute one of $\iota$ (bottom) as a function of $\iota$ for various
    distances for an ET-like detector (left) and for a single CE one (right),
    for 300 simulations at each distance. Continuous lines are averages over intervals
    of $0.1$ radians in $\iota$.
    Note the dip in $d_L$ uncertainty for $\iota \to\pi/2$ in the ET case.
    The points where CE outperforms ET in $\Delta d_L$ are due to the better
    spectral noise sensitivity of the detector, see Figure \ref{fig:reach3G},
    hence higher SNR, see Figure \ref{fig:dd_1dT}.
    Most of the recovered $\iota$ for CE present bimodality ($\Delta\iota \sim 60^o$), bimodality that happens far more rarely for ET, as shown
    by the red dots in the left plot of Figure \ref{fig:scatter_es_3G} compared
    to the majority of gray points in the bottom right plot here clustering
    around $\Delta\iota\sim 60^o$.
    For ET-like detector there is no dip in $d_L$ uncertainty for $z\sim 1$ as
    the $SNR$ decrease for $\iota\to\pi/2$ moves the signal below the $SNR=8$
    threshold, whereas for CE injections at $z=1$ are just above threshold.}
  \label{fig:dd_1dL}
  \end{center}
\end{figure}

\begin{figure}
  \begin{center}
    \includegraphics[width=.48\linewidth]{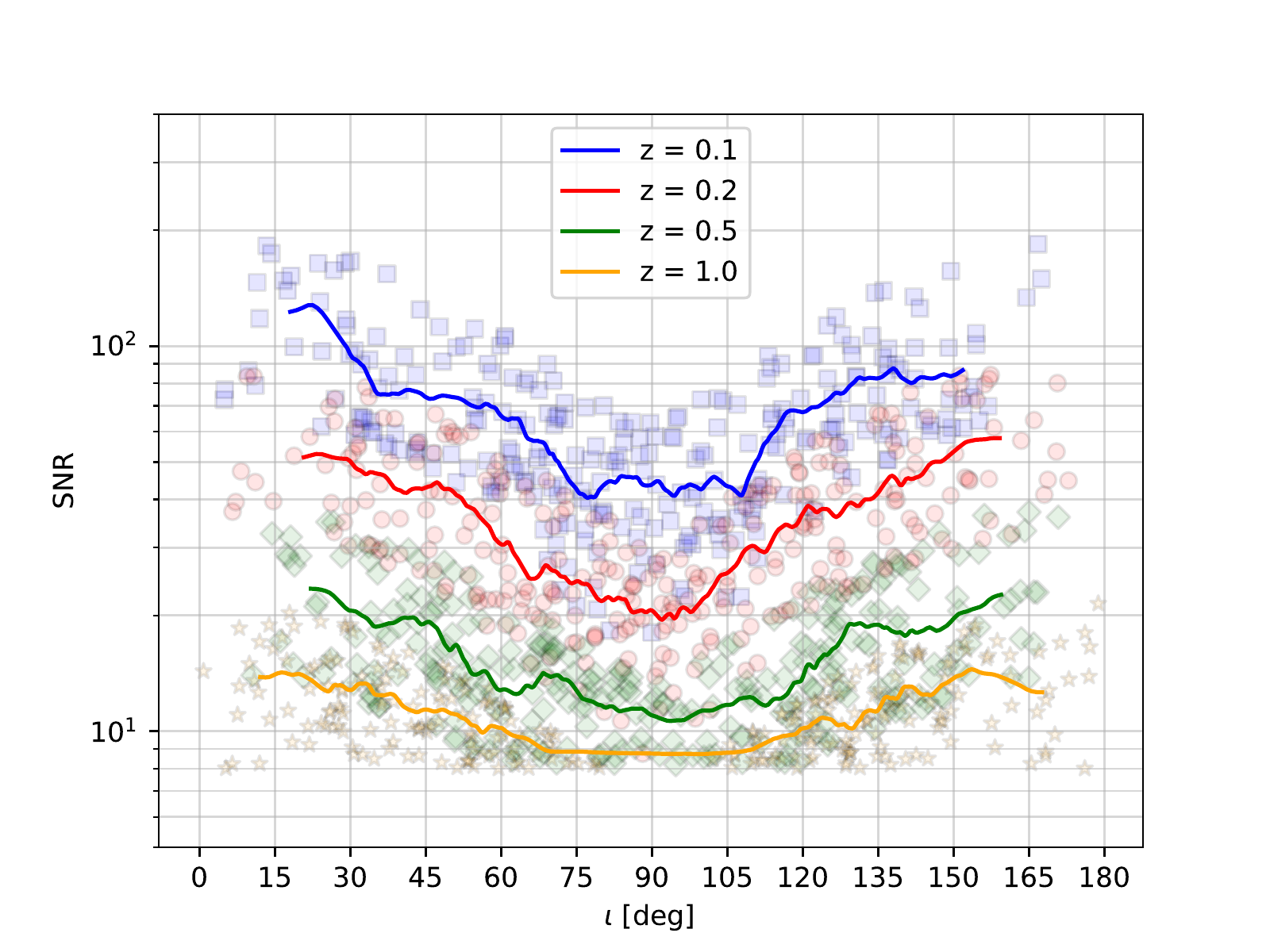}
    \includegraphics[width=.48\linewidth]{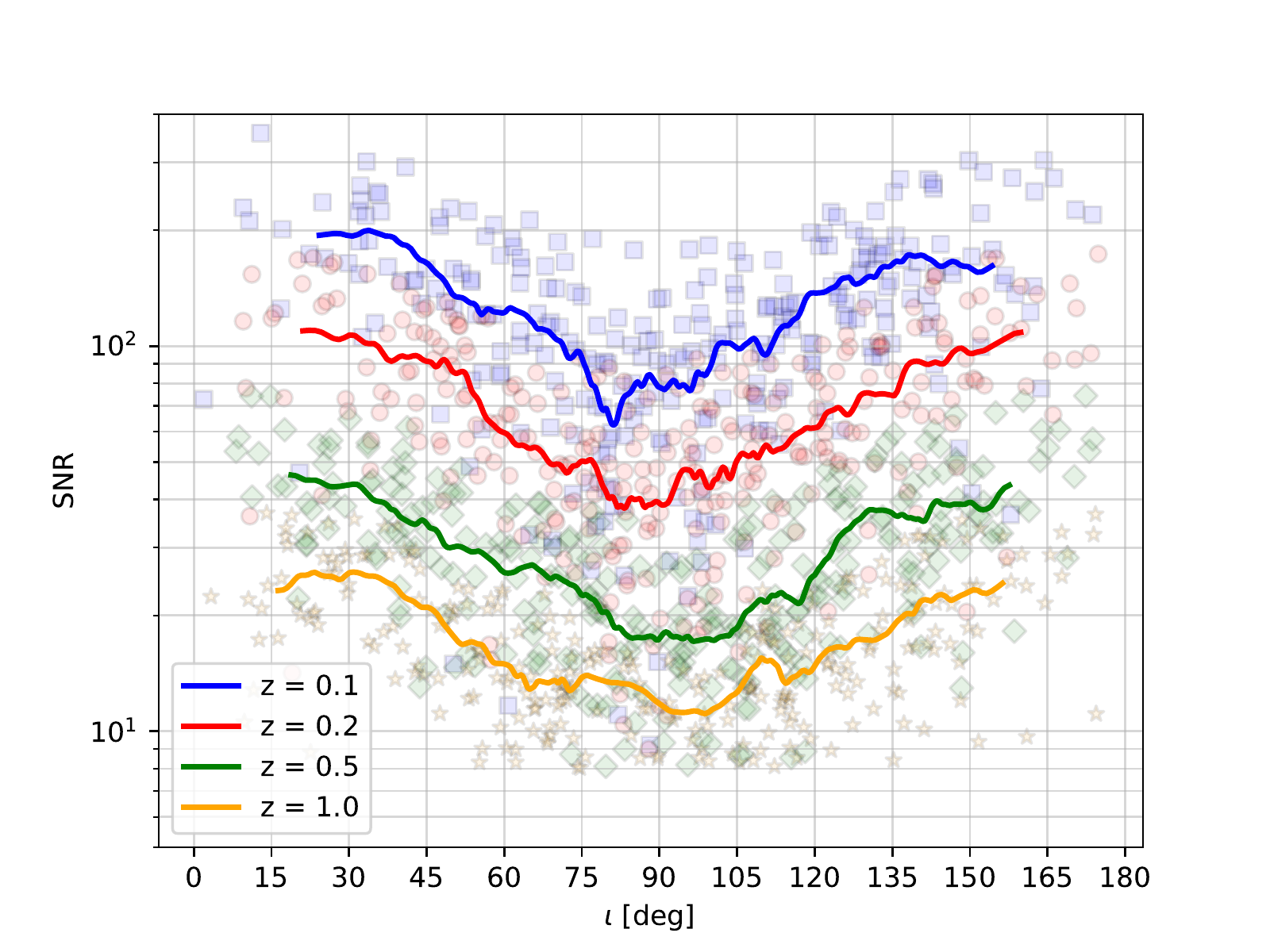}\\
    \caption{$SNR$ as a function of $\iota$ for various
    distances for a triangle interferometer (left) and for a
    single $L$-shaped detector (right).}
  \label{fig:dd_1dT}
  \end{center}
\end{figure}
  
\begin{figure}
  \begin{center}
    \includegraphics[width=.193\linewidth]{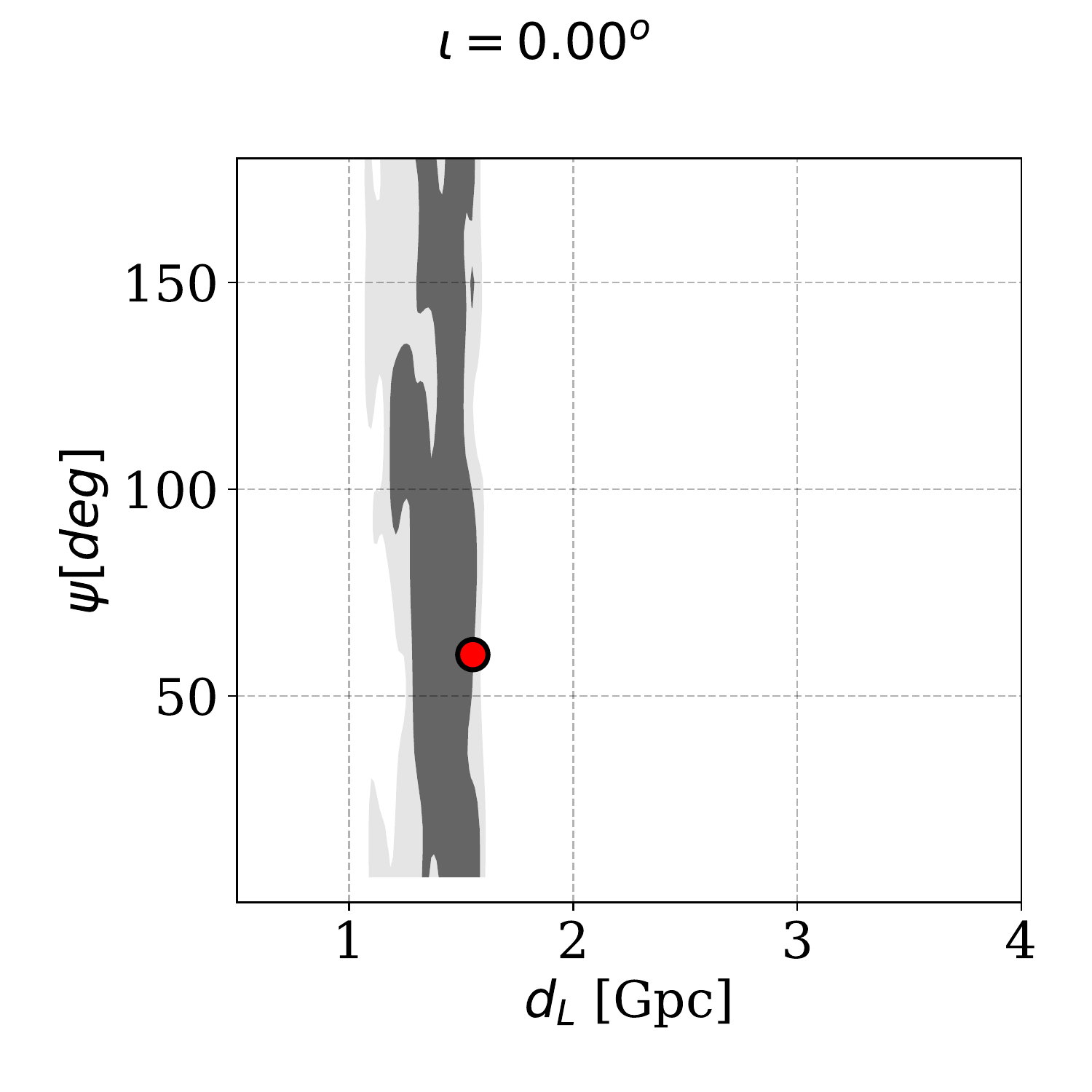}
    \includegraphics[width=.193\linewidth]{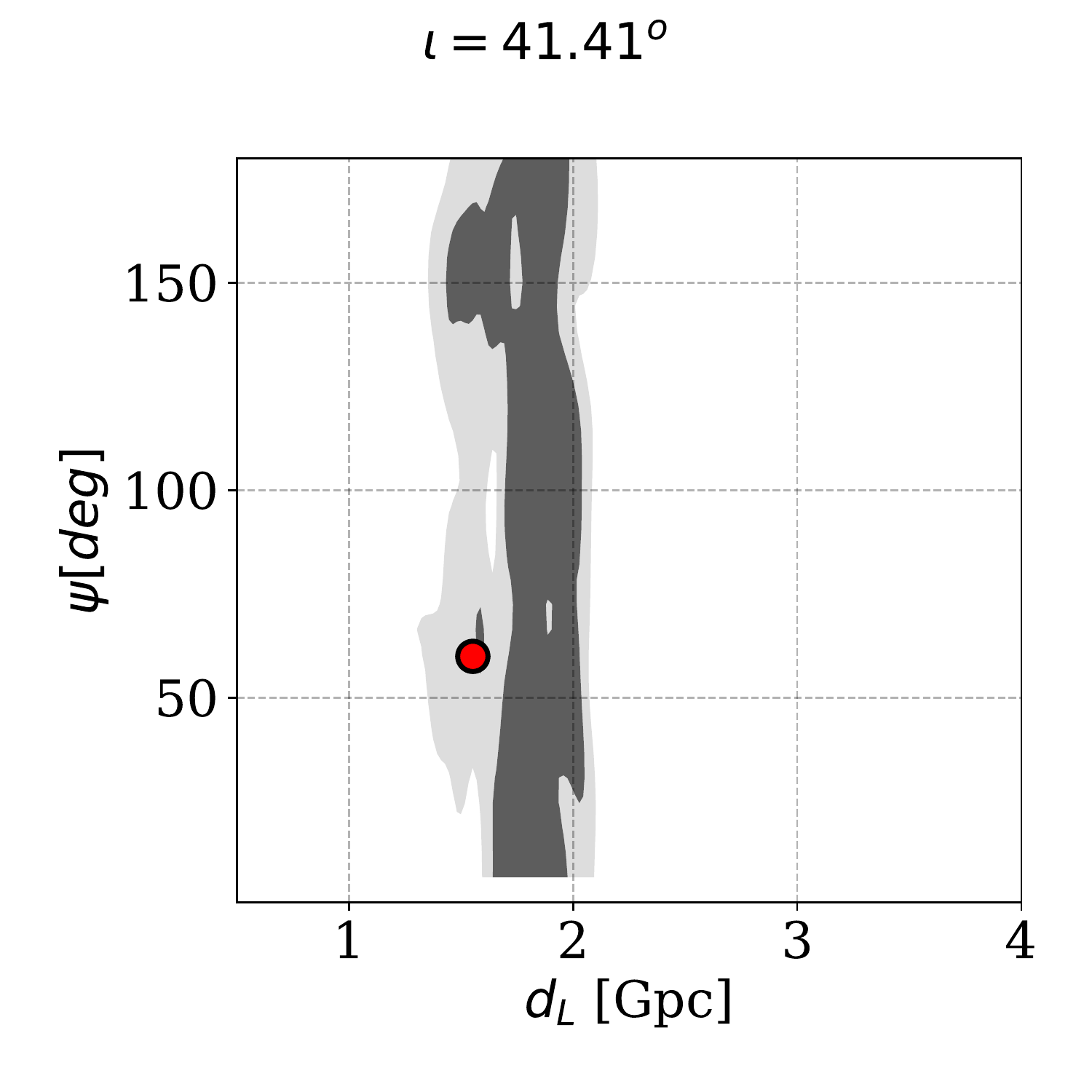}
    \includegraphics[width=.193\linewidth]{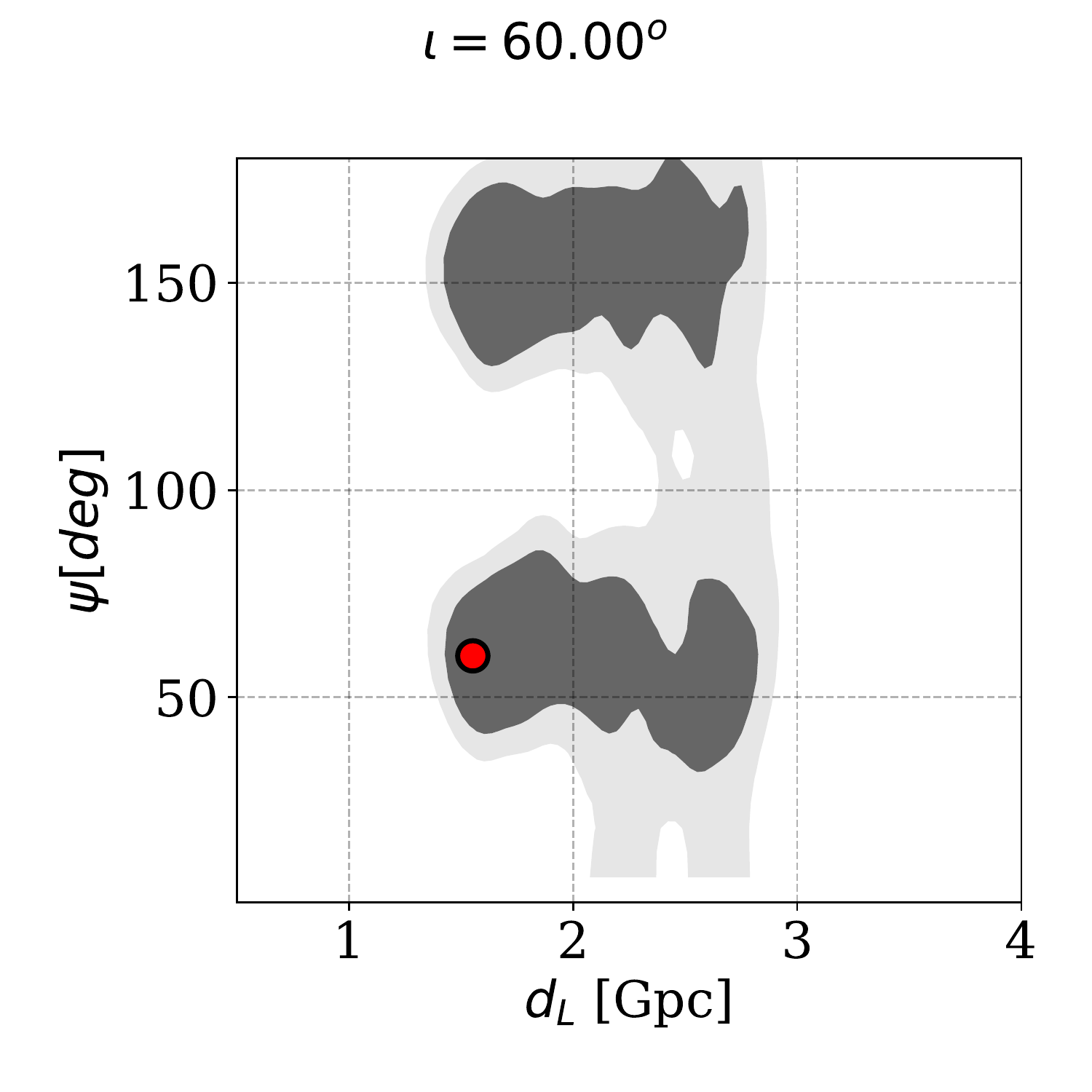}
    \includegraphics[width=.193\linewidth]{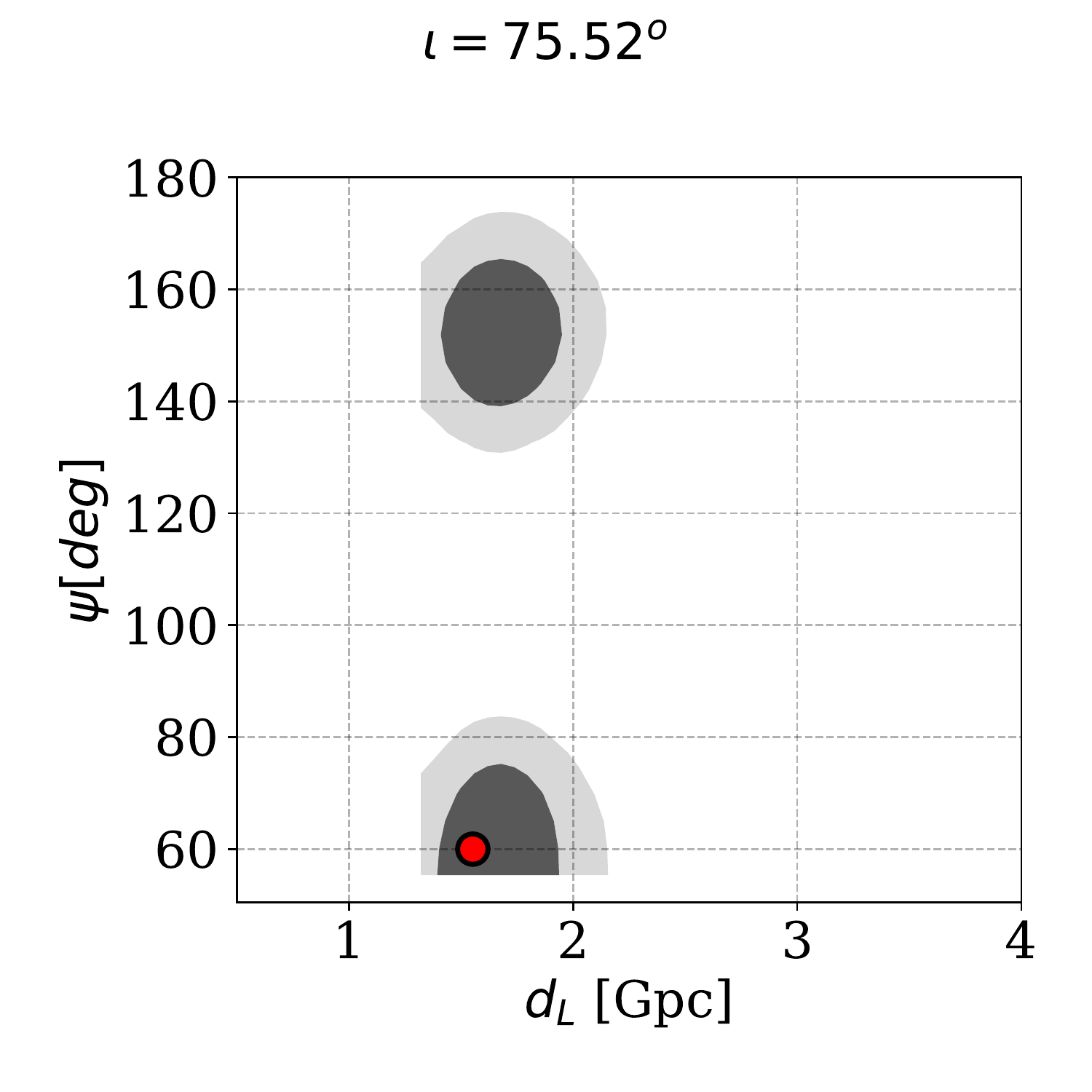}
    \includegraphics[width=.193\linewidth]{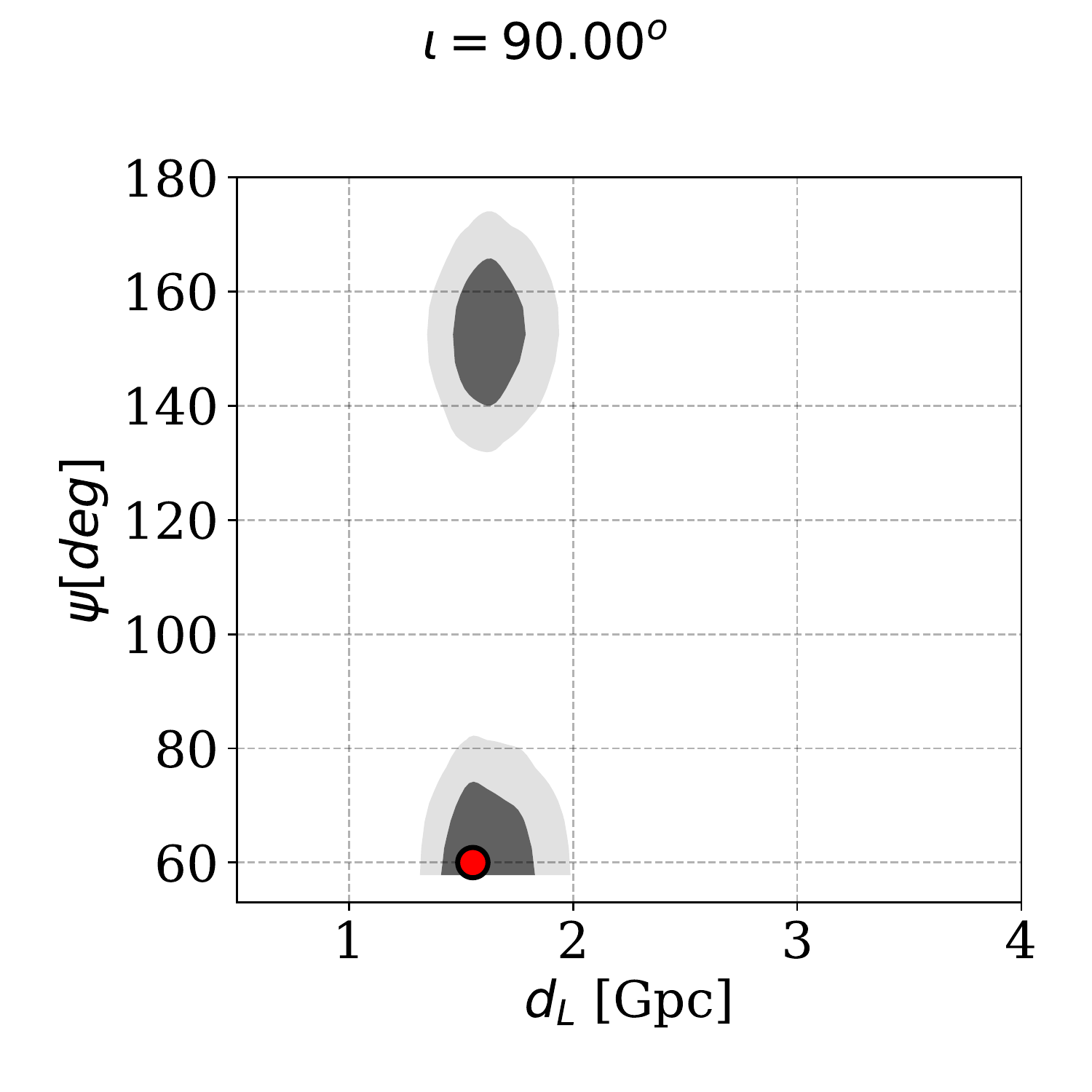}\\
    \caption{Examples of two-dimensional PDF for $\psi$ vs. $d_L$ for a single triangle-shaped interferometer showing that $\psi$ determination accuracy
      improves as $\iota\to \pi/2$, as expected from Equation (\ref{eq:cfn}), whose
      $\psi$-dependent term is maximum for $\upsilon=1$, which is $\iota=\pi/2$.
      The value for $\epsilon$ is the same as in Figure \ref{fig:di_deg}.}
    \label{fig:psid_deg}
  \end{center}
\end{figure}
  
\end{enumerate}

\subsection{Impact of detector relative orientation and localisation on
  \boldmath$d_L$ uncertainty}

\subsubsection{Two detectors}

\begin{figure}
  \begin{center}
    \includegraphics[width=.6\linewidth]{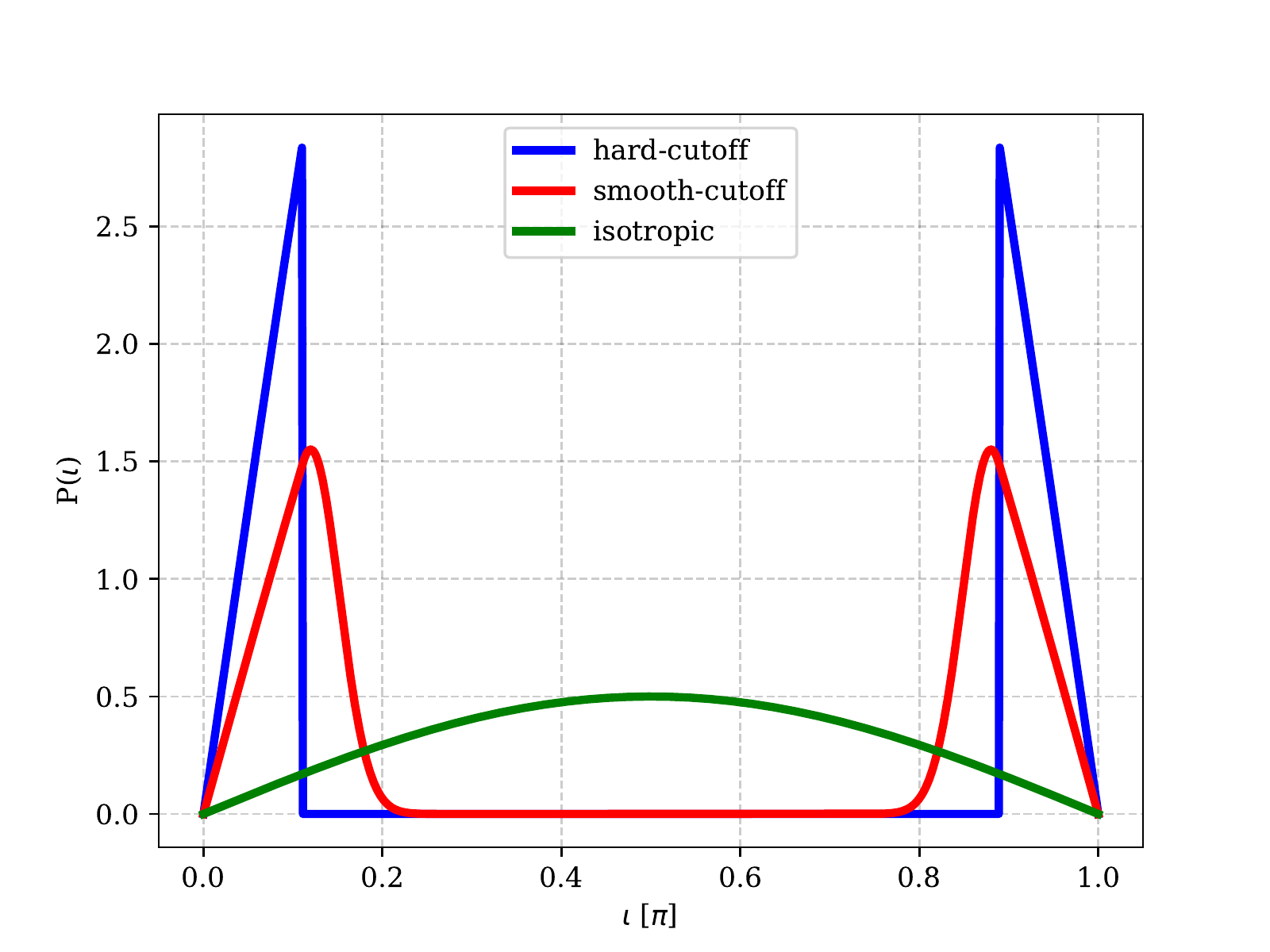}
    \caption{The three inclination angle distributions for $\iota$ used in
        injections, dubbed ``isotropic'', ``smooth cutoff'', ``hard cutoff''.}
    \label{fig:iotad_net}
  \end{center}
\end{figure}

\begin{figure}
  \begin{center}
    \includegraphics[width=.48\linewidth]{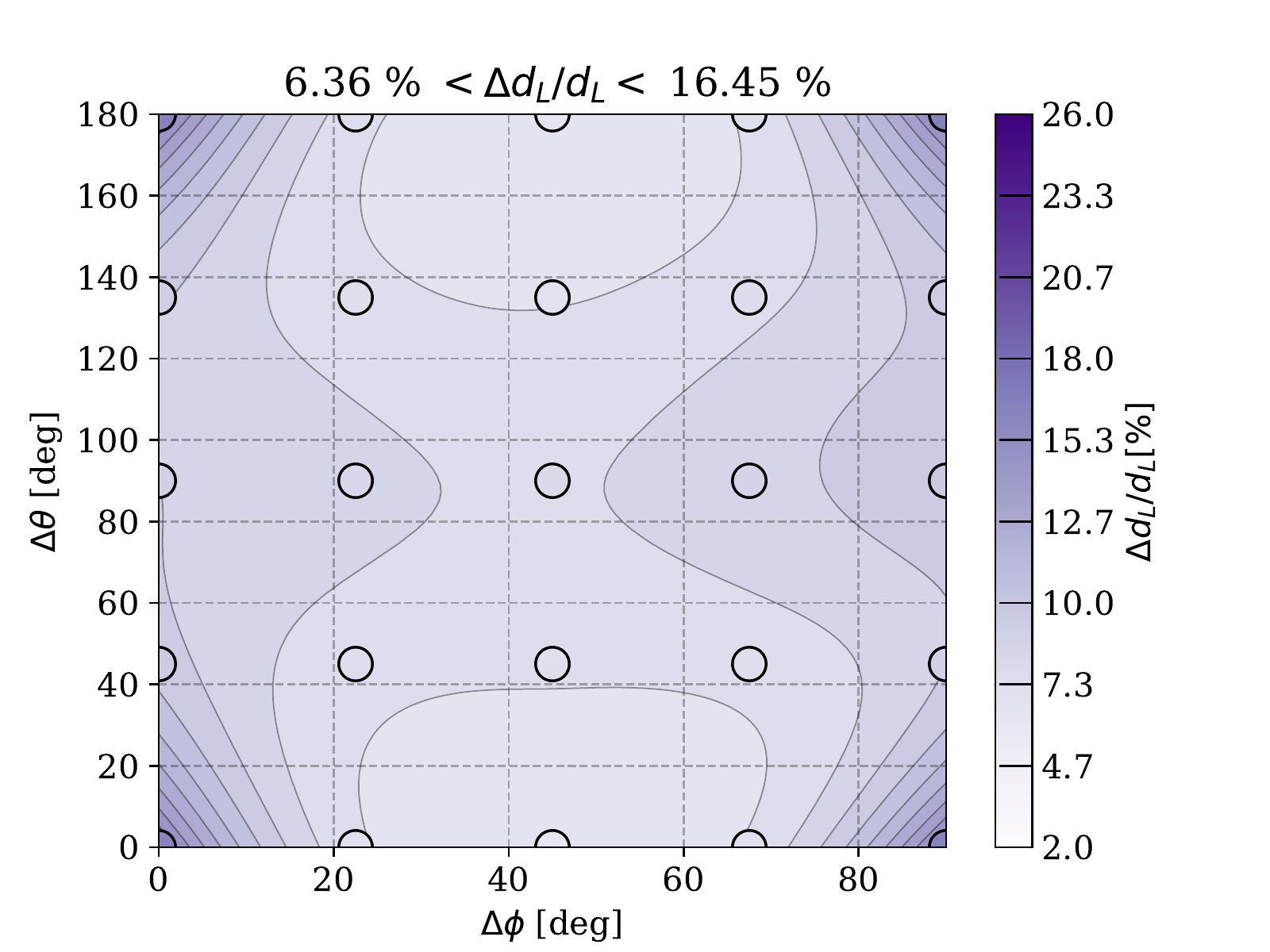}
  \includegraphics[width=.48\linewidth]{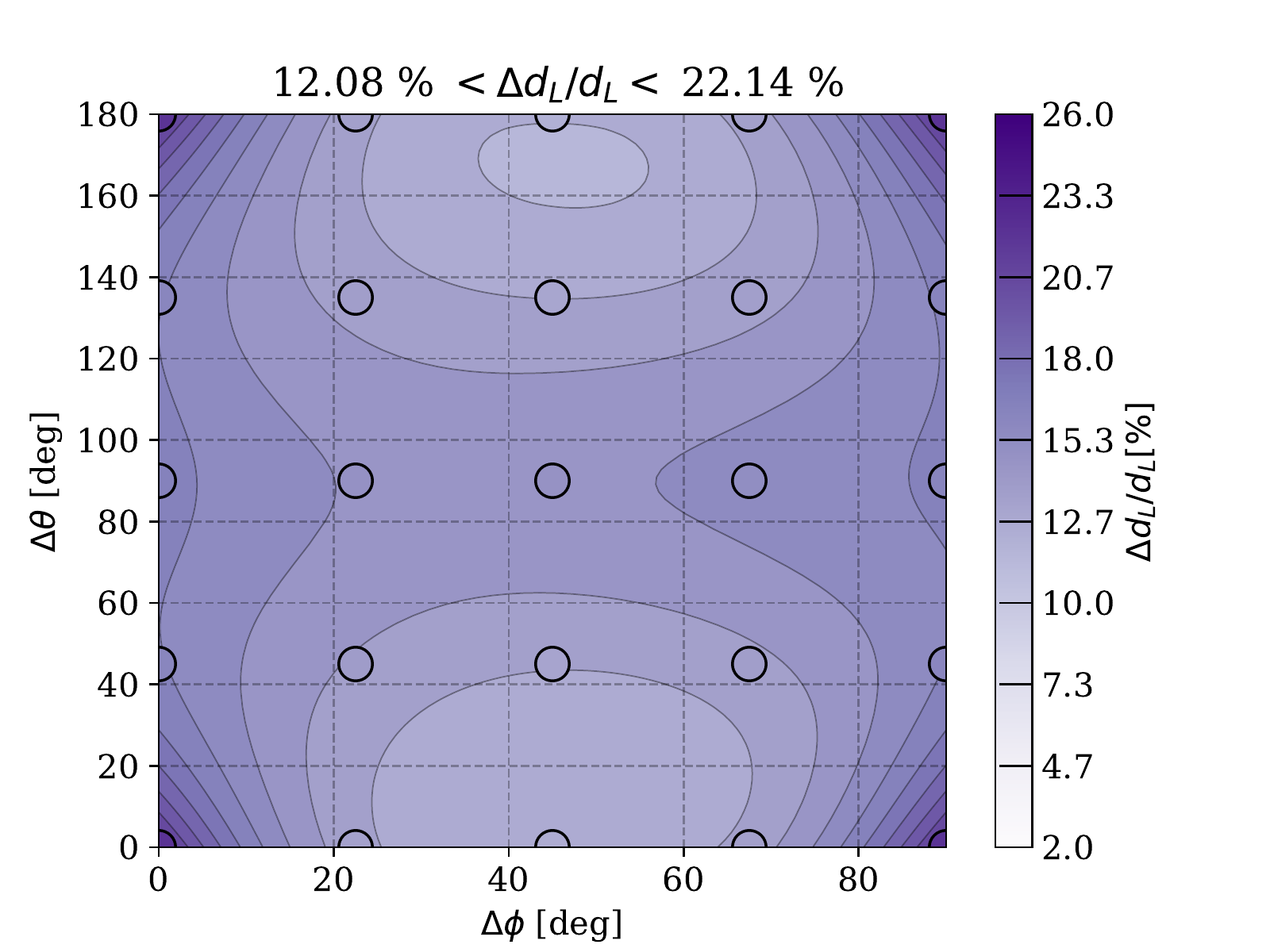}\\
  \includegraphics[width=.48\linewidth]{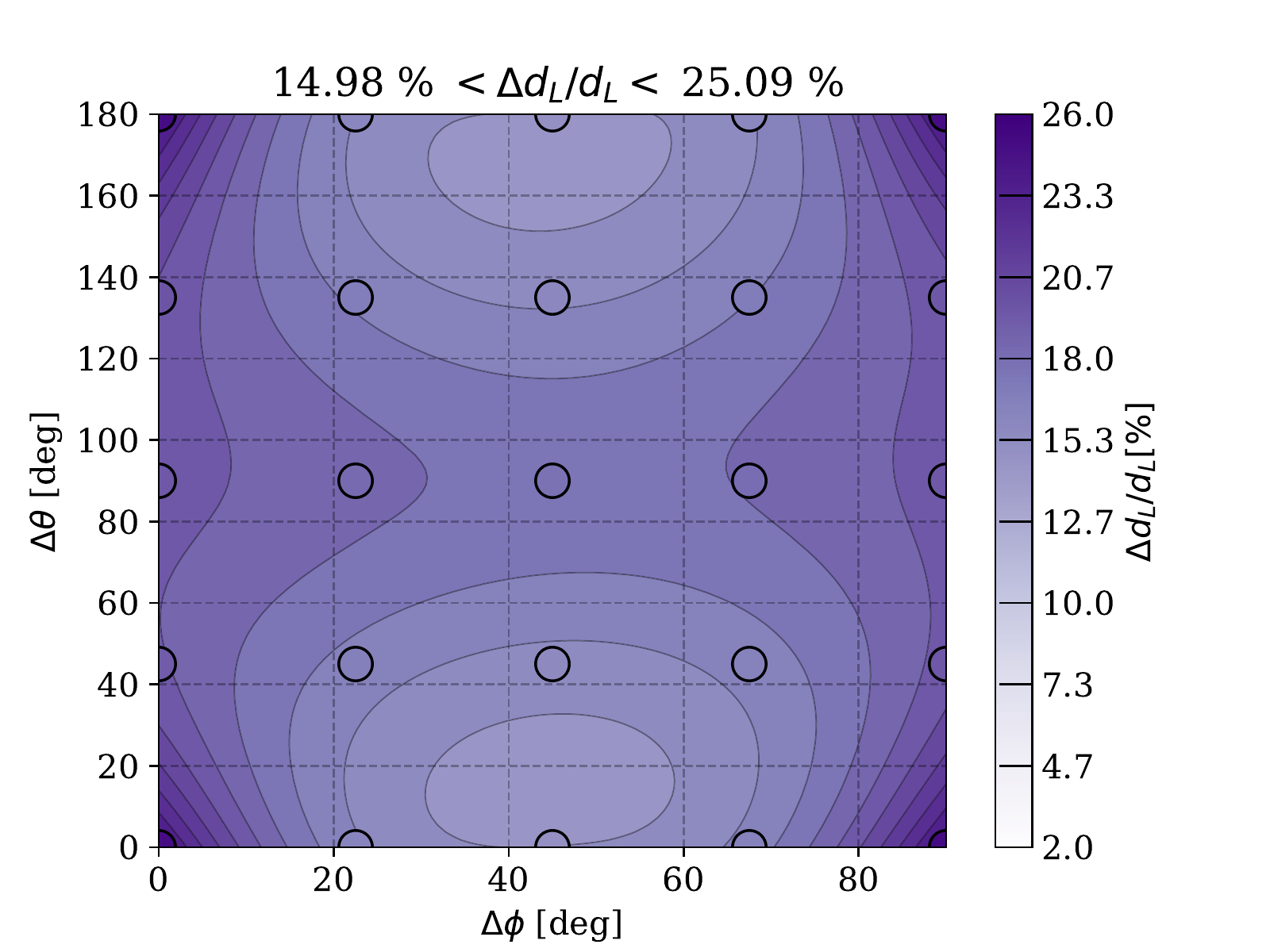}
  \includegraphics[width=.48\linewidth]{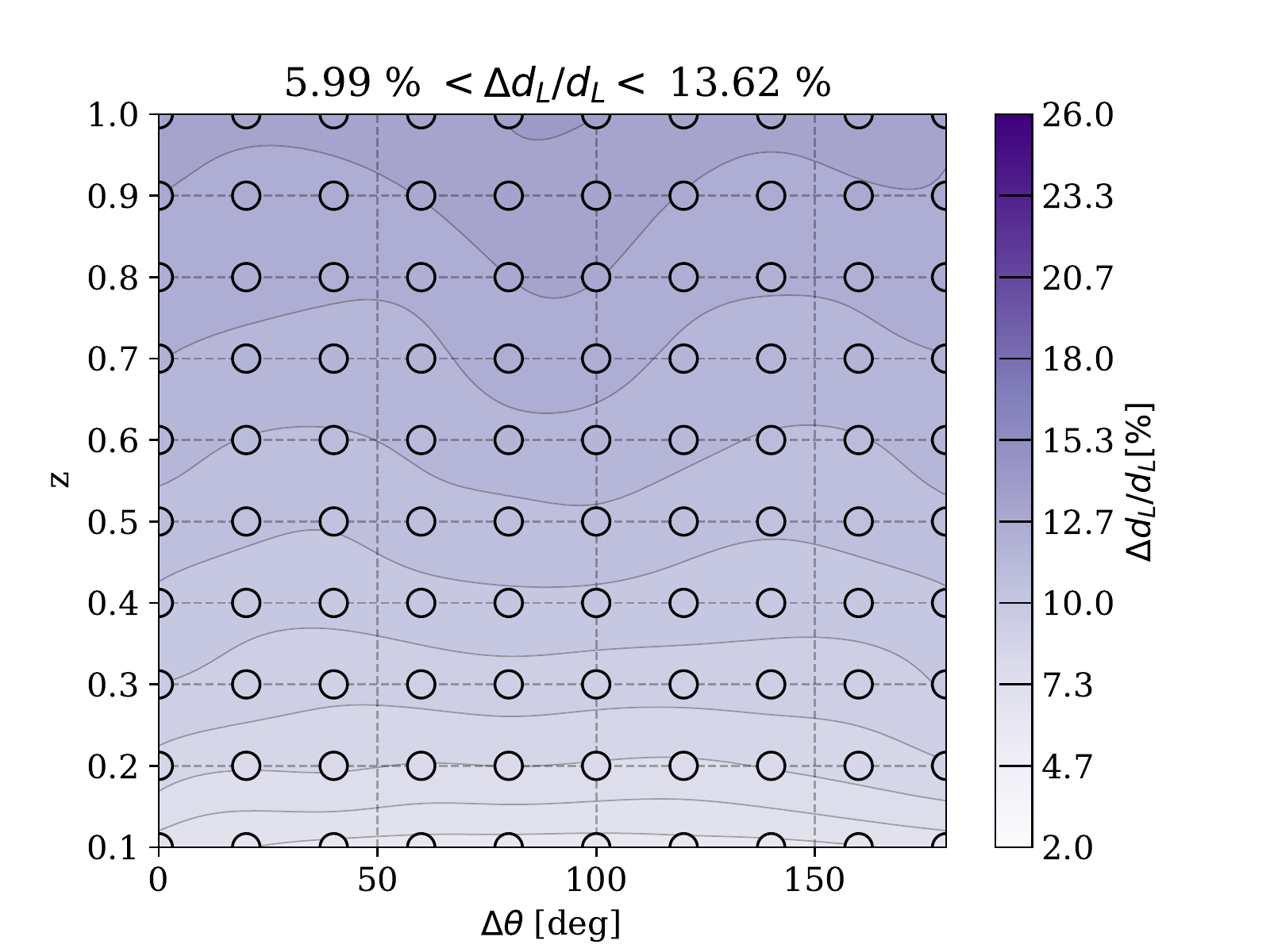}\\
  \caption{Error in distance determination averaged over source location,
    as a function of the angular distance between
    two 3G detectors. Sources are distributed isotropically
    in the sky before the $SNR_i>8$ cut in each detector.
    Source inclinations are distributed according to
    \emph{smooth-cutoff} function,
    see Figure \ref{fig:iotad_net}, Bayesian prior for $\iota$ at recovery is isotropic.
  The bottom right plot refers to two ETs, the others to two CEs with sources respectively at $z=0.1,0.55,1$.}
  \label{fig:2dets}
  \end{center}
\end{figure}

To investigate the best relative location and orientation of two detectors
we place two CE-like detectors on the earth surface at an angular distance
$\Delta\theta$ one from the other and relative axis orientation $\Delta\phi$,
with the result shown in the first three plots of Figure \ref{fig:2dets}.
The signals are simulated with three different distributions of inclination
  angles, all symmetric for $\iota\to\pi-\iota$,
  as reported in Figure \ref{fig:iotad_net} (dubbed \emph{isotropic}, \emph{smooth cutoff} and \emph{hard cutoff}) corresponding to
  $\hat L$ direction being isotropic in the 2-sphere, or to
  $\iota$ values concentrated around $0$ and $\pi$.
    Signals are produced for sources at three sample values of redshift: $z=0.1,0.55,1$

\begin{figure}
  \begin{center}
      \includegraphics[width=.48\linewidth]{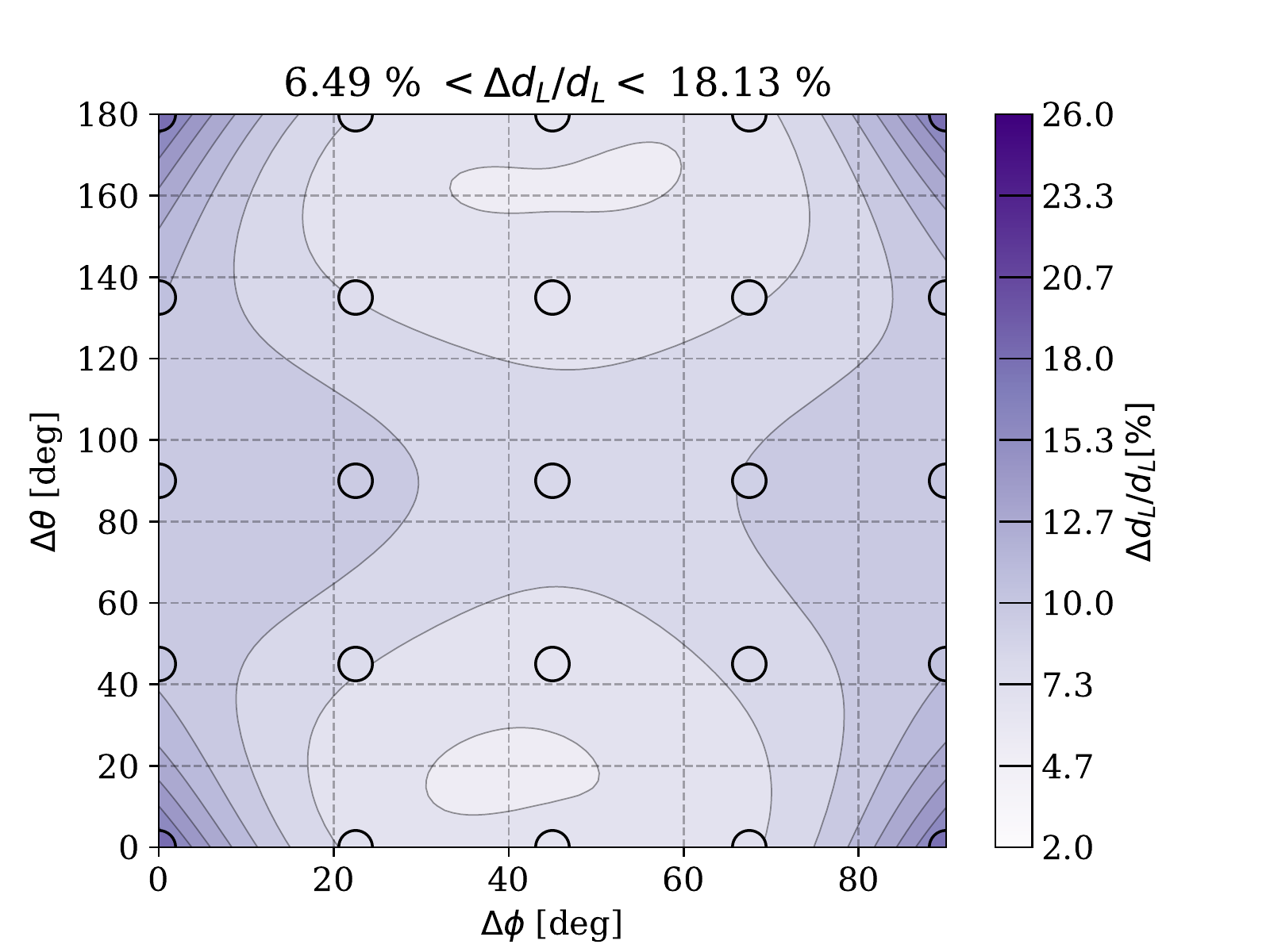}
      \includegraphics[width=.48\linewidth]{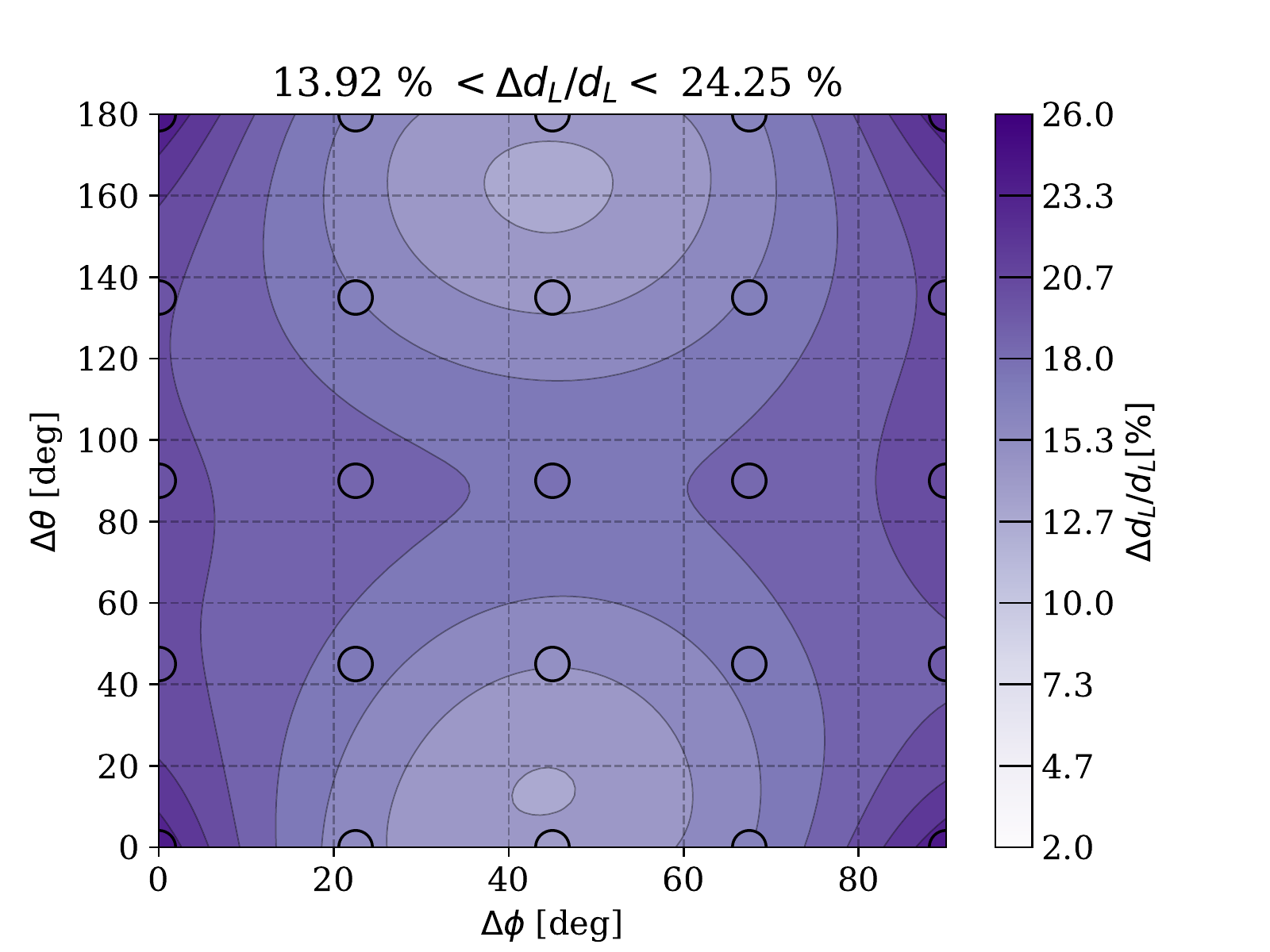}\\
      \includegraphics[width=.48\linewidth]{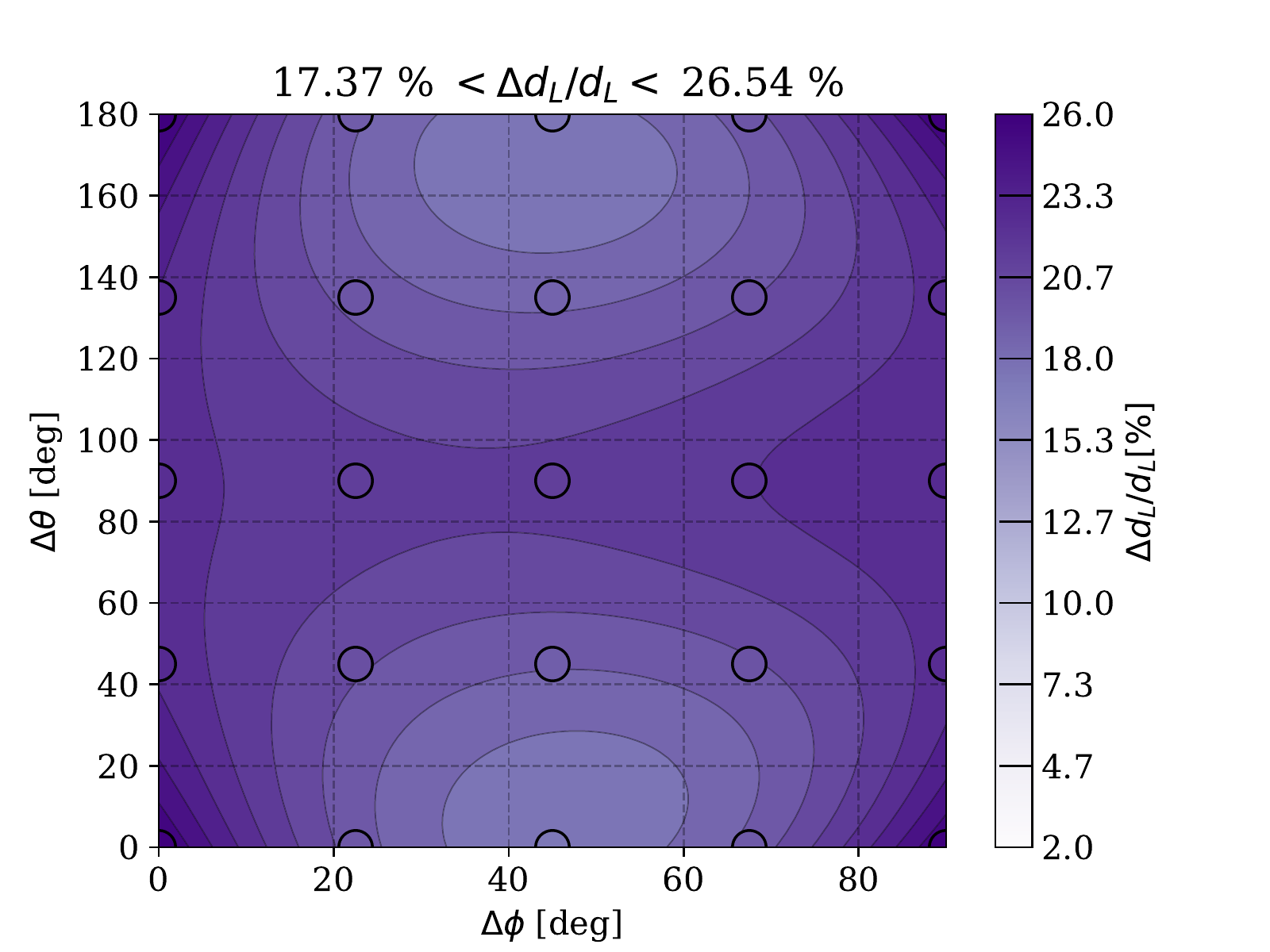}
      \includegraphics[width=.48\linewidth]{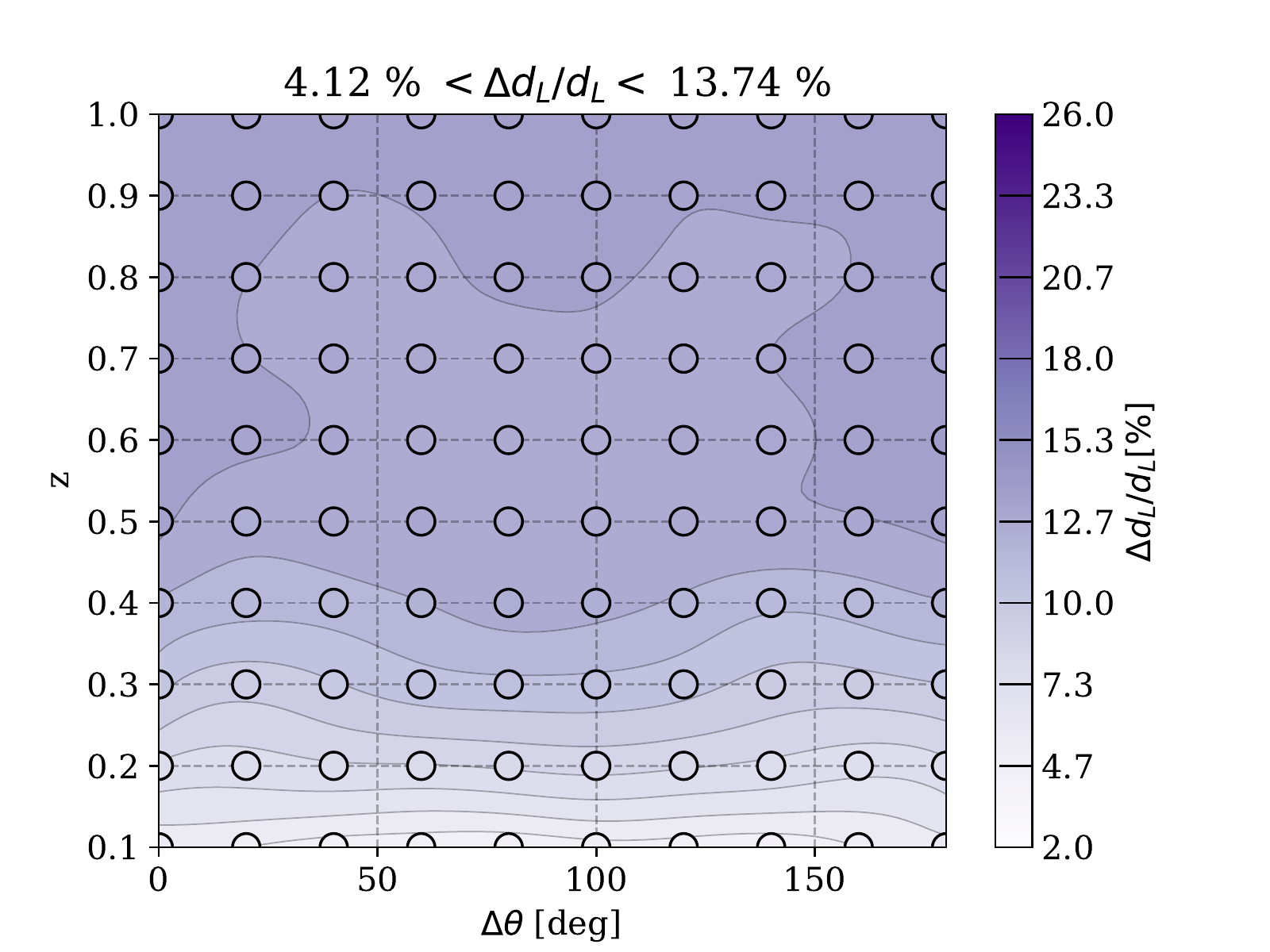}
  \caption{Same as in Figure \ref{fig:2dets} with source inclinations distributed
    isotropically on the 2-sphere.}
  \label{fig:2dets_iso}
  \end{center}
\end{figure}

The lowest uncertainty
is given by detectors either co-located or at antipodal sites, i.e. located
on parallel planes, and at $\Delta\phi=45^o$ degree, so that
such a network will have no blind spots, see Figure \ref{fig:2dets},\ref{fig:2dets_iso}, which refers respectively to $\iota$ of source distributed
according to ``smooth cutoff'' or ``isotropic'' (there is no difference in the
results between ``smooth'' and ``hard cutoff''), and recovered in both cases with an
isotropic prior in $\iota$.

In the same Figures \ref{fig:2dets},\ref{fig:2dets_iso} we also report the result of an analogue exercise with
two ET-like detectors, suppressing
the coordinate $\Delta\phi$ that does not affect the result.
In this case we find a moderate gain (a few percents) for angular separation
$\Delta\theta\simeq 40^o$ (or $\Delta\theta\simeq 140^o$), which becomes more
pronounced at larger redshift, where $SNR$s are smaller and uncertainties
larger.

The differences between Figures \ref{fig:2dets} and \ref{fig:2dets_iso} are
minimal, showing that
when the prior at recovery is isotropic in the inclination angle $\iota$,
the injection distribution in $\iota$ has little impact on $d_L$ recovery
precision.

\subsubsection{Three detectors}
\label{ssec:three_3G}

\begin{figure}
  \begin{center}
    \includegraphics[width=.49\linewidth]{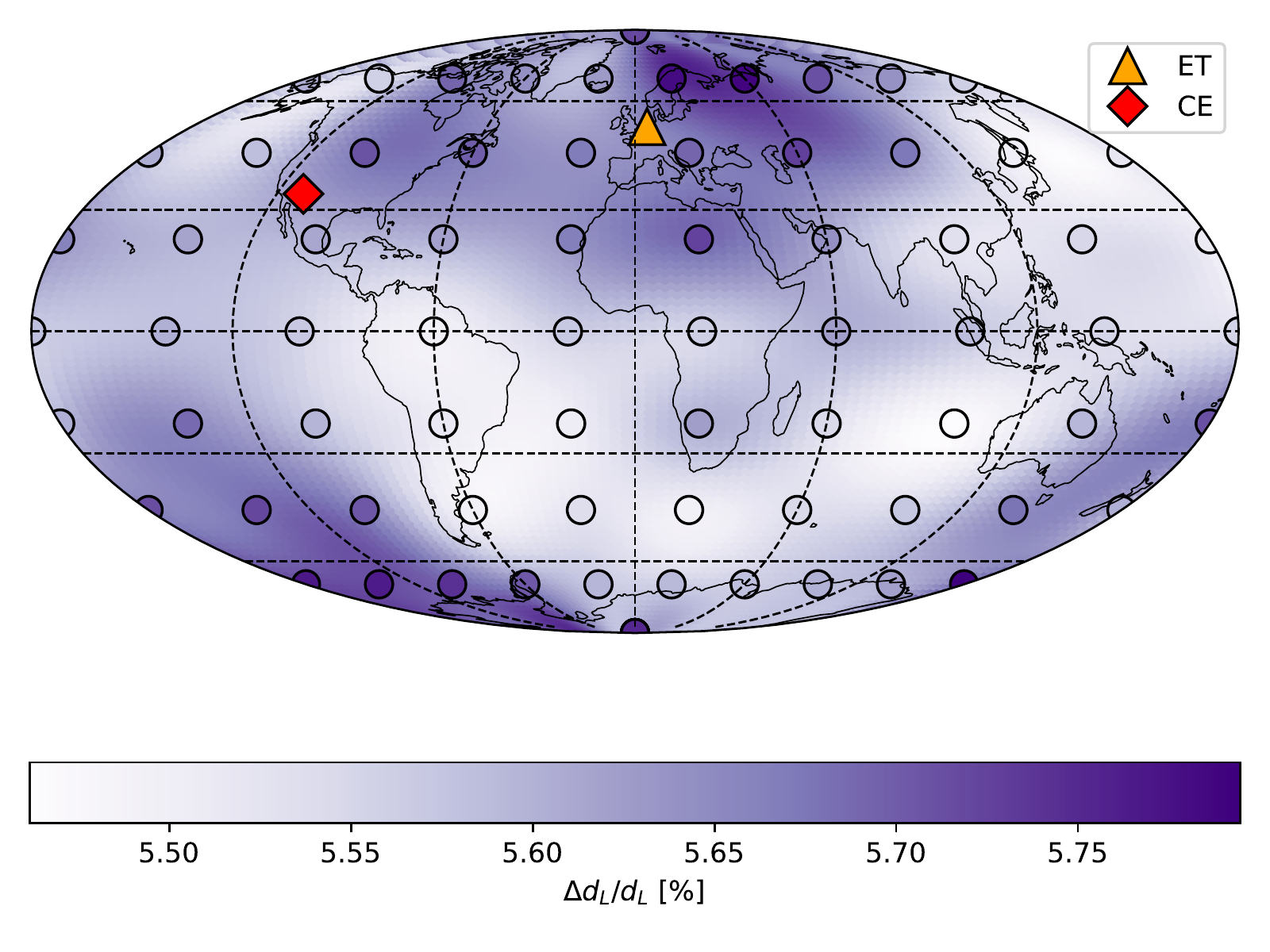}
    \includegraphics[width=.49\linewidth]{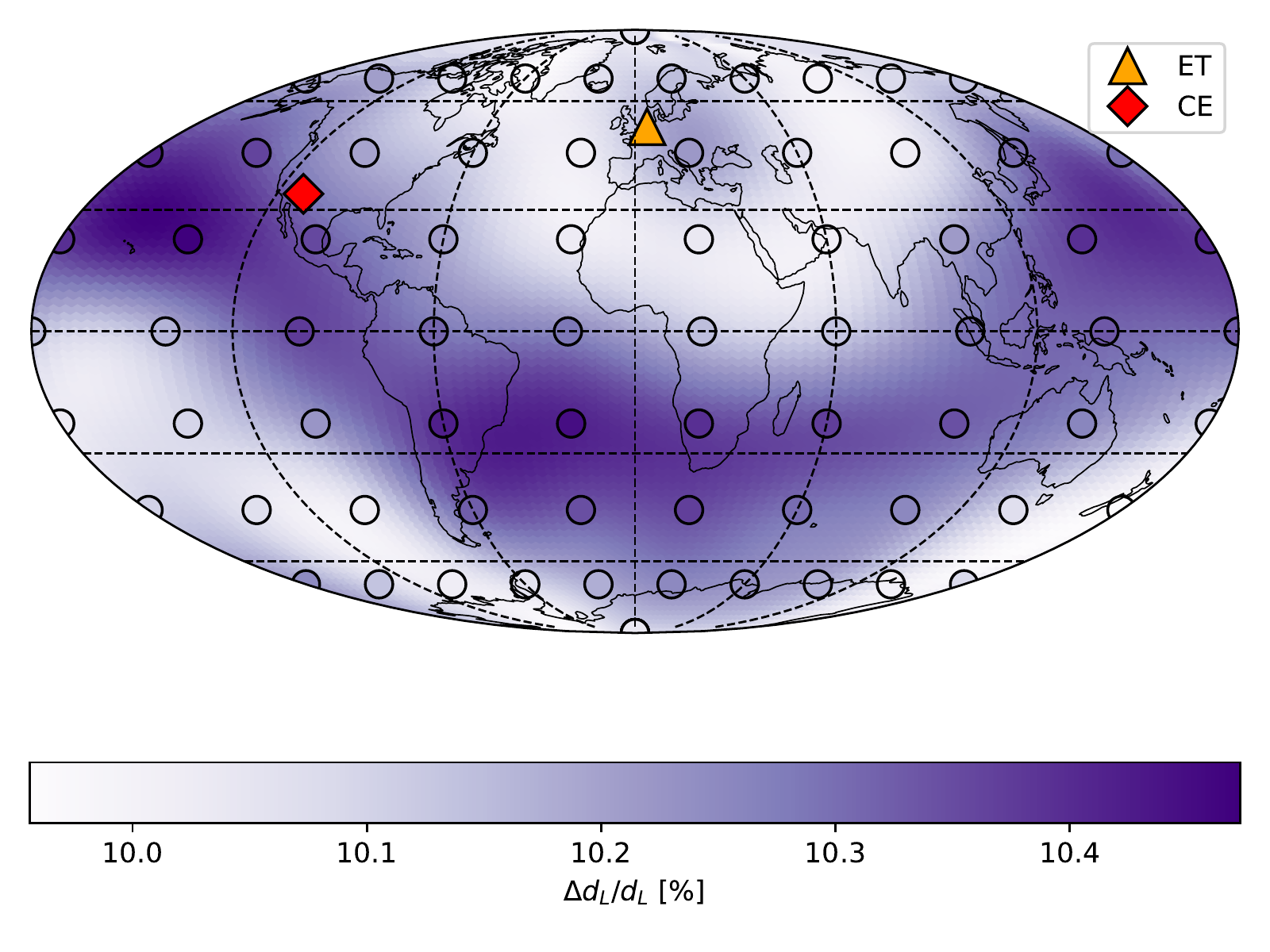}\\
    \includegraphics[width=.7\linewidth]{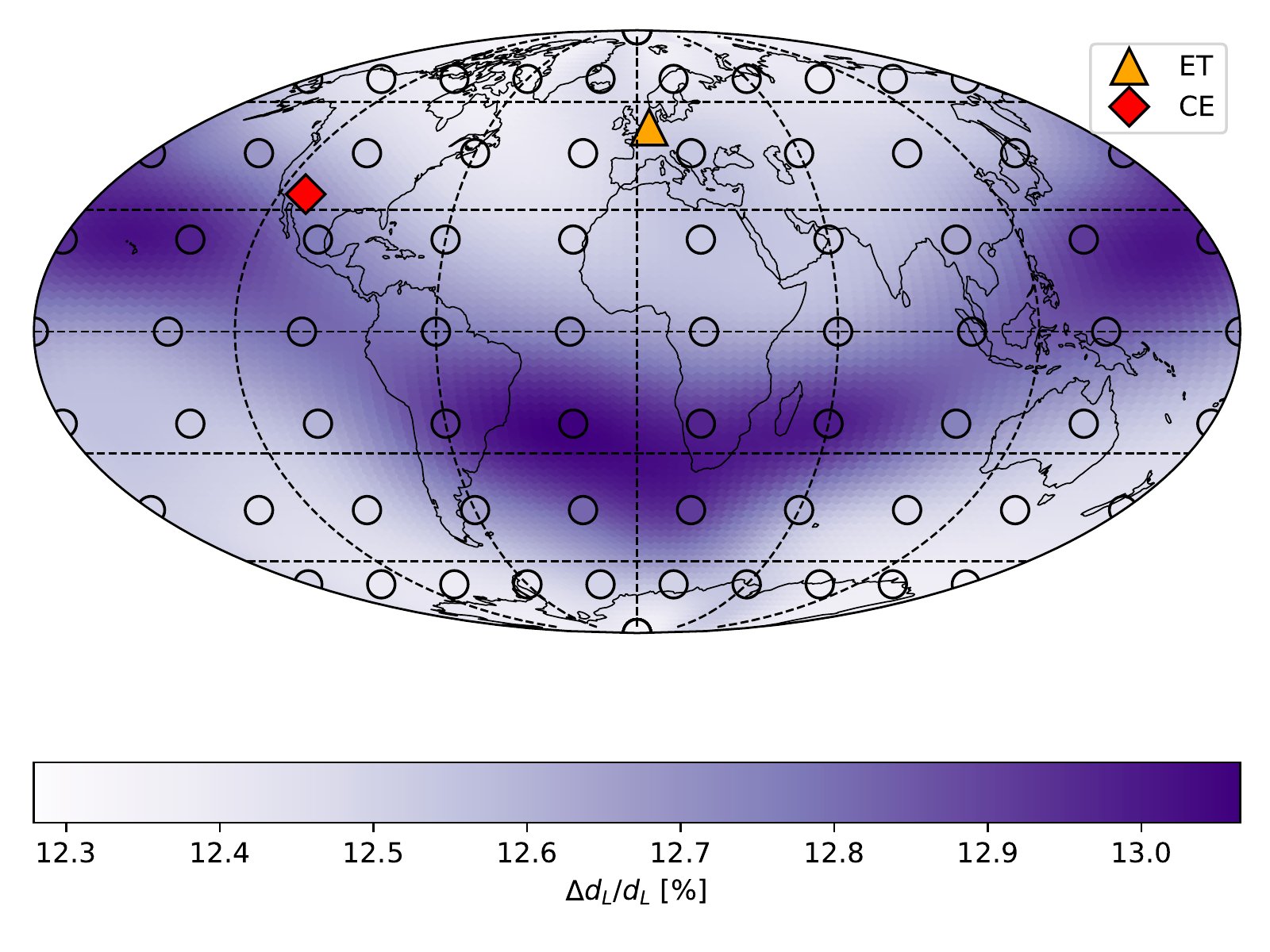}
  \end{center}
  \caption{Luminosity distance uncertainty of a network of 3 detectors (ET+CE+ET) averaged over source location, 
    depending on the location of the second ET-like detector.
    Empty circles denote the 90 trial
    location of the third detector, while the first ET and the CE detector
    are denote respectively by a yellow triangle and red diamond. The three figures refer to redshift $z=0.1,0.5,1$, respectively, moving clockwise from top left.}
\label{fig:3dets}
\end{figure}

Finally we fix the location of a ET-detector and a CE-one, corresponding
to an angular distance of $77^o$. In this case we verified how the relative luminosity
distance uncertainty averaged over source sky location varies with the position of a third ET-like detector,
with results displayed in Figure \ref{fig:3dets} for redshift $z=0.1,0.5,1$,
showing overall a mild (sub-percent) dependence on the location of the third
detector. When dealing with three detectors the relative measurement error
on $\Delta d_L$ depend very mildly on the source sky location, as shown
by Figure \ref{fig:three_3Giso}.

\subsection{Impact of \boldmath{$\iota$} distribution and type of network on \boldmath{$d_L$} uncertainty}

Beside the obvious importance of the number and quality of detectors in the
network, another important feature in the forecast of luminosity distance
uncertainty is the \emph{source distribution of inclination angles}, for
  which we adopted the three distribution function in Figure \ref{fig:iotad_net}.

Note that it is not a priori clear what distribution will be seen by 3G detector.
While 2G ones are likely to see a distribution of small values for
$min(\iota,\pi-\iota)$, which give larger $SNR$s, since 3G detector will have a
much larger reach, they should in principle see a distribution closer to the
isotropic one, as observed in \cite{Schutz:2011tw,Vitale:2016aso}.
However the selection of \emph{bright} standard sirens may bias the observation
towards GW signals accompanied by short GRB, which are
expected to be somehow beamed \cite{Metzger:2011bv} hence more likely to be
observed for small $\iota$ or $\pi-\iota$.
On the other hand, short GRBs do not have good sky localisation, which can be achieved
at high degree of accuracy with optical counterparts like kilonovae, that
are broadly expected to be isotropically emitting
\cite{Cowperthwaite:2017dyu}, hence can support the expectation of 
a $\sin\iota$, isotropic distribution of sources.

\begin{figure}
  \includegraphics[width=.8\linewidth]{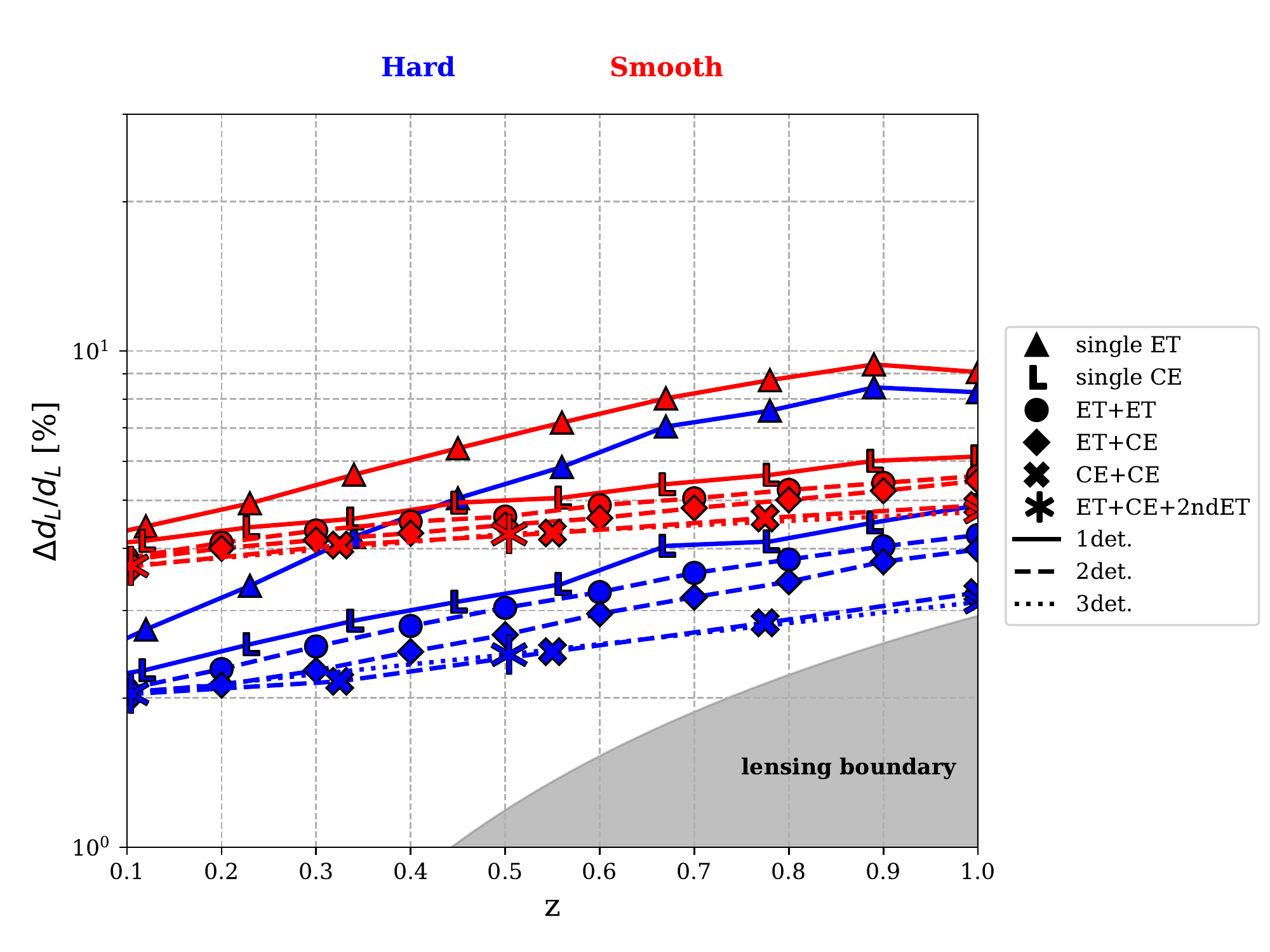}
  \includegraphics[width=.8\linewidth]{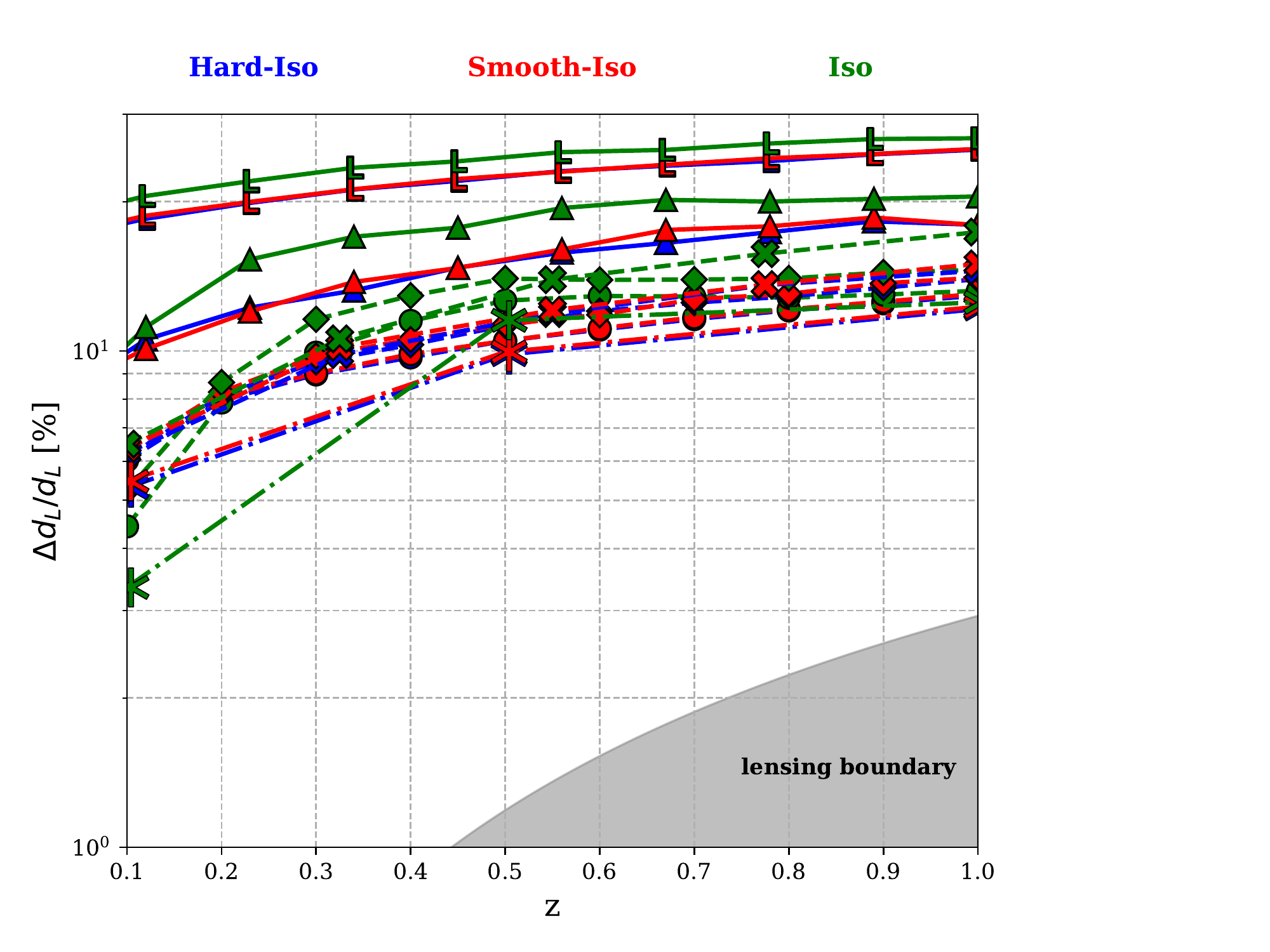}\\
  \caption{Average distance measure error for various network of ET- and CE-like
    detectors as a function of redshift. The top graph shows the case of anisotropic distribution
    of inclination angles with prior at recovery. The bottom one shows the case of
    isotropic prior at recovery for all cases of $\iota$ injection distribution. See Figure \ref{fig:iotad_net} for $\iota$ injection distributions (\emph{Hard}, \emph{Smooth}, \emph{Iso}).}
\label{fig:err_iotad_net}
\end{figure}

We then summarize the result for the average $d_L$ uncertainty
for 6 different network of detectors: \{ET, CE, CE+CE, ET+CE,ET+ET, ET+ET+CE\}
given the three different $\iota$ source distributions.
The $\iota$ prior at recovery is chosen to be equal to the injected
cutoff distribution in the top plot of Figure
\ref{fig:err_iotad_net}, and equal to an isotropic distribution in the bottom
plot of the same Figure.

From Figure \ref{fig:err_iotad_net} we can draw interesting conclusions
about the impact of the underlying inclination
angle distributions and on the network features:
\begin{enumerate}
\item In the case GW sources are characterized by ``polar'' inclination angles,
  i.e. small values of either $\iota$ or $\pi-\iota$,
  folding in such information at recovery leads to
  a significant improvement (up to a factor $\sim 5$) in $d_L$ uncertainty
  determination. This is shown by the comparison
  of the two plots in Figure \ref{fig:err_iotad_net}, where on the top plot
  results are shown for injections distributed at small angles only
  (see ``hard-'' and ``smooth-cutoff'' in Figure \ref{fig:iotad_net}, using
  recovery prior equal to injection distribution),
  and the bottom plot has been obtained using
  an isotropic prior at recovery, i.e. $\propto \sin\iota$.
  In particular, in the top plot there is virtually no difference between the
  cases of hard- and smooth-cutoff, and also little difference between injections
  isotropic on the sphere or concentrated near the poles.

\item 
  As expected, adding a 3G detectors to an existing network is beneficial to the
  $d_L$ recovery precision,
  but less obvious is the effect of adding a CE-like detector instead of an ET-like one.
  A CE-CE network can ensure roughly the same precision as ET+ET+CE, and it is slightly better
  in terms of $d_L$ precision than a ET+CE system, showing that the better design sensitivity we
  adopted for CE compared to ET, see left plot in Figure \ref{fig:reach3G}, has a non-negligible
  effect when it comes to forming a network which has already good sky coverage, like
  e.g. a network of two $L$-shaped detectors.
   
\end{enumerate}

For the dispersion of $\Delta d_L/d_L$ values see Figure \ref{fig:disp_dl}
in the 3 detector case of Figure \ref{fig:err_iotad_net}.

Overall uncertainty in $d_L$ smoothly increases with redshift, with results in
broad agreement with the cumulative distributions shown
in \cite{Iacovelli:2022bbs}, see also \cite{Vitale:2016icu} for binary black holes, even if a more detailed comparison is not possible
as here, differently from there, we present results broken at specific redshift.
For the reader interested in comparing with present observations, we report in
Appendix \ref{app:pan} the same result of Figure \ref{fig:err_iotad_net} superimposed to scatter plots of luminosity distance uncertainties obtained with 2G
detectors Advanced LIGO and Virgo in their first three observation runs
\cite{LIGOScientific:2018mvr,LIGOScientific:2020ibl,LIGOScientific:2021djp},
and with the \emph{standard candle} catalogue \cite{Pan-STARRS1:2017jku}.

Note that for a wide and sensitive enough detector network (ET+ET+CE) or CE+CE
it is possible at moderately high redshift ($\sim 0.7$) to almost reach the limit
on $d_L$ uncertainty imposed by the  \emph{lensing} intervening between source and observer, whose approximate value can be
found in \cite{Zhao:2010sz}. Also in \cite{Shan:2020esq} it is
argued that with two or three 3G detectors working at design sensitivity
it may be worth including de-lensing in the analysis of the signals,
anticipating to 3G detectors what was foreseen for space interferometers \cite{Tamanini:2016zlh}. Note while some present estimate indicate that
  having an EM counterpart of a GW detection may be challenging for source
  at redshift larger than $\sim 0.7$ \cite{Belgacem:2019tbw}, there exist
  large uncertainties for the expected reach of next decades
  target-of-opportunity searches.

\section{Conclusions}
\label{sec:conclusions}

With the goal of contributing to the use of gravitational wave signals from coalescing binaries
as standard sirens to reconstruct the cosmic expansion history of the Universe,
we analyzed the projected uncertainty measures of luminosity distances of third generation detectors.
Observatories like the triangular Einstein Telescope, with arms at 60$^o$, and the $L$-shaped
Cosmic Explorer, with arms at 90$^o$, are currently planned to start taking data
just over a
decade from now, but some of their features, like the exact location and topology, have not
been finalized yet.

Luminosity distance precision measurement affects directly the measure of cosmological parameters,
but luminosity distance correlates with a relatively large number of angular variables defining the relative location and
orientation of source and detector.
Such correlations can degrade the expected precision measurements, e.g.
the one relying on Fisher matrix approximation, thus requiring a numerical Bayesian framework for
a consistent analysis.
For the sake of definiteness, we focused on bright standard sirens
of binary neutron stars, for which a host galaxy can be identified
and sky localization obtained with negligible error, thus reducing the extrinsic angular variables
to correlate with distance, to inclination, polarisation and phase-shift.
We neglect arrival time and extrinsic parameters like masses which
can be constrained with high accuracy from the GW phasing, and spins that are expected to be small
for neutron star binaries giving rise to bright standard sirens.

We have neglected completely the effect of possible tidal deformation of
neutron stars, which demand an accurate modelling
of the waveform close to the merger, that is way beyond the scope of our work.

Our main results can be summarised as follows:

\begin{enumerate}

\item While the presence of bimodality in the luminosity distance ($d_L$) versus inclination ($\iota$) angle distribution is a
  well known feature of detections by single $L$-shaped interferometers,
  we have quantitatively linked
  such impossibility to separate individual polarisation contributions to a single scalar parameter, the
  $\epsilon_D$ first introduced in \cite{Cutler:1994ys} (simply $\epsilon$ in this work). This parameter relates to the information of how much the sub-dominant polarisation is present in the combined detector output.
  In particular we have shown that detectors like the ones
  forming a triangular interferometer, which cover all sky localisations without
  blind directions,
  have $\epsilon<1$ for virtually all of the sky, and they can present
  bimodality in $d_L$-$\iota$ recovery only for specific
  directions with $\epsilon\sim 1$.
  
\item Another well-known feature of $d_L$ vs. $\iota$ uncertainty is the reduced error uncertainty
  for $\iota\to\pi/2$ for triangular interferometers. We found
  that this is a generic feature ascribable to an improved precision in the
  determination of the \emph{polarisation} angle, whose better constrained
  values are correlated with $d_L$ measures.
  
\item We have shown quantitative measures of $d_L$ uncertainties for a
  variety of networks
  made of up to three third generation detectors. Beside qualitative results
  presented in plots, we showed that given
  a network of at least two misaligned detectors, which then have
  virtually no blind spots in the sky, the best way to increase the precision
  measurement is to add a more sensitive detector, rather than adding an equally performing one, even if with more arms. Moreover we have shown
    that with three detectors one can almost reach the measurement error
    level set by lensing, which start to be at per-cent level from $z\gtrsim 0.6$.

  \item We have quantified how the inclination angle distribution affect the $d_L$
  uncertainty measures, with the result that knowing the underlying $\iota$
  distribution can improve up to a factor $5$ the luminosity distance
  uncertainty.
  
\end{enumerate}

Obvious generalisations of the present work include to explore the 
non symmetric mass and spinning case. However, apart for the case of precessing
binaries which however are not expected for bright standard sirens \cite{Vitale:2014mka},
this features are expected to induce quantitative, rather than qualitative
changes of the results obtained here.
One feature that could change the picture substantially is instead given by
matter/tidal effect of neutron star, which are relevant close to the merger
phase. Such effects are not only relevant for understanding the state of matter
at high density inside the neutron stars, but have a non-trivial impact on
cosmological parameter estimation, as they introduce into the phasing
of the gravitational waveform a term that depends explicitly on redshift
\cite{Messenger:2011gi}.
This would give a handle to estimate redshift with gravitational
information alone, which however require accurate development of accurate
and exhaustive matter waveforms, see e.g. \cite{Gonzalez:2022mgo} for a
database.

\appendix
\section{Polarisation angle}
\label{app:psi}

The radiation frame relative to the $i$-th detector is defined
taking the $\hat z_{rad}$ axis along the line pointing from the source to the
observer ($\hat N$)
and the $\hat x_{rad}$ axis in the $\hat z_i$-$\hat z_{rad}$ plane,
being $\hat z_i$ the unit vector normal to the plane of the detector, see Figure \ref{fig:det_angles} for detector and radiation geometry.

The polarisation angle $\psi_i$ is then conventionally defined as the angle
from $\hat x_{rad}$ to the line of \emph{ascending nodes}, which is determined
by the unit vector parallel to $\hat N\times \hat L$, being $\hat L$ the unit
vector parallel to the source angular momentum.
One then has
\be
\label{eq:xrad}
\ba{rcl}
\hat x_{i,rad}&\equiv&\ds \frac{\hat z_i-(\hat z_i\cdot \hat N)\hat N}
     {|\hat z_i-(\hat z_i\cdot \hat N)\hat N|}\,,\\
\hat y_{i,rad}&\equiv&\ds\hat N\times \hat x_{i,rad}\,,
\ea
\ee
and
\be
\ba{rcl}
\cos\psi&=&\ds\hat x_{i,rad}\cdot \frac{\hat N\times\hat L}{|\hat N\times\hat L|}\,,\\
\sin\psi&=&\ds\hat y_{i,rad}\cdot \frac{\hat N\times\hat L}{|\hat N\times\hat L|}\,,
\ea
\ee
from which it follows
\be
\label{eq:pol}
\ba{rcl}
\ds\tan \psi_i &=&\ds
\frac{(\hat N\times\hat z_i)\cdot(\hat N\times\hat L)}
     {z_i\cdot(\hat N\times \hat L)}\\
     &=&\ds
\frac{\hat L\cdot \paq{\hat z_i-\pa{\hat z_i\cdot \hat N}\hat N}}
     {\hat z_i\cdot\pa{\hat N\times \hat L}}\,.
     \ea
     \ee
     In the specific case when $\hat L\parallel \hat N$, the polarisation angle
     $\psi_i$ is not defined, as it is degenerate with a rotation in the plane
     of the orbit.
The polarisation angle is also undefined if $\hat z\parallel \hat N$, in which case one cannot define the radiation triad.

Note that the angles $\iota,\psi_i$, with $cos\iota\equiv \hat L\cdot \hat N$,
determine the polar angles of $\hat L$, whose explicit components in the
reference frame where $\hat N\parallel \hat z$ and $\hat z_i$ is in the $\hat x$-$\hat z$ plane, are:
\be
\hat L=(\sin\iota\sin\psi_i,-\sin\iota\cos\psi_i,\cos\iota)\,,\qquad \hat N\parallel \hat z\,.
\ee
We denote by $\alpha,\beta$ the polar angles defining $\hat N$,
($\alpha$ is the right ascension and the declination angle $\delta$ is related
to $\beta$ via $\delta=\pi/2-\beta$) in the frame in which the reference vector
$\hat z_i=(0,0,1)$:
\be
\label{eq:Npole}
\hat N=(\sin\beta\cos\alpha,\sin\beta\sin\alpha,\cos\beta)\,,\qquad \hat z_i\parallel \hat z\,.
\ee

The transformation taking from the source frame to the radiation frame is
$R_z(\psi-\pi/2)R_y(\iota)R_z(\phi)$, and the one taking $\hat N$ from the form
(\ref{eq:Npole}) to the canonical form  $(0,0,1)$ is
$\paq{R_z(\alpha)R_y(\beta)}^{-1}$.

For a detector at latitude $\lambda$ and longitude $u$, with $\hat N$ given by
Equation (\ref{eq:Npole}), one has
\be
\ba{rcl}
\ds \hat z_i&=&\ds \pa{\cos\lambda \cos u,\cos\lambda\sin u,\sin \lambda}\,,\\
\ds \hat x_i&=&\ds \left(\cos\lambda \cos u-\cos\alpha\cos\beta\sin\lambda\sin\beta-\cos\lambda\cos\pa{u-\alpha}\cos\alpha\sin^2\beta,\right.\\
&&\ds\, \cos\lambda \sin u-\sin\alpha\cos\beta\sin\lambda\sin\beta-\cos\lambda\cos\pa{u-\alpha}\sin\alpha\sin^2\beta,\\
&&\ds\left.\sin\lambda\sin^2\beta-\cos\lambda\cos\pa{u-\alpha}\cos\beta\sin\beta\right)/{\cal N}\,,\\
\ds{\cal N}^2&\equiv&\ds 1-\paq{\cos\beta\sin\lambda+\cos\lambda\cos\pa{u-\alpha}\sin\beta}^2\,,
\ea
\ee
and one can then find that $\psi_i=\psi_0+\delta\psi_i$ where $\delta\psi_i$ is
determined as
\be
\ba{rcl}
{\cal N}\cos\delta\psi_i&=&\ds \sin\lambda\sin\beta-\cos\lambda\cos\pa{u-\alpha}\cos\beta\,,\\
{\cal N}\sin \delta\psi_i&=&\ds -\cos\lambda\sin\pa{u-\alpha}\,,
\ea
\ee
which shows that $\delta\psi_i$ depends only on the location of the source and
not on the reference polarisation angle $\psi_0$, as stated in Section
\ref{sec:method}.

We conclude this Appendix by reporting the explicit expression of the angle
$\bar\psi$ defined in Equation (\ref{eq:xi_diag}):
\be
\label{eq:bpsi}
\ba{rcl}
\ds\cos 4\bar\psi &=&\ds\frac{\Xi_{++}-\Xi_{\times\times}}{\Xi_0}\,,\\
\ds\sin 4\bar\psi &=&\ds\frac{2\Xi_{+\times}}{\Xi_0}\,,\\
\ds\Xi_0^2&\equiv&\Xi_{++}^2+\Xi_{\times\times}^2+4\Xi_{+\times}^2-2\Xi_{++}\Xi_{\times\times}\,,
\ea
\ee
and finally the relationship between ${\Xi_{dt_i}}_{AB}$ and its diagonal
version (\ref{eq:xibar})
\be
\bar{\Xi}{{}_{d_it}}_{AB}=M^{-1}_{AC}{\Xi_{d_it}}_{CD}M_{DB}\,,
\ee
with
\be
M_{AB}=\pa{\ba{cc}
  \frac{F_+\sin\pa{2\psi_t}-F_\times\cos(2\psi_t)}
       {F_+\cos\pa{2\psi_t}+F_\times\sin(2\psi_t)}&
       \frac{F_+\cos\pa{2\psi}+F_\times\sin(2\psi)}
       {F_\times\cos\pa{2\psi}-F_+\sin(2\psi)}\\
       1 & 1
  \ea}\,.
\ee

\section{Degeneracy between {\boldmath$d_L$} and {\boldmath$\iota$}}
\label{app:dL_iota}

\begin{figure}
  \begin{center}
    \includegraphics[width=.48\linewidth]{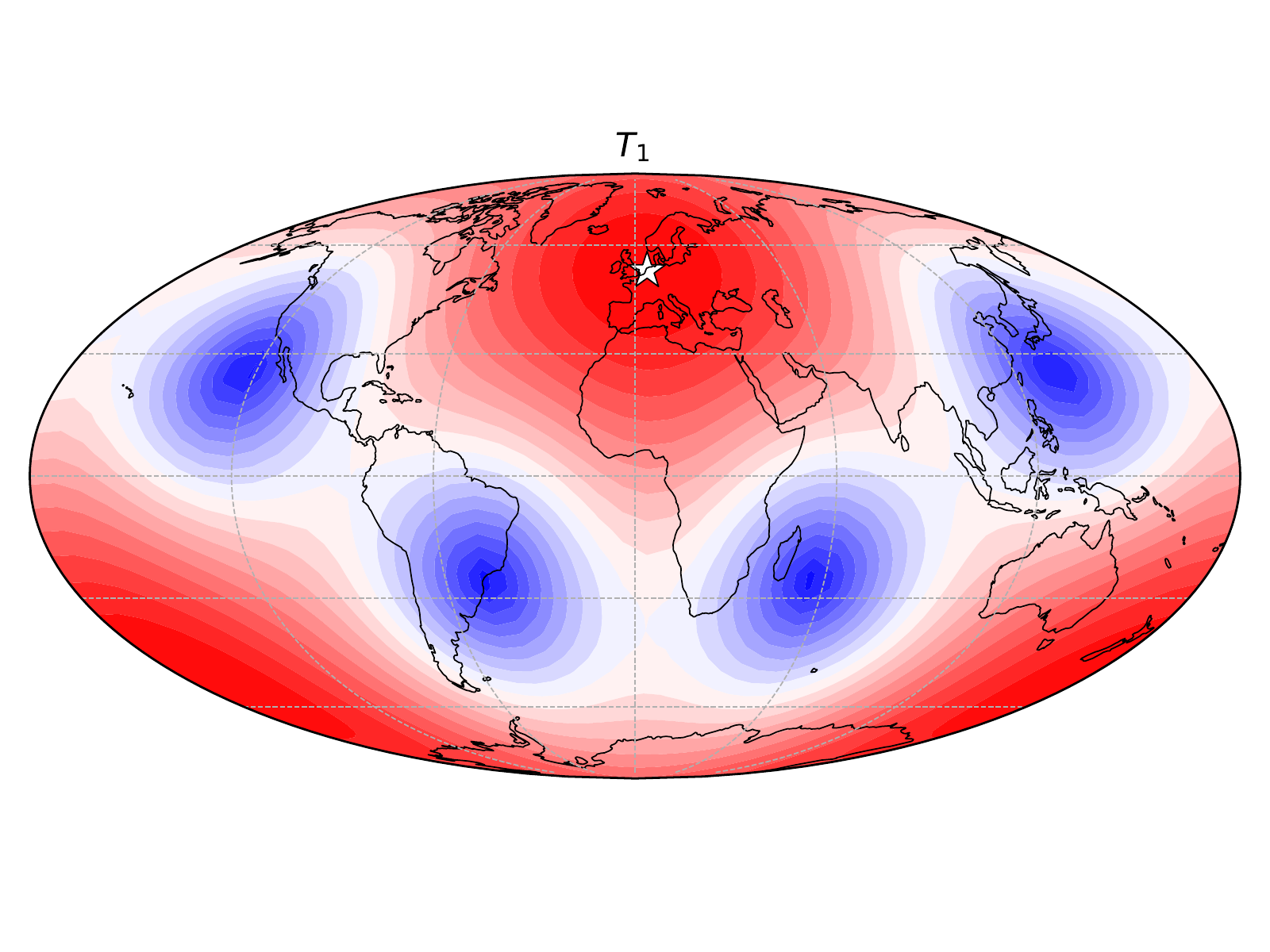}
    \includegraphics[width=.48\linewidth]{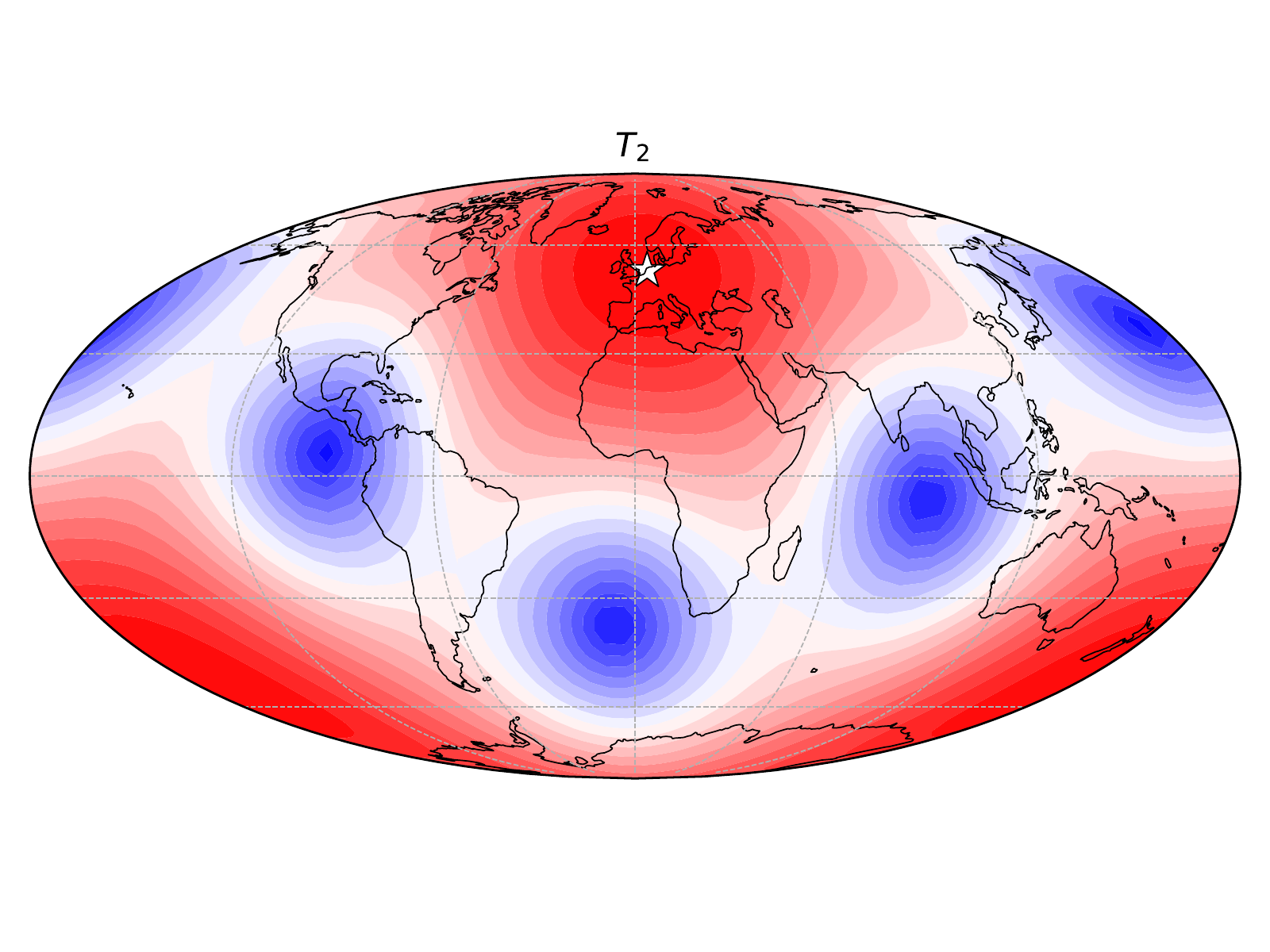}\\
    \includegraphics[width=.48\linewidth]{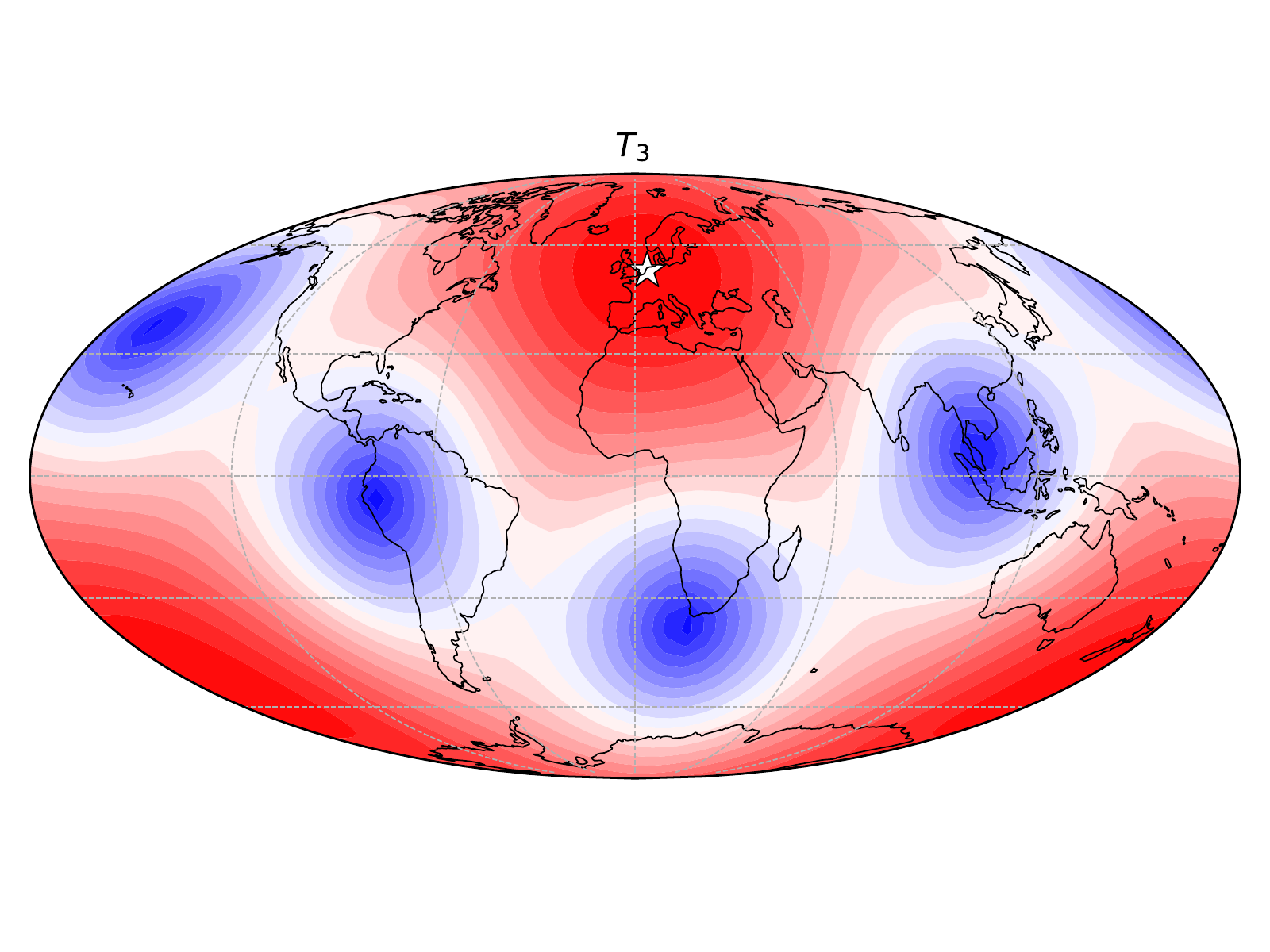}\\
    \caption{Quadratic sum of pattern function $\pa{f^2_++f_\times^2}^{1/2}$ for
      each of the three component of a triangular interferometer with arms
      at 60$^o$. The blind spots of each individual iterferometer lie
    in the plane of the detector.}
  \label{fig:pattern_T}
  \end{center}
\end{figure}

According to the explanation given in Section \ref{sec:method}, see point 3 below
Equation (\ref{eq:cfn}), the presence of bimodality is unavoidable (for $\iota$
sufficiently distant from the value $\pi/2$) for $\epsilon\sim 1$.
Note that the blind zones of the individual interferometers making the ET
are very close together, see Figure \ref{fig:pattern_T}, so that
for those specific source position the response of ET is not too
dissimilar from the response of $L$-shaped detector.

\begin{figure}
  \begin{center}
    \includegraphics[width=.9\linewidth]{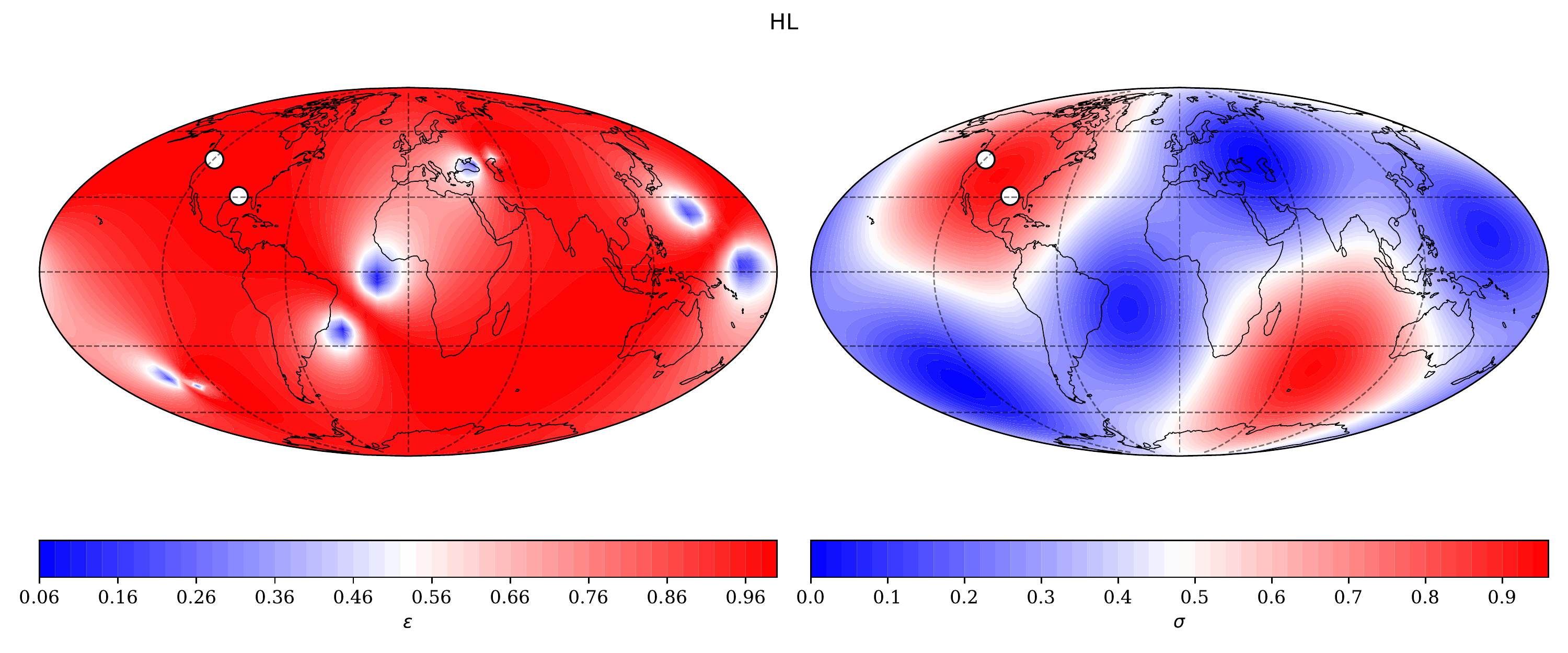}\\
    \includegraphics[width=.9\linewidth]{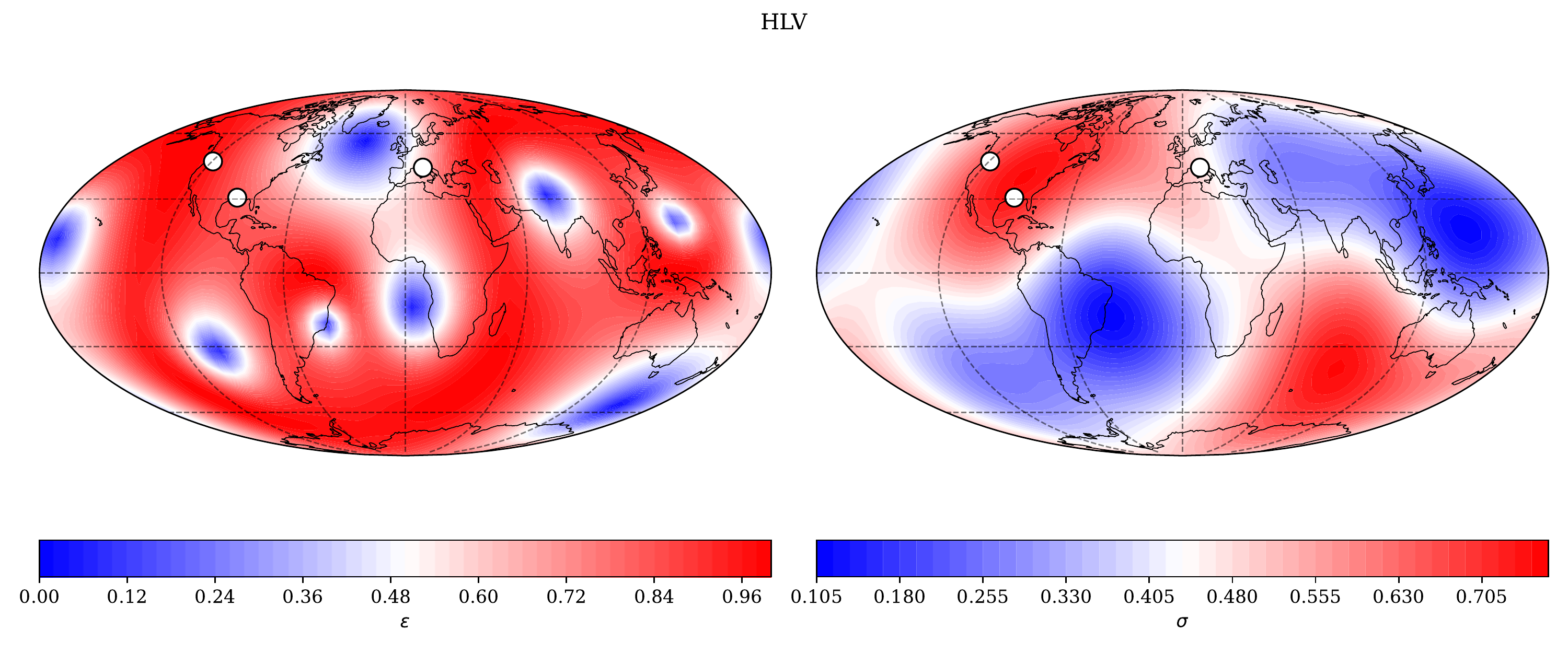}\\
    \includegraphics[width=.9\linewidth]{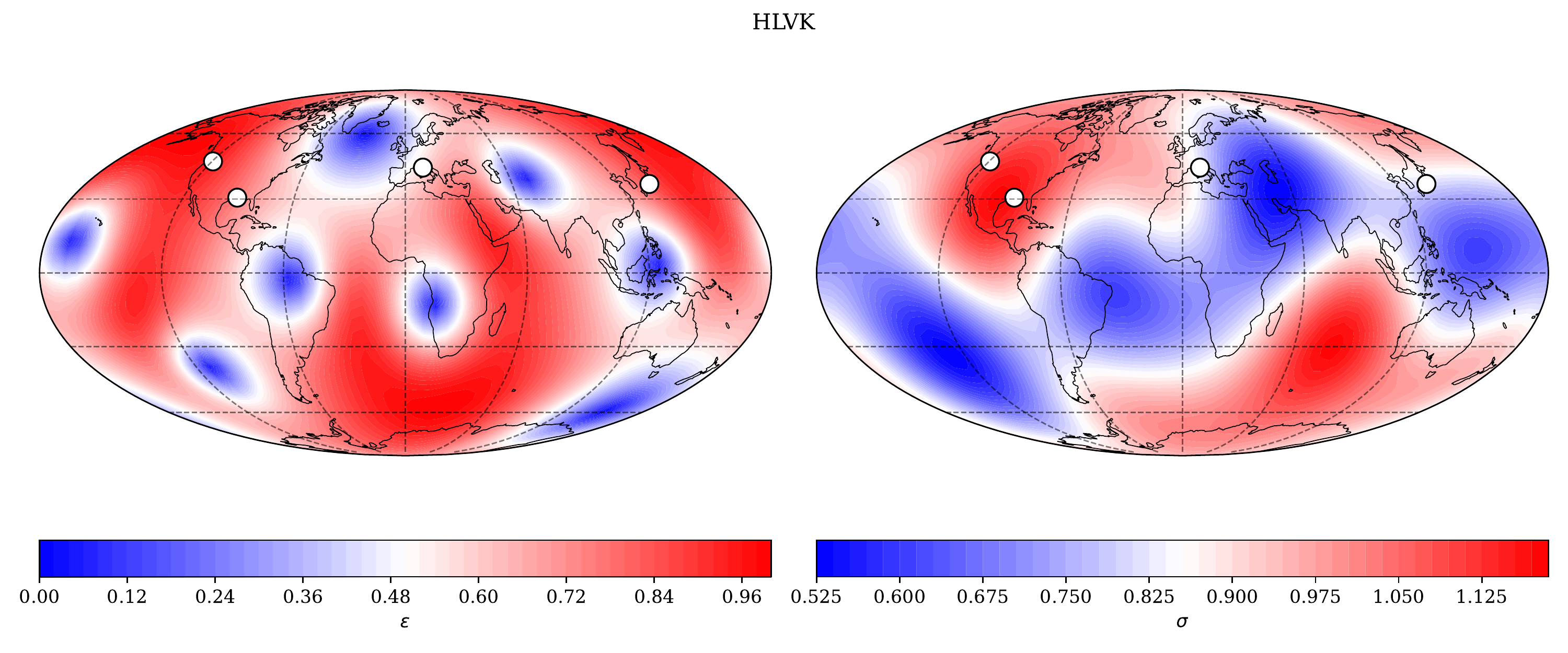}
    \caption{Values of $\sigma$ and $\epsilon$ for various networks of 2G
      detectors, spectral noise curves used are displayed in Figure \ref{fig:2Gnoise}.}
    \label{fig:es_2G}
  \end{center}
\end{figure}

\begin{figure}
  \begin{center}
    \includegraphics[width=.48\linewidth]{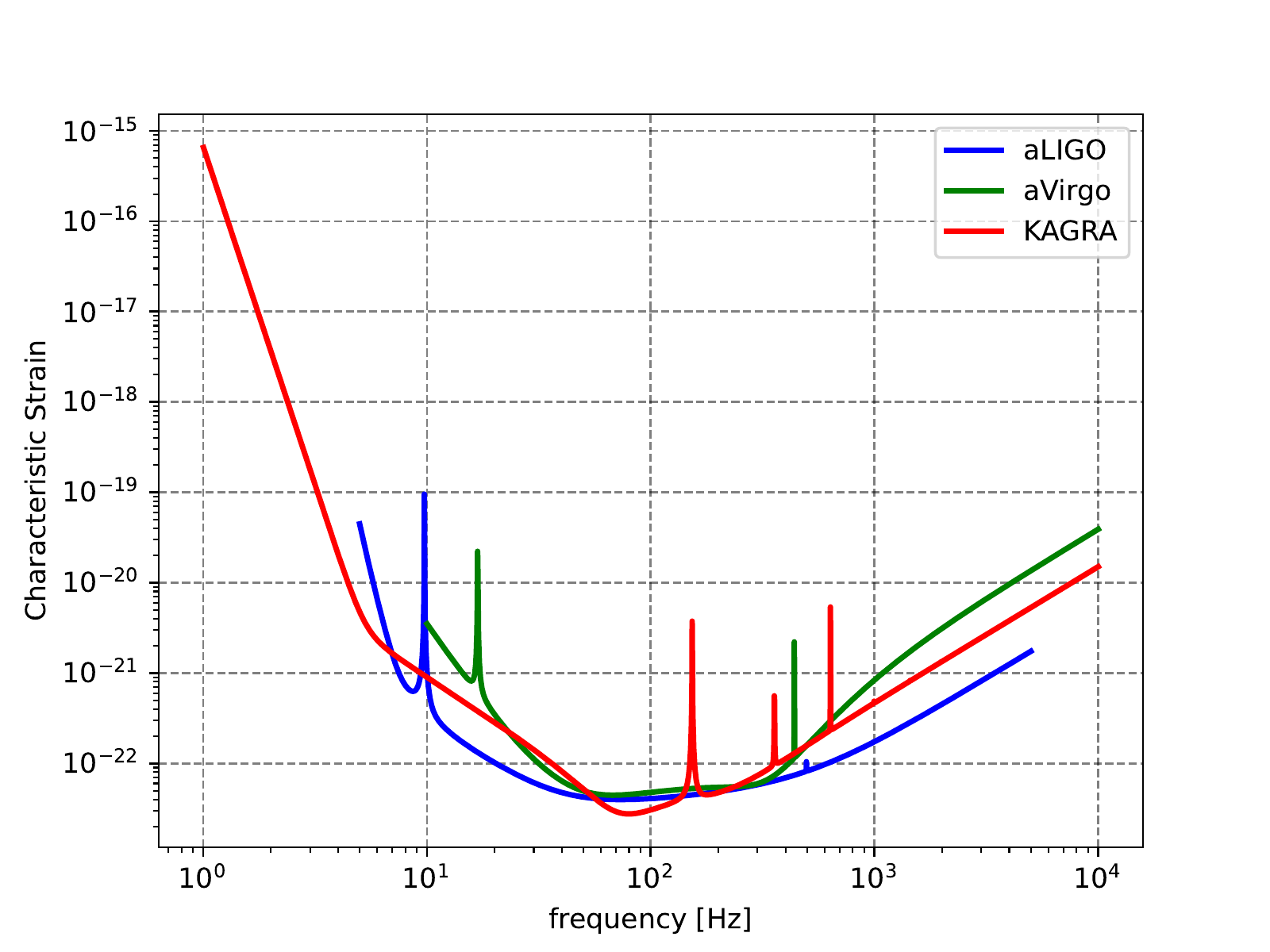}
    \includegraphics[width=.48\linewidth]{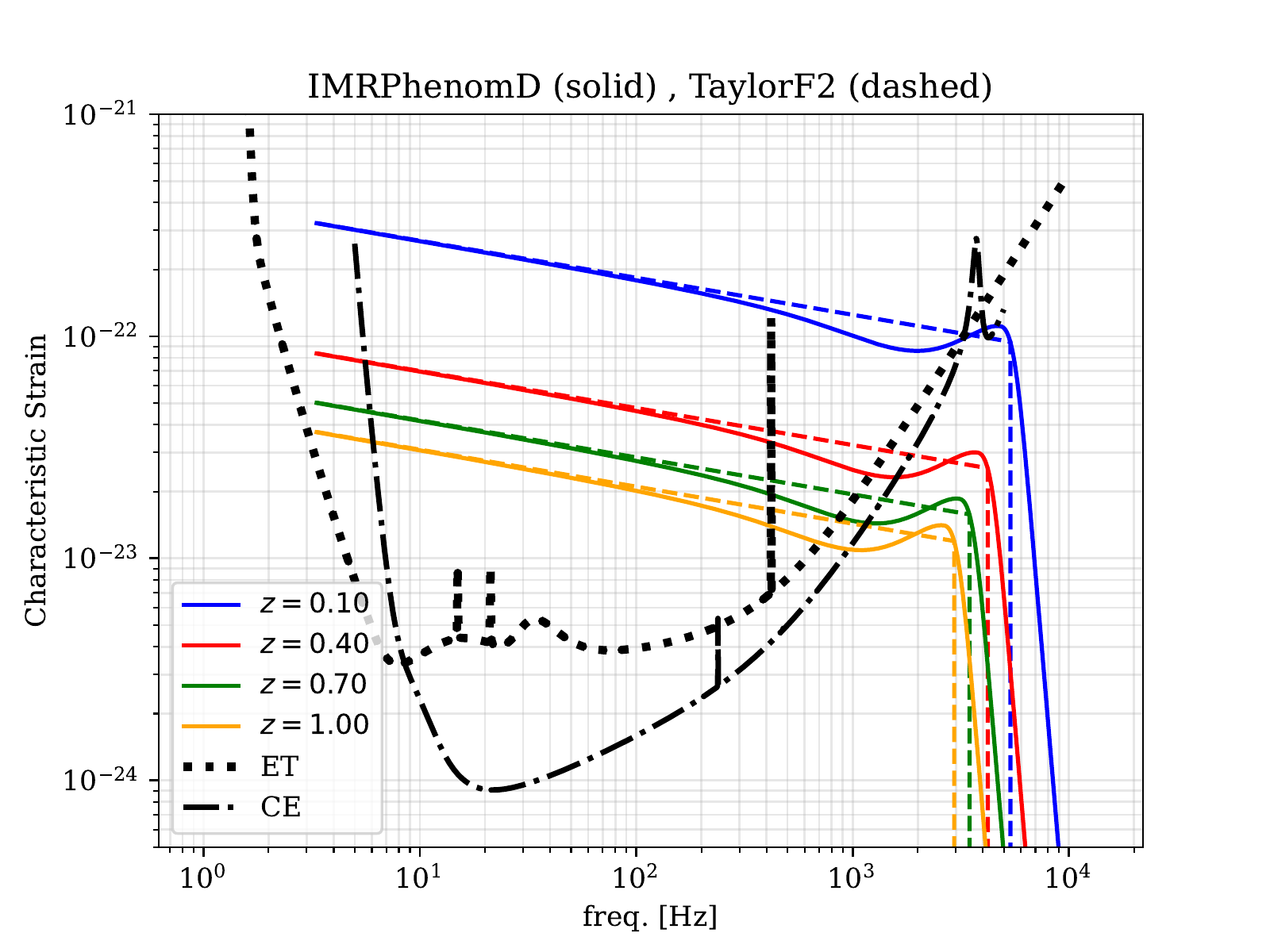}
    \caption{(Left) Characteristic strain $\sqrt{fS_n}$ used to generate maps in
      Figure \ref{fig:es_2G}, from \cite{2gnoise}. (Right) Examples of TaylorF2
      (dashed) and IMRPhenomD (solid) waveforms for total mass $3M_\odot$ and equal binary component masses.}
    \label{fig:2Gnoise}
  \end{center}
\end{figure}

Taking e.g. the case of 2G detectors, the two LIGOs are almost perfectly
aligned,
making $\epsilon\sim 1$ for most of the sky, the addition of Virgo and
KAGRA will not change drastically the situation as they have larger
spectral noise sensitivity, see Figure \ref{fig:es_2G} for the $\sigma$ and
$\epsilon$ maps and Figure \ref{fig:2Gnoise} for 2G detector design spectral
noise sensitivities.

\section{Relation to dominant polarisation frame}
\label{app:dpf}

In this work we relied on the parameterization leading to expression
(\ref{eq:cfn}) that we borrowed from \cite{Cutler:1994ys}.
In \cite{Klimenko:2005xv} a similar parameterization has been
introduced, identifying the \emph{dominant polarisation frame},
i.e., the \emph{radiation} frame for which the detector network is maximally
sensitive to the $+$ polarisation, by using the general property that different
radiation frames are related by a shift in the polarisation angle, i.e. a
rotation around the propagation direction.
The detector signal is then parameterized in \cite{Klimenko:2005xv}
as
\be
h_{det}=g_{Ref.\text{\cite{Klimenko:2005xv}}}\pa{h_++\epsilon_{Ref.\text{\cite{Klimenko:2005xv}}} h_\times}
\,,
\ee
leading to the following mapping of these coefficients into our
$\sigma,\epsilon$ a
\be
\ba{rcl}
g_{Ref.\text{\cite{Klimenko:2005xv}}}&=&\sigma\sqrt{1+\epsilon}\,,\\
\epsilon_{Ref.\text{\cite{Klimenko:2005xv}}}&=&\ds\sqrt{\frac{1-\epsilon}{1+\epsilon}}\,.
\ea
\ee

\section{3G luminosity distance uncertainty comparison with 2G detectors and standard candles}
\label{app:pan}

\begin{figure}
  \includegraphics[width=.49\linewidth]{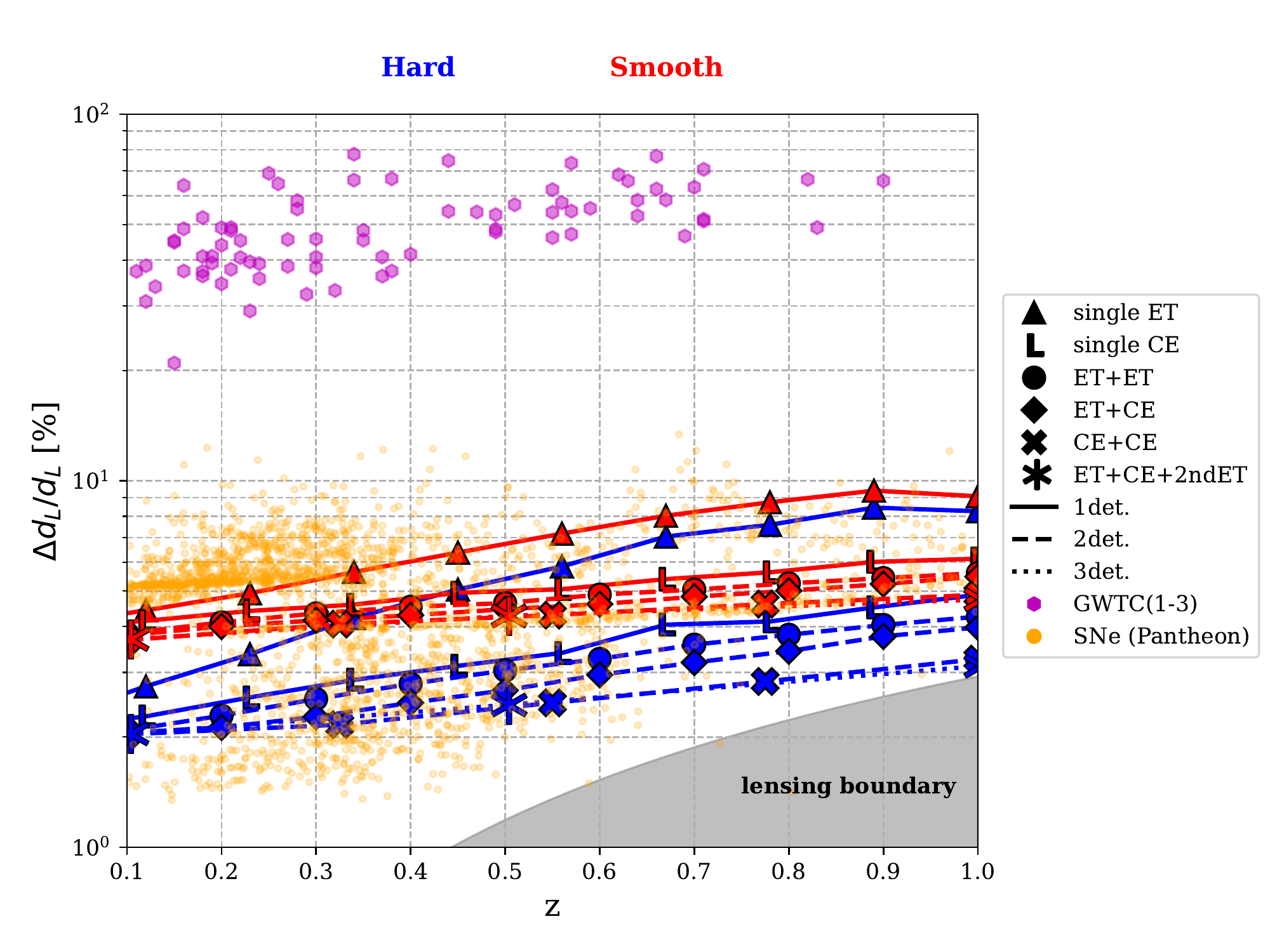}
  \includegraphics[width=.49\linewidth]{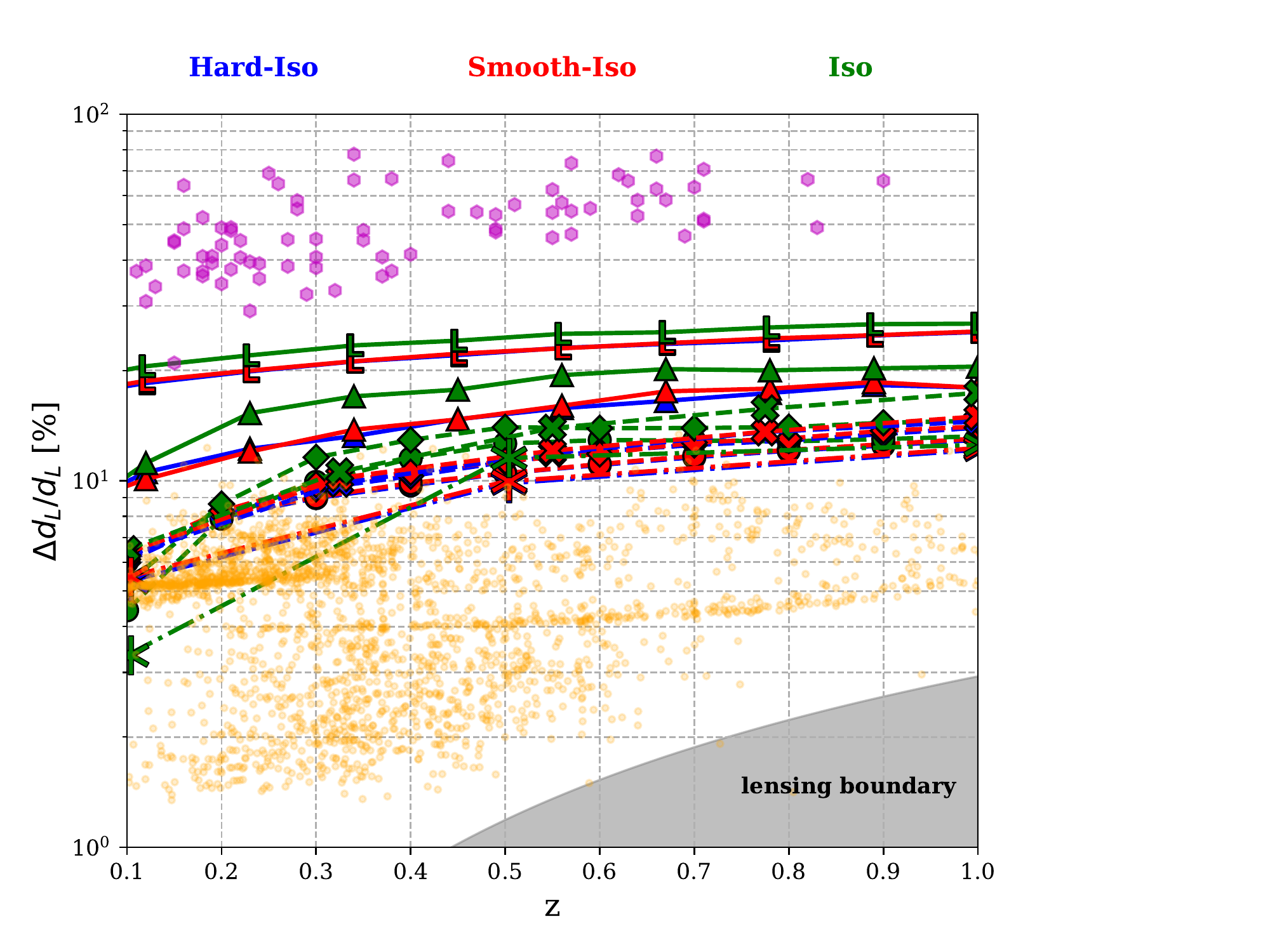}
  \caption{Same as in Figure \ref{fig:err_iotad_net}, with uncertainties
    in luminosity distance of 2G detections from \cite{LIGOScientific:2018mvr,LIGOScientific:2020ibl,LIGOScientific:2021djp} and standard sirens from \cite{Pan-STARRS1:2017jku} added.}
  \label{fig:err_iotad_net_pan}
\end{figure}

As a comparison with luminosity distance uncertainties obtained with 2G GW
detectors and standard candles, we report in Figure \ref{fig:err_iotad_net_pan} 
luminosity distance uncertainties from our 3G projections superimposed with
the catalogues in \cite{LIGOScientific:2018mvr,LIGOScientific:2020ibl,LIGOScientific:2021djp}
and \cite{Pan-STARRS1:2017jku}.

\begin{figure}
  \includegraphics[width=.45\linewidth]{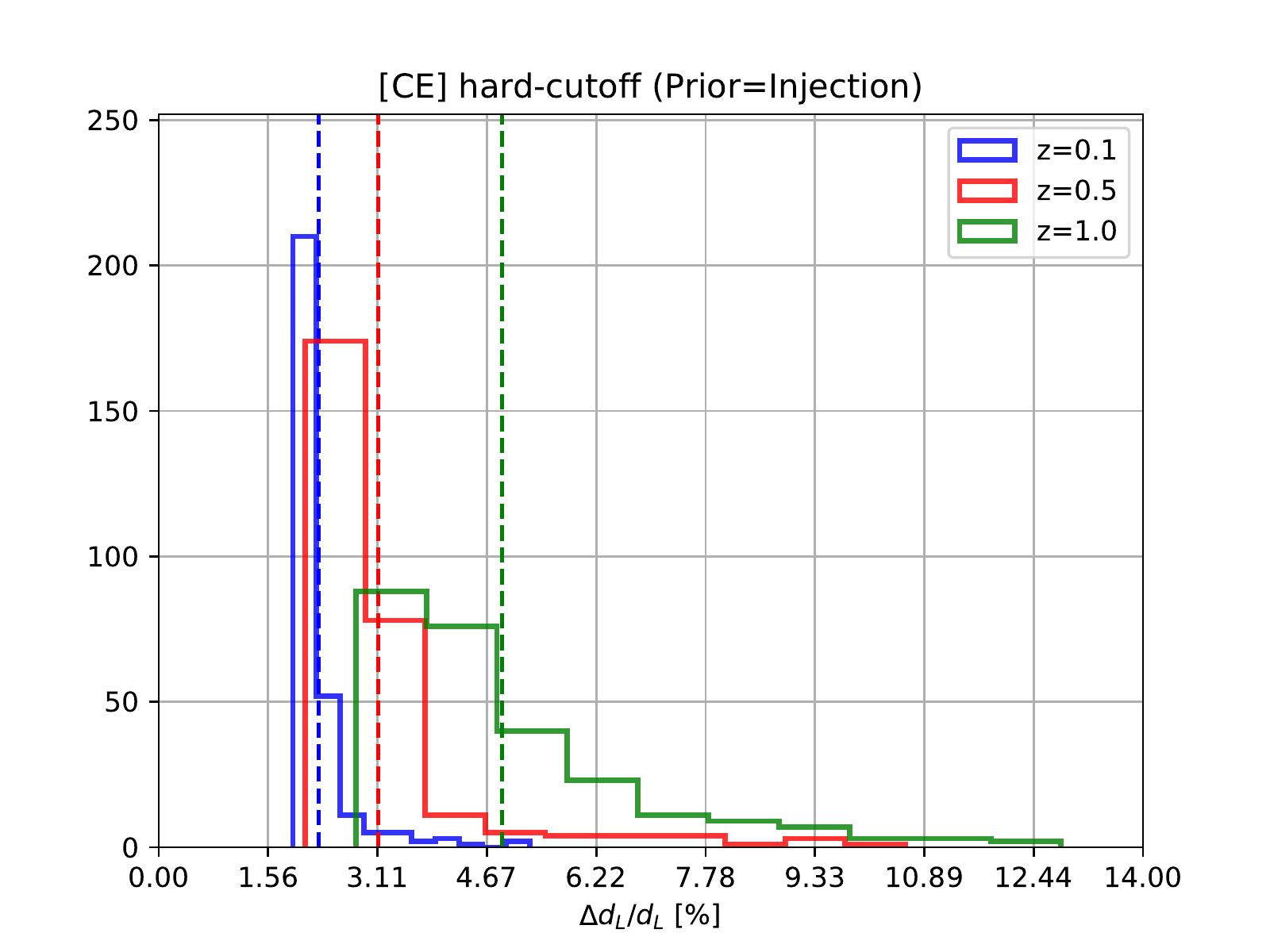}
  \includegraphics[width=.45\linewidth]{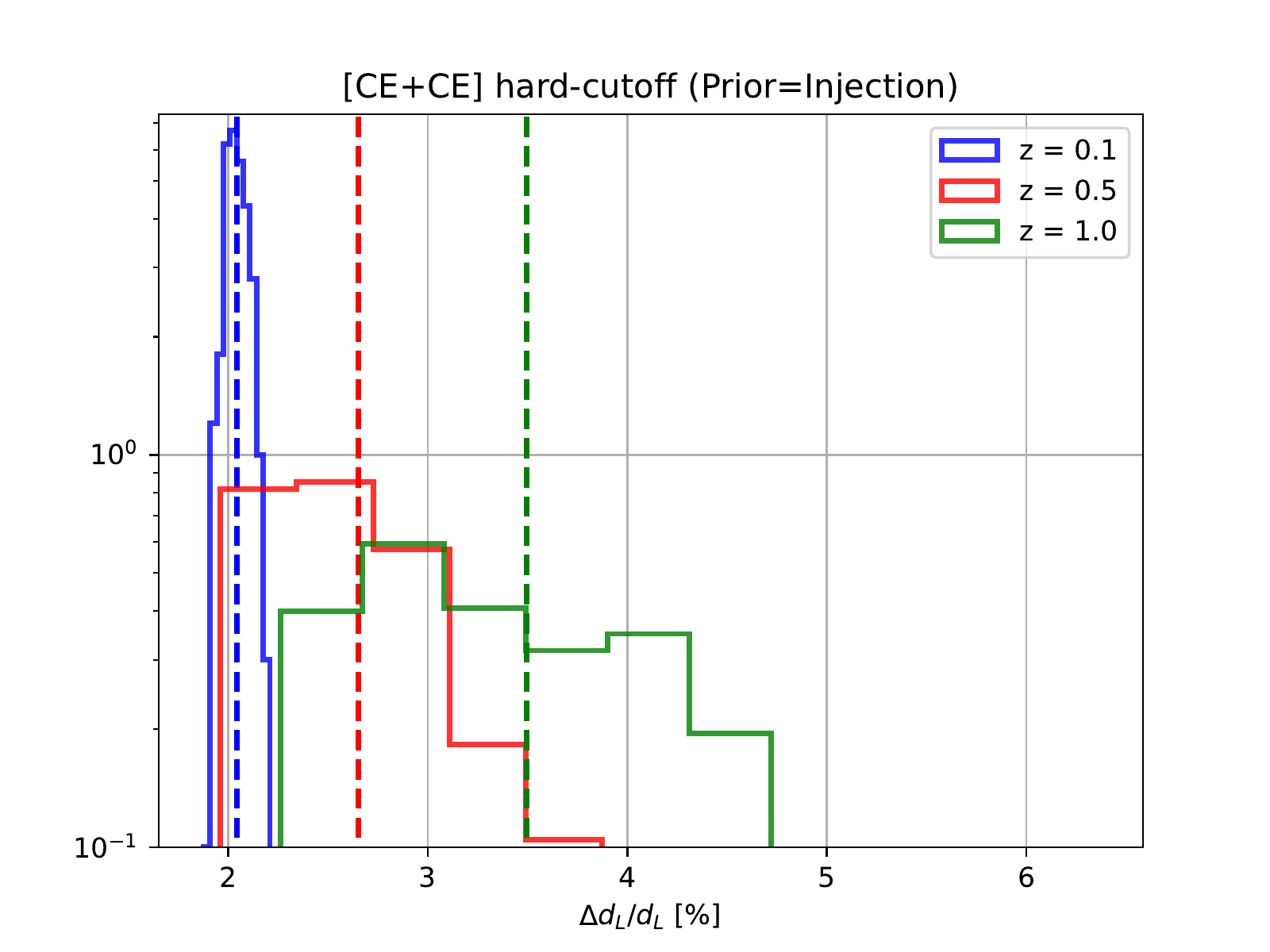}\\
  \includegraphics[width=.45\linewidth]{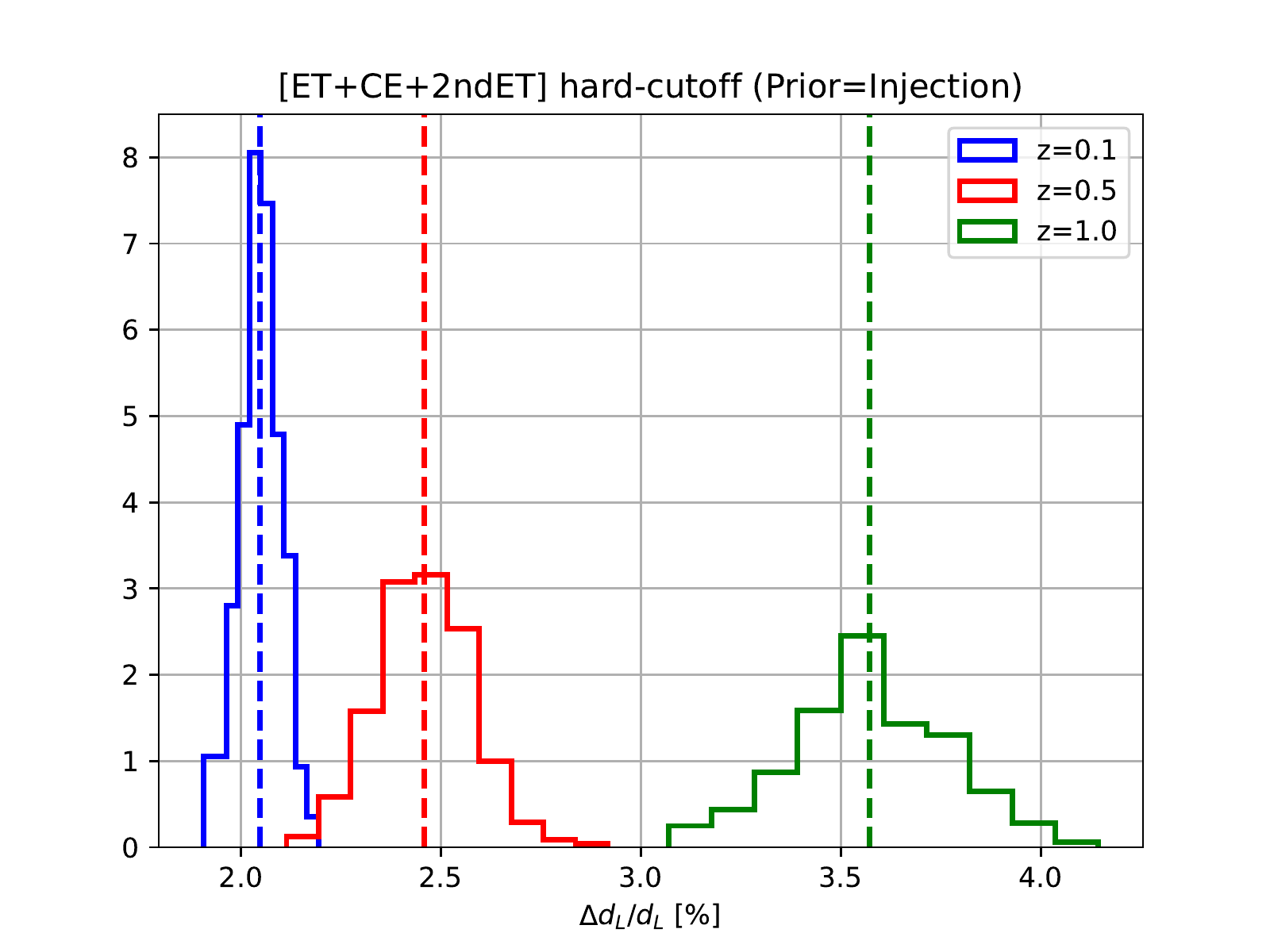}
  \caption{Histograms representing the dispersions of $\Delta d_L/d_L$ measurements for sample 1,2 and 3 detector case case of Figures \ref{fig:err_iotad_net}. \emph{Hard-cutoff} refers to the $\iota$ distributions of the injections
      as per Figure \ref{fig:iotad_net}.}
  \label{fig:disp_dl}
\end{figure}

Finally we report here the sky distribution of the injections used for the
three detector analysis of Section \ref{ssec:three_3G} (ET+CE+ET), which
highlight the location selection effect of the SNR threshold at large distances.
\begin{figure}
  \includegraphics[width=.45\linewidth]{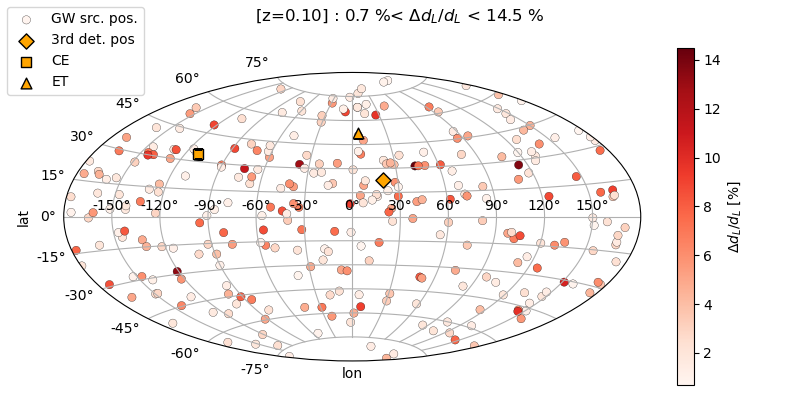}
  \includegraphics[width=.45\linewidth]{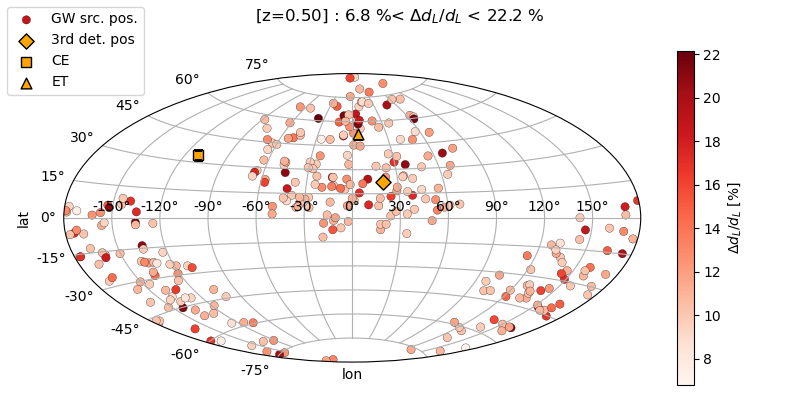}
  \includegraphics[width=.45\linewidth]{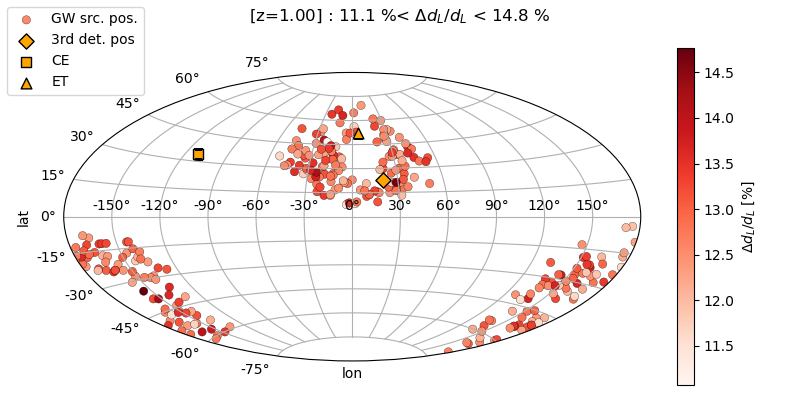}
  \caption{Sky distribution of above threshold events for the 3 detector
      network (ET+CE+ET) of Section \ref{ssec:three_3G}, showing how
      threshold cut selects source sky regions.}
  \label{fig:three_3Giso}
\end{figure}

\begin{acknowledgments}
  The authors thank Viviane Alfradique and Miguel Quartin for useful discussions.
  JMSdS is supported  by the Coordena\c{c}\~ao de Aperfei\c{c}oamento de Pessoal de
  N\'\i vel
  Superior (CAPES)  -- Graduate Research Fellowship/Code 001.
  The work of RS is partially supported by CNPq under grant 310165/2021-0 and
  by FAPESP grants 2021/14335-0 and 2022/06350-2.
  The authors thank the High Performance Computing Center (NPAD) at UFRN for
  providing computational resources that made the present work possible.
\end{acknowledgments}

\end{document}